\documentclass[numbers,webpdf,imaiai]{ima-authoring-template}

\usepackage{cite}

\usepackage{verbatim}

\graphicspath{{figures/}}

\usepackage{booktabs}
\usepackage{calc}
\usepackage{xcolor}
\usepackage{dsfont}
\label{comments}
\usepackage{amsmath}
\usepackage{amssymb}
\usepackage{amsthm}
\usepackage{mathtools}
\usepackage{tabularx}
\newcommand{\minus}{\scalebox{0.75}[1.0]{$-$}}

\DeclareMathOperator*{\argmin}{arg\,min}
\usepackage{multirow}

\usepackage[switch]{lineno}

\usepackage[inline]{enumitem}

\theoremstyle{thmstyletwo}
\newtheorem{definition}{Definition}
\newtheorem{proposition}{Proposition}
\newtheorem{theorem}{Theorem}
\newtheorem{corollary}{Corollary}
\newtheorem{lemma}{Lemma}
\newtheorem{remark}{Remark}

\newtheorem{innercustomgeneric}{\customgenericname}
\providecommand{\customgenericname}{}
\newcommand{\newcustomtheorem}[2]{%
	\newenvironment{#1}[1]
	{%
		\renewcommand\customgenericname{#2}%
		\renewcommand\theinnercustomgeneric{##1}%
		\innercustomgeneric
	}
	{\endinnercustomgeneric}
}

\newcustomtheorem{customthm}{Theorem}
\newcustomtheorem{customlemma}{Lemma}
\newcustomtheorem{customDef}{Definition}

\usepackage{stackengine}

\usepackage{booktabs}

\definecolor{myBlue}{RGB}{219, 48, 122}

\begin{document}
	
	\DOI{DOI HERE}
	\copyrightyear{2021}
	\vol{00}
	\pubyear{2021}
	\access{Advance Access Publication Date: Day Month Year}
	\appnotes{Paper}
	\copyrightstatement{Published by Oxford University Press on behalf of the Institute of Mathematics and its Applications. All rights reserved.}
	\firstpage{1}
	
	
	\title[Phase Retrieval for Radar]{Phase Retrieval for Radar Waveform Design}
	
	\author{Samuel Pinilla
		\address{\orgname{Universidad Industrial de Santander}, \orgaddress{Bucaramanga, \postcode{680002}, \state{Santander}, \country{Colombia}}}}
	\author{Kumar Vijay Mishra* and Brian M. Sadler
		\address{\orgname{United States DEVCOM Army Research Laboratory}, \orgaddress{\street{Adelphi}, \postcode{20783}, \state{MD}, \country{USA}}}}
	\author{Henry Arguello
		\address{\orgname{Universidad Industrial de Santander}, \orgaddress{Bucaramanga, \postcode{680002}, \state{Santander}, \country{Colombia}}}}

	\authormark{Pinilla et al.}
	
	\corresp[*]{Corresponding author: \href{email:kvm@ieee.org}{kvm@ieee.org}}
	
	\received{Date}{0}{Year}
	

	\abstract{
		The ability of a radar to discriminate in both range and Doppler velocity is completely characterized by the \textit{ambiguity function} (AF) of its transmit waveform. Mathematically, it is obtained by correlating the waveform with its Doppler-shifted and delayed replicas. We consider the inverse problem of designing a radar transmit waveform that satisfies the specified AF magnitude. This process may be viewed as a signal reconstruction with some variation of phase retrieval methods. We provide a trust-region algorithm that minimizes a smoothed non-convex least-squares objective function to iteratively recover the underlying signal-of-interest for either time- or band-limited support. The method first approximates the signal using an iterative spectral algorithm and then refines the attained initialization based on a sequence of gradient iterations. Our theoretical analysis shows that unique signal reconstruction is possible using signal samples no more than thrice the number of signal frequencies or time samples. Numerical experiments demonstrate that our method recovers both time- and band-limited signals from sparsely and randomly sampled, noisy, and noiseless AFs. 
		}
	
	\keywords{Ambiguity function, band-limited signals, non-convex optimization, phase retrieval, radar waveforms.}
	
	\maketitle
	
	\section{Introduction}
	The mathematical theory of radar and sonar sensing widely employs cross-correlation technique for signal detection and parameter estimation. The transmit waveform correlated with its Doppler-shifted and delayed replicas yields the \textit{ambiguity function} (AF) that plays a key role in the detection and resolution capabilities of the system \citep{levanon1988radar,peebles1998radar,glisson1970sonar}. Through the AF, the transmit waveform enters into the performance analyses related to detection, target parameter accuracy, and resolution of multiple closely-spaced targets. 
	
	The AF was first introduced by Ville \citep{ville1948theorie} and its significance as a signal design metric in mathematical radar theory is credited to Woodward \citep{woodward1965probability,woodward1967radar}, later expounded in detail by Siebert \citep{siebert1956radar}. A radar system radiates a known waveform - characterized by its amplitude, frequency and polarization state - toward the targets-of-interest. When the radiated wave from the radar interacts with moving objects, all three characteristics of the scattered wave change. The detection and estimation of these changes helps to infer the targets' size/shape, location relative to the radar, and radial Doppler velocity. Woodward observed that a suitable radar transmit signal should discriminate between the returning echoes from different targets and the AF provides a measure of the uncertainty in such a discrimination. 
	
	Woodward \citep{woodward1965probability} defined a mean-squared error metric between a known waveform $x(t)$ and its replica with frequency-shift $f$ and time-delay $\tau$ as
	\begin{align*}
		\Omega(\tau, f) = \int_{\mathbb{R}} \left\lvert x(t) - x(t-\tau)e^{-\mathrm{i}2 \pi f t} \right\rvert^{2}dt.
	\end{align*}
	In the expansion of this error metric, the only term that depends on the parameters is the inner product between the original waveform and its time-delayed/frequency-shifted version. The magnitude-squared inner product yields the narrow-band AF
	\begin{align}
		A(\tau, f) = \left \lvert \int_{\mathbb{R}} x(t)x^{*}(t-\tau)e^{-\mathrm{i}2 \pi f t} dt \right \rvert^{2}.
		\label{eq:narrow}
	\end{align}
	The AF is not uniquely defined \citep{baylis2016myths,abramovich2008bounds}, including in the works of Woodward \citep{woodward1967radar}. Later works have generalized this definition to handle larger bandwidth signals \citep{swick1966ambiguity,speiser1967wide}, long duration signals \citep{swick1968wideband}, volumetric scatterers \citep{doviak2006doppler,george2010implementation}, targets with high velocity \citep{gassner1967note,boashash2015time}, and angular domains \citep{de1967extension,san2007mimo,ilioudis2019generalized}. The AF is also used in other application areas such as optics \citep{guigay1978ambiguity,bartelt1984misfocus,ojeda1988ambiguity}. It has been implied that the AF is related to the uncertainty principle of quantum theory \citep{auslander1984characterizing}. In this paper, we restrict the discussion to the above-mentioned classical narrowband definition in the context of radar remote sensing.
	
	The AF represents the noise-free output of a matched filter at the receiver. It describes the distortion caused by the range and/or Doppler of a target when compared to a reference target of equal radar cross-section with no delay or Doppler shift. The ambiguity function evaluated at $\left(\tau, f\right)=(0,0)$ is equal to the matched filter output that correlates perfectly to the signal reflected from the target of interest \citep{mahafza2005radar}. In other words, returns from the nominal target are located at the origin of the ambiguity function. Thus, the ambiguity function at nonzero $\tau$ and $f$ represents returns from some range and Doppler different from those for the nominal target. Therefore, the AF may be used to scale the signal-to-noise ratio (SNR) in detection analyses when the target and detector are mismatched \citep{soliman1988spread}. 
	
	The radar AF may be used as an aid to select suitable radar waveforms. An ideal AF would be non-zero only at a single point \citep[Chapter 3]{jankiraman2007design}, although the probability of the target lying in such a precise response region would be approaching zero \citep{rihaczek1996principles}. In the absence of an ideal AF, significant theoretical efforts have been devoted to the problem of finding functions of $\tau$ and $f$ that are legitimate as AFs and have suitable radar performance. The resulting \textit{waveform design} has traditionally focused on achieving AFs that approximate a `thumbtack' shape with a sharp central spike and low sidelobes in the $(\tau$-$f)$ plane. Classic all-purpose parameterized and easily generated conventional radar signals include a narrowband pulse, linear frequency modulated (LFM) waveform, and stepped frequency pulse train \citep{jankiraman2007design}. Depending on the application, the AF may also be expressly shaped \citep{delong1967design,stutt1968best}. For example, when the returning echo comprises multiple scatterers with varying strengths, the AFs of all targets are superimposed and shifted only by the relative differences in range and Doppler velocities. It is likely that the sidelobe of a stronger scatterer masks the main lobe of the weaker scatterer thereby hindering detection of the latter. In that case, AF shaping is beneficial to avoid such masking in specific regions of the $(\tau$-$f)$ plane. 
	
	In this paper, we focus on the inverse problem of synthesizing waveforms that satisfy a given, desired, or specifically shaped ambiguity function. This has been studied extensively in the past, beginning with the least-squares approximations of Wilcox \citep{wilcox1960synthesis} and Sussman \citep{sussman1962least} over the entire $\tau$-$f$ plane to more recent formulations \citep{arikan1990time,sen2009adaptive} for specific regions in the $\tau$-$f$ plane. 
	
	In practice, these problems employ discrete formulations. Assume the sampling periods in time and Doppler domains are $\Delta t$ and $\Delta f$, respectively, such that the discrete indices are $p=\tau/\Delta t$ and $k=f/\Delta f$. The AF of a discrete signal $\mathbf{x}\in \mathbb{C}^{N}$ is a map $\mathbb{C}^{N}\rightarrow \mathbb{R}_{+}^{N\times N}$ is the matrix $\mathbf{A}$ with $(p,k)$-th element 
    \begin{equation}
	    \mathbf{A}[p,k]:= \left| \sum_{n=0}^{N-1} \mathbf{x}[n] \overline{\mathbf{x}[n-p]}e^{-\mathrm{i}2\pi nk/N} \right|^2,\;p,k=0,\cdots,N-1.
	    \label{eq:Ambiguity}
    \end{equation}
	The AF of the resulting waveform must satisfy the following properties \citep[Chapter 3]{jankiraman2007design}:
	\begin{description}
		\item[P1]{\hspace{0.5em} The maximum value, say $M$, of the AF occurs at $(p,k)=(0,0)$, i.e., 
			\begin{align}
				\max {\mathbf{A}[p,k]} &= \mathbf{A}[0,0] = M, \nonumber\\
				\mathbf{A}[p,k] &\leq A[0,0].
				\label{eq:property1}
		\end{align}}
		\item[P2]{\hspace{0.5em}The total volume under the AF is the constant
			$\sum_p\sum_k \lvert \mathbf{A}[p,k] \rvert^{2}$.}
		\item[P3]{\hspace{0.5em}AF is symmetric in both $p$ and $k$, i.e., 
			$
			\mathbf{A}[p,k] = \mathbf{A}[-p,-k]$.}
		\item[P4]{\hspace{0.5em}If the AF of $\mathbf{x}[n]$ is $\mathbf{A}[p,k]$, then the AF of $\mathbf{x}[n]\exp(i\pi s n^{2}/N)$ is 
			$\mathbf{A}[p, k - sp]$.} 
	\end{description}
	For normalized signals ($M = 1$), it follows from \textbf{P1} that the maximum value of $\mathbf{A}[p,k]$ is unity at the origin. Further, \textbf{P2} states that the volume of $\mathbf{A}[p,k]$ for normalized signals is also unity. Therefore, if $\mathbf{A}[p,k]$ is squeezed to a narrow peak near to the origin, then that peak cannot exceed unity. The property \textbf{P3} indicates that the AF is symmetrical with respect to the origin. Finally, \textbf{P4} implies that multiplying the envelope of any signal by a quadratic phase (linear frequency) shears the AF. We refer the reader to \citep{levanon1988radar} for the proofs of these properties.
	
	The inverse waveform design problem exploits these properties to find a pulse that best fits the desired AF. For instance, \citep{de1970signals} uses boundedness (\textbf{P1}) to show that when the (continuous) AF is bounded by a \textit{Hermite function}, i.e. a polynomial times the standard normal distribution with variance $\sigma^{2}=1/(2\pi)$, then the unknown signal is also a Hermite function where the polynomial is found from its AF by comparing the coefficients. The non-convex optimization employed in \citep{sen2009adaptive} employs the volume property (\textbf{P2}) to adaptively construct a Hermite function that best approximates the volume of a desired AF. More recent works such as \citep{alhujaili2019quartic,nuss2020frequency,rodriguez2020improved} have proposed to numerically design the signal to improve the delay resolution of the AF in order to better determine the position of objects.

	\begin{figure}[t!]
		\centering
		\includegraphics[width=0.6\textwidth]{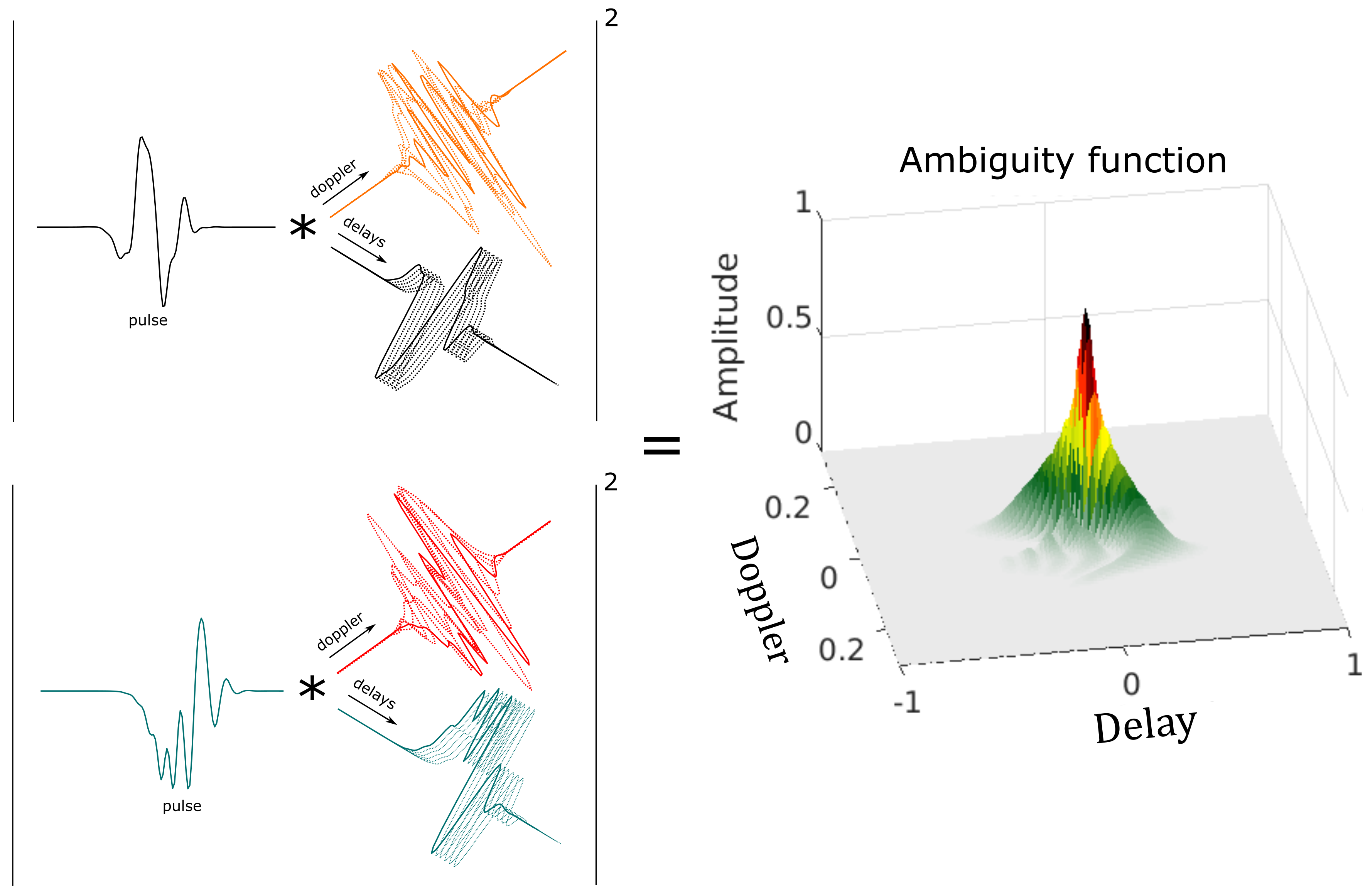}
		\caption{Illustration of multiple signals with trivial ambiguities mapping to the same AF. On the left, two different signals on top and bottom are correlated with their Doppler- and delay-shifted replicas followed by the magnitude-squared operation.}
		\label{fig:intro}
	\end{figure}
	It follows from \eqref{eq:Ambiguity} that $\mathbf{A}[p,k]$ is a phaseless mapping. Note that \textbf{P1}-\textbf{P4} imply that the following \textit{trivial ambiguities} or transformed versions of $\mathbf{x}[n]$ lead to the same $\mathbf{A}[p,k]$. 
    \begin{description}
	    \item[T1]{\hspace{0.5em}Rotated signal $e^{\mathrm{i}\phi}\mathbf{x}$ for some $\phi\in \mathbb{R}$.}
	    \item[T2]{\hspace{0.5em}Translated signal $\mathbf{x}[n-a]$ for some $a\in \mathbb{R}$.}
	    \item[T3]{\hspace{0.5em}Conjugated and reflected $\overline{\mathbf{x}}[-n]$.}
	    \item[T4]{\hspace{0.5em}Scaled signal $e^{\mathrm{i}bn}\mathbf{x}[n]$ for some $b\in \mathbb{R}$.}
	\label{eq:ambiguities1}
    \end{description}
	Fig.~\ref{fig:intro} illustrates how different signals with these transformations map into the same AF. It follows that recovering a signal from a given AF, such that the transformed versions result in the same AF, requires a phase retrieval (PR) step from the phaseless function $\mathbf{A}[p,k]$ \citep{jaming2010phase}. This is similar to the PR problem encountered in optics when a square-law intensity detector is used \citep{candes2015phase}. 
	In this work, we focus on a PR approach to AF-based waveform design.

	\subsection{Prior art}
	\label{sec:radarPR}
	The PR-based waveform design using a given AF is relatively unexamined when compared to AF-shaping optimization approaches \citep{alhujaili2019quartic}. The earliest instance is the seminal work of Rudolf De Buda \citep{de1970signals} that showed recovery of Hermite functions from their (phaseless) AF. A few later works attempted to design code sequences with impulse-like correlations using the cyclic algorithm-new (CAN) and its variants (see \citep{he2012waveform} for a summary of this line of research). These methods are closely related to the classical least-squares recovery suggested by Sussman \citep{sussman1962least} and follow the alternating projection method detailed in the classical Gerchberg-Saxton algorithm (GSA) \citep{gerchberg1972practical} for PR. However, the convergence of GSA is usually to a local minimum and, therefore, its recovery abilities are limited (even in a noiseless setting). Moreover, the CAN-based waveform design in \citep{he2012waveform} does not produce a signal with specified AF but rather focuses on minimizing various sidelobe metrics (e.g., peak or integrated sidelobe levels) in the signal correlation.

	More recently, \citep{jaming2010phase,jaming2014uniqueness} applied the PR approach to the problem by exploiting the relationship between the AF and fractional Fourier transform (FrFT) \citep{almeida1994fractional}. 
	Specifically, the Fourier magnitude of the product of the unknown signal with a conjugate time-shifted version of itself for several different shifts is equivalent to a frequency rotation using FrFT \citep{jaming2010phase}. Mathematically, this equivalent definition is obtained for a time-shift $\tau=0$ and frequency $\zeta=f\sin(\alpha)$, where $\alpha$ is the rotation angle. 
	It was shown in \citep{jaming2010phase,jaming2014uniqueness} that Hermite function and rectangular pulse trains are uniquely identified from their respective FrFT-based AF. Extending this FrFT formulation 
	to functions other than these two classes of signals is an open problem of high interest \citep{arikan1990time,sen2009adaptive}. Further, \citep{de1967extension,jaming2010phase,jaming2014uniqueness} do not provide a computable procedure to estimate the signal from its AF. In this paper, we propose a non-convex PR algorithm that addresses these drawbacks.

	\begin{table}[h!]
		\centering
		\caption{Comparison with Prior Art}
		\begin{tabular}{p{1.3cm} p{7.8cm} p{5.0cm}}
			\hline \\
			\multirow{2}{*}{\textbf{PR model}} \vspace{0.5em}& 		\multirow{2}{*}{\textbf{Measurements}} \vspace{0.5em}& \textbf{Uniqueness}\hrule\\& & \textbf{Algorithm} \vspace{0.5em}\\
			\hline 
			\multirow{3}{*}{STFT} & \multirow{2}{*}{$\displaystyle \mathbf{A}[p,k]:= \left \lvert \sum_{n=0}^{N-1} \mathbf{x}[n]\mathbf{g}[pL-n]e^{\frac{-2\pi \mathrm{i}kn}{N}}\right \rvert^{2}$, $L<N$}
			& Uniqueness (up to a global) phase for almost all signals for some $L$s and non-vanishing signal $\mathbf{x}$ and window function $\mathbf{g}$ \citep{jaganathan2016stft}; Uniqueness if the first $L$ samples of $\mathbf{x}$ are known a priori for some $L$s and $\mathbf{g}$ is non-vanishing \citep{nawab1983signal}; Uniqueness (up to a global phase) for some $L$s, $N$s and mild conditions on $\mathbf{g}$ \citep{eldar2014sparse}. \hrule \hspace{0.1em}\\
			& & Non-convex algorithm \citep{bendory2018non} that employs an initialization followed by a gradient descent update rule.\hspace{0.1em}\\
			\hline
			\multirow{3}{*}{FROG} & \multirow{2}{*}{$\displaystyle \mathbf{A}[p,k]:= \left \lvert \sum_{n=0}^{N-1} \mathbf{x}[n]\mathbf{x}[pL+n]e^{\frac{-2\pi \mathrm{i}kn}{N}}\right \rvert^{2}$, $L<N$}
			& Uniqueness (up to global phase, translated and reflected signal) for some $L$s and band-limited pulses \citep{bendory2017uniqueness}. \hrule \hspace{0.1em}\\
			& & Non-convex algorithm \citep{pinilla2019frequency} that employs an initialization followed by a gradient descent update rule.\\
			\hline
			
			\multirow{3}{*}{FrFT} & \multirow{3}{*}{$\displaystyle\mathbf{A}[\alpha,\zeta]:= \left\lvert \sum_{n_{1}=0}^{N-1} \left\lvert \sum_{n_{2}=0}^{N-1}\mathbf{x}[n_{2}]e^{\frac{-2\pi i n_{2}n_{1} -i\pi (n_{2})^2\cos(\alpha)}{\sin(\alpha)}} \right\rvert^{2}e^{\frac{-2\pi i n_{1}\zeta}{N}} \right\rvert^2$}
			& Uniqueness (up to global phase, translated, reflected, and scaled signal) for Hermite and compact support functions, rectangular pulse trains and linear combinations of Gaussians \citep{jaming2014uniqueness}. \hrule \hspace{0.1em}\\
			& & No known algorithm.
			\\
			\hline
			\multirow{3}{*}{This paper} & \multirow{3}{*}{$\displaystyle\mathbf{A}[p,k]:= \left\lvert \sum_{n=0}^{N-1} \mathbf{x}[n] \overline{\mathbf{x}[n-p]}e^{\frac{-2\pi i nk}{N}} \right\rvert^2$}
			& Uniqueness (up to global phase, translated, reflected and scaled signal) for band/time-limited signals.\hrule \hspace{0.1em}\\
			& & Non-convex algorithm that employs an initialization followed by a gradient descent update rule.\\	
			\hline
		\end{tabular}
		\label{tab:1}
	\end{table}
	
	Recall that PR constitutes an instance of non-convex programming, that is generally known to be NP-hard \citep{shechtman2015phase}. For instance, in the case of real-valued signals, this is a combinatorial optimization problem because it seeks a series of signs over the entries of the target signal that obey a given AF. For complex-valued signals, the procedure becomes more complicated and, instead of a set of signs, one must determine a collection of unimodular complex scalars that obey the given AF. 
	
	Substantial work has been done and is still ongoing to overcome the ill-posedness of the PR problem, and the literature is too large to summarize here (see, e.g., \citep{vaswani2020nonconvex} for a contemporary survey, and references therein; also \citep{balan2006signal,alexeev2014phase,bandeira2014saving,fickus2014phase}). Overall, two major approaches have emerged: the first harnesses prior knowledge of the signal structure, such as sparse support \citep{ranieri2013phase}, non-vanishing behavior \citep{bendory2018non}, or band-limitedness \citep{pinilla2021banraw}, while the second exploits technology and assumes it is possible to make additional measurements of the magnitude via, for example, coded diffraction patterns \citep{candes2015phase}, masks \citep{jaganathan2016phaseless}, and short-time Fourier transform (STFT) \citep{pinilla2019frequency}. These approaches employ traditional optimization strategies such as gradient descent \citep{candes2015wirtinger} and semidefinite relaxations \citep{candes2015phase}. 
	
	In particular, the AF-based signal recovery belongs to the class of one-dimensional (1-D) PR problems \citep{hayes1982reconstruction}, wherein uniqueness of the retrieved signal cannot be ensured unless some information about the signal structure or additional measurements are available. Note that a real 2-D signal is uniquely specified (up to the trivial ambiguities) by the magnitude of its continuous Fourier transform, with the exception of a set of signals of measure zero \citep{hayes1982reconstruction}. 
	The 1-D AF-based signal design is bivariate in parameters $p$ and $k$. Among prior works, other similar bivariate 1-D problems include STFT \citep{bendory2018non} and frequency-resolve optical gating (FROG) PR \citep{pinilla2019frequency}, and these are distinct from the previously mentioned FrFT formulation \citep{jaming2014uniqueness}. Table~\ref{tab:1} summarizes and compares the discrete models, guarantees, and algorithms for various bivariate PR problems. 
	Among these, the radar PR is the most challenging because there are more trivial ambiguities than the other formulations. 
	In the real signal case, AF-based radar PR reduces to FROG. However, radar signals are complex and the conjugation in the AF expression brings a total of four ambiguities (\textbf{T1}-\textbf{T4}) versus only three ambiguities (\textbf{T1}-\textbf{T3}) in FROG \citep{beinert2018enforcing}. We note that \textbf{T1}-\textbf{T4} have been mentioned for the FROG formulation in \citep{bendory2017uniqueness} but this work provides a uniqueness result for only the first three ambiguities. The proof of identifiability in \citep{bendory2017uniqueness} is based on the methods developed in \citep{beinert2018enforcing}. This holds because FROG does not involve a conjugation, and it follows that these algorithms do not directly address AF-based PR. Our work fills this gap and also advances the theory of AF-based radar waveform design.
    
    While mathematically the AF-based waveform synthesis is similar to PR, the problems are distinct. Uniqueness and identifiability are essential in PR. On the other hand, if multiple waveforms have the same AF, this leads to more design choices. For example, in \citep{de1970signals}, the goal is to obtain a \textit{family} of functions with the same AF. 

	\subsection{Our contributions}
In general, radar systems employ a wide variety of signals. Therefore, previous radar PR studies that were restricted to Hermite functions \citep{de1970signals} or rectangular pulses \citep{jaming2010phase} provide strong steps forward in theory but are not very useful in practice. 
In this paper, we extend AF-based waveform design to arbitrary signals that are time- or band-limited. We present a uniqueness result that states that a band-limited (time-limited) signal can be recovered from its AF using at least $3B-1$ ($3S-1$) measurements, where $B$ ($S$) is the signal's bandwidth (pulse-width). To this end, we develop a trust region algorithm that minimizes a smoothed non-convex least-squares objective function to iteratively estimate the band-limited signal of interest. 

Our algorithm defines a region around the current iterate within which it trusts the model to be an adequate representation of the objective function. Then, it chooses the step to be an approximate minimizer of the model in this region. In particular, our proposed method consists of two steps: approximating the signal via an iterative spectral algorithm and subsequent refining of the attained initialization based upon a sequence of gradient iterations.


Our numerical experiments suggest that our proposed algorithm is able to estimate a band- or time-limited signal from its AF for both complete and incomplete measurements; here we consider the AF incomplete if only a few time-shifts or Fourier frequencies are considered in its formulation. Preliminary results of this work appeared in our conference publication \citep{pinilla2021banraw}, where we formulated AF-based waveform design of band-limited signals as a non-convex PR problem and solved it using the \textit{ba}nd-limited \textit{ra}dar \textit{w}aveform design via PR (BanRaW) algorithm. However, details of the uniqueness results, extension to time-limited signals, and extensive numerical experiments were excluded from \citep{pinilla2021banraw}. In this paper, we include these details, present cases with incomplete/missing AF measurements and investigate additional practical constraints on signals such as linear/non-linear frequency modulated (LFM/NLFM) waveforms.



The rest of the paper is organized as follows. In the next section, we introduce the necessary background on the radar PR problem. In Section~\ref{sec:algorithm}, we develop our iterative procedure to refine the solution by minimizing a smooth least-squares objective. The spectral initialization step of this procedure is detailed in Section~\ref{sec:initialization}. We validate our models and methods through extensive numerical experiments in Section~\ref{sec:numerical}. We conclude in Section~\ref{sec:conclusion}.

Throughout this paper, we denote the sets of positive and strictly positive real numbers by $\mathbb{R}_{+}:=\{w\in \mathbb{R}: w\geq 0\}$ and $\mathbb{R}_{++}:=\{w\in \mathbb{R}: w>0\}$, respectively. We use boldface lowercase and uppercase letters for vectors and matrices, respectively. The sets are denoted by calligraphic letters and $\lvert \cdot \rvert$ represents the cardinality of the set. The conjugate and conjugate transpose of the vector $\mathbf{w}\in \mathbb{C}^{N}$ are denoted as $\overline{\mathbf{w}}\in \mathbb{C}^{N}$ and $\mathbf{w}^{H}\in \mathbb{C}^{N}$, respectively. The $n$-th entry of a vector $\mathbf{w}$, which is assumed to be periodic, is $\mathbf{w}[n]$. The $(k,l)$-th entry of a matrix $\mathbf{A}$ is $\mathbf{A}[k,l]$. We denote the Fourier transform of a vector and its conjugate reflected version (that is, $\hat{\mathbf{w}}[n] := \overline{\mathbf{w}}[-n]$) by $\tilde{\mathbf{w}}$ and $\hat{\mathbf{w}}$, respectively. The notation $\text{diag}(\mathbf{W},\ell)$ refers to a column vector with entries $\mathbf{W}[j,(j+\ell)\mod N]$ for $j=0,\cdots,N-1$. For vectors, $\|\mathbf{w}\|_p$ is the $\ell_p$ norm. Additionally, we use $\odot$ and $*$ for the Hadamard (point-wise) product, and convolution, respectively; $\sqrt{\cdot}$ is the point-wise square root; superscript within parentheses as $(\cdot)^{(r)}$ indicates the value at $r$-th iteration; $\lVert \cdot \rVert_{\mathcal{F}}$ denotes the Frobenius norm of a matrix; $\textrm{Tr}(\cdot)$ denotes the matrix trace function; $\sigma_{\max}(\cdot)$ represents the largest singular value of its matrix argument; $\mathcal{R}(\cdot)$ denotes the real part of its complex argument; $\mathbb{E}[\cdot]$ represents the expected value; and $\omega=e^{\frac{2\pi i}{n}}$ is the $n$-th root of unity.

\section{Problem Formulation}
\label{sec:problem}
Our goal is to uniquely estimate the signal $\mathbf{x}$, up to trivial ambiguities, from the AF $\mathbf{A}$ in \eqref{eq:Ambiguity}. Here, the total number of AF measurements $m$ is the product of number of Doppler frequencies and time delays. To this end, we first show that such a recovery of $\mathbf{x}$ is possible under rather mild conditions for a band-limited signal. We then extend this result to time-limited signals.


\subsection{Uniqueness for band-limited signals}
We introduce the following definition of a band-limited signal.
\begin{definition}[$B$-band-limitedness]
	A signal $\mathbf{x}\in \mathbb{C}^{N}$ is defined to be $B$-band-limited if its Fourier transform $\tilde{\mathbf{x}}\in \mathbb{C}^{N}$ contains $N-B$ consecutive zeros. That is, there exists $k$ such that $\tilde{\mathbf{x}}[k]=\cdots=\tilde{\mathbf{x}}[N+k-B-1]=0$.
	\label{def:bandlimitedSignal}
\end{definition}
Recall two auxiliary lemmata. The following Lemma~\ref{lem:uniqueMeas} states conditions to uniquely determine the signal $\mathbf{x}$ from the knowledge of its power spectrum $\lvert \tilde{\mathbf{x}}[k] \rvert$.
\begin{lemma}(\citep[Theorem 2]{beinert2018enforcing})
	Let $s$ be an arbitrary integer between $0$ and $N-1$. Then, almost every $\mathbf{x}\in \mathbb{C}^{N}$ can be uniquely recovered from $\{ \lvert \tilde{\mathbf{x}}[k] \rvert \}_{k=0}^{2N-2}$ and $\mathbf{x}[s]$. If $s=(N-1)/2$, then the reconstruction is up to conjugate reflection.
	\label{lem:uniqueMeas}
\end{lemma}

The conditions to uniquely determine the signal $\mathbf{x}$ from the knowledge of its power spectrum $\lvert \tilde{\mathbf{x}}[k] \rvert$ and its absolute value $\lvert \mathbf{x}[n] \rvert$ are specified by the following Lemma~\ref{lem:uniqueFourier}.
\begin{lemma}(\citep[Corollary 2]{beinert2018enforcing})
	Almost every complex-valued signal $\mathbf{x}\in \mathbb{C}^{N}$ can be uniquely recovered from $\{\lvert \tilde{\mathbf{x}}[k] \rvert \}_{k=0}^{N-1}$ and $\{\lvert \mathbf{x}[n] \rvert \}_{n=0}^ {N-1}$ up to rotations.
	\label{lem:uniqueFourier}
\end{lemma}
In Lemma~\ref{lem:uniqueFourier}, the phrase \textit{almost every} or \textit{almost all} means that the set of signals that cannot be uniquely determined, up to trivial ambiguities, is contained in the vanishing locus of a nonzero polynomial. In the following Proposition~\ref{prop:uniqueness}, we state the conditions for recovering band-limited signals from their AF.
\begin{proposition}
	Assume $\mathbf{x}\in \mathbb{C}^{N}$ is a $B$-band-limited signal for some $B\leq N/2$. Then, for $N\geq 4$, almost all signals are uniquely determined from at least $3B\minus1$ measurements of their AF $\mathbf{A}[p,k]$, up to trivial ambiguities. If, in addition, we have access to the signal's power spectrum and $N \geq 3$, then $2B\minus1$ measurements suffice.
	\label{prop:uniqueness}
\end{proposition}
\begin{proof}	
	We reformulate the measurement model to a more convenient structure. From the inverse Fourier transform, $\mathbf{x}[n]=\frac{1}{N}\sum_{k=0}^{N-1}\tilde{\mathbf{x}}[k]e^{2\pi i kn/N}$. The discrete form of inner product between the waveform and its time-delayed, frequency-shifted version is 
	\begin{align}
		\mathbf{S}[p,k]
		&= \sum_{n=0}^{N-1} \mathbf{x}[n] \overline{\mathbf{x}[n-p]}e^{\frac{-2\mathrm{i}\pi nk}{N}} \nonumber\\
		&=\frac{1}{N^{2}}\sum_{n=0}^{N-1} \left(\sum_{\ell_{1}=0}^{N-1}\tilde{\mathbf{x}}[\ell_{1}]e^{\frac{2\pi i \ell_{1}n}{N}}\right) \left(\sum_{\ell_{2}=0}^{N-1}\overline{\tilde{\mathbf{x}}[\ell_{2}]}e^{\frac{-2\pi \mathrm{i} \ell_{2}n}{N}}e^{\frac{2\pi i \ell_{2}p}{N}}\right) e^{\frac{-2\pi \mathrm{i}kn}{N}} \nonumber\\
		&=\frac{1}{N^{2}}\sum_{\ell_{1},\ell_{2}=0}^{N-1} \tilde{\mathbf{x}}[\ell_{1}]\overline{\tilde{\mathbf{x}}[\ell_{2}]}e^{\frac{2\pi i \ell_{2}p}{N}} \sum_{n=0}^{N-1}e^{\frac{2\pi i n(\ell_{1}- \ell_{2} - k)}{N}} \nonumber\\
		&=\frac{1}{N}\sum_{\ell=0}^{N-1}\tilde{\mathbf{x}}[\ell+k]\overline{\tilde{\mathbf{x}}[\ell]}e^{\frac{2\pi i \ell p}{N}},
		\label{eq:newModel}
	\end{align}
	where the last equality follows because the sum $\sum_{n=0}^{N-1}e^{\frac{2\pi i n(\ell_{1}- \ell_{2} - k)}{N}}=N$, if $\ell_{1} = \ell_{2} + k$ and zero otherwise. Obviously, 
	$\mathbf{A}[p,k] = \lvert \mathbf{S}[p,k] \rvert^{2}$.
	
	Assume that $B = N/2$, $N$ is even, that $\tilde{\mathbf{x}}[n] \not= 0$ for $k = 0\dots,B\minus1$, and that $\tilde{\mathbf{x}}[n] = 0$ for $k = N/2,\dots, N-1$. If the signal's nonzero Fourier coefficients are not in the interval $0,\dots, N/2-1$, then we can cyclically re-index the signal without affecting the proof. If $N$ is odd, then we replace $N/2$ by $\lfloor N/2 \rfloor$ everywhere in the sequel. Clearly, the proof carries through for any $B \leq N/2$. From \eqref{eq:newModel}, the cross product $\tilde{\mathbf{x}}[\ell+k]\overline{\tilde{\mathbf{x}}[\ell]}$ forms an ``inverted pyramid" structure as a consequence of the band-limitedness of the signal $\tilde{\mathbf{x}}$. Then, for each fixed $k$ (as row) and varying $\ell$ (as column), where $\ell,k=0,\dots,N/2-1$, this pyramid is
	\begin{align}
		&\lvert \tilde{\mathbf{x}}[0]\rvert^{2},\lvert \tilde{\mathbf{x}}[1]\rvert^{2},\dots,\lvert \tilde{\mathbf{x}}[B-1]\rvert^{2},0,\dots,0 \nonumber\\
		&\overline{\tilde{\mathbf{x}}[0]}\tilde{\mathbf{x}}[1],\overline{\tilde{\mathbf{x}}[1]}\tilde{\mathbf{x}}[2],\dots,\overline{\tilde{\mathbf{x}}[B-2]}\tilde{\mathbf{x}}[B-1],0,\dots,0 \nonumber\\
		&\overline{\tilde{\mathbf{x}}[0]}\tilde{\mathbf{x}}[2],\overline{\tilde{\mathbf{x}}[1]}\tilde{\mathbf{x}}[3],\dots,\overline{\tilde{\mathbf{x}}[B-3]}\tilde{\mathbf{x}}[B-1],0,\dots,0 \nonumber\\
		&\hspace{10em}\vdots\nonumber\\
		&\overline{\tilde{\mathbf{x}}[0]}\tilde{\mathbf{x}}[B-1],0,\dots,0,0,\dots,0\hspace{5em} \nonumber\\
		&0,0,\dots,\tilde{\mathbf{x}}[0]\overline{\tilde{\mathbf{x}}[B-1]},0,\dots,0\hspace{5em} \nonumber\\
		&0,\dots,\tilde{\mathbf{x}}[0]\overline{\tilde{\mathbf{x}}[B-2]},\tilde{\mathbf{x}}[1]\overline{\tilde{\mathbf{x}}[B-1]},0,\dots,0\hspace{5em}\nonumber\\
		&\hspace{10em}\vdots \nonumber\\
		&0,\mathbf{x}[0]\overline{\tilde{\mathbf{x}}[1]},\tilde{\mathbf{x}}[1]\overline{\tilde{\mathbf{x}}[2]},\dots,\tilde{\mathbf{x}}[B-2]\overline{\tilde{\mathbf{x}}[B-1]},0,\dots,0.
		\label{eq:piramid}
	\end{align}
	Thus, each term $\mathbf{S}[p,k]$ from \eqref{eq:newModel} is recovered from the above equation by computing a zero-padded Fourier transform of each pyramid's rows. 
	
	\textbf{Step 0: }From the $(B\minus1)$-th row of \eqref{eq:piramid}, we have 
	\begin{align}
		\lvert \mathbf{S}[p,B] \rvert = \lvert \tilde{\mathbf{x}}[0]\rvert \lvert\tilde{\mathbf{x}}[B-1]\rvert, \forall p=0,\dots,N-1.
		\label{eq:var1}
	\end{align}
	The parameter $a$ in the ambiguity \textbf{T2} is real. 
	Therefore, we set $\tilde{\mathbf{x}}[0]$ to be real and, without loss of generality, assume $\tilde{\mathbf{x}}[0]=1$ \citep{pinilla2021banraw}. Then, from \eqref{eq:var1}, 
	\begin{align}
		\lvert \mathbf{S}[p,B-1] \rvert = \lvert\tilde{\mathbf{x}}[B-1]\rvert,\; \forall p=0,\dots,N-1.
	\end{align}
	
	\textbf{Step 1: } It follows from the first row of \eqref{eq:piramid} that 
	\begin{align}
		\lvert \mathbf{S}[p,0] \rvert = \frac{1}{N}\left\lvert \sum_{\ell=0}^{B-1}\lvert \tilde{\mathbf{x}}[\ell] \rvert^{2}e^{2\pi i \ell p/N} \right\rvert, p=0,\dots,N-1.
		\label{eq:var2}
	\end{align}
	From \textbf{Step 0}, the entries $\lvert\tilde{\mathbf{x}}[0]\rvert$, and $\lvert\tilde{\mathbf{x}}[B-1]\rvert$ are known. Note that the support of $\lvert \tilde{\mathbf{x}} \rvert$ is $\{0,\cdots,B-1\}$ with $B\leq N/2$, \eqref{eq:var2} is its Fourier transform of size $2B$, and $\lvert\tilde{\mathbf{x}}[0]\rvert$ is known. Therefore, Lemma \ref{lem:uniqueMeas} is applicable. Hence, for almost all signals, it follows that $\lvert\tilde{\mathbf{x}}[1]\rvert, \dots, \lvert\tilde{\mathbf{x}}[B-2]\rvert$ are uniquely determined from \eqref{eq:var2} and the knowledge of $\lvert\tilde{\mathbf{x}}[0]\rvert$. This analysis does not imply that $\tilde{\mathbf{x}}[1],\dots,\tilde{\mathbf{x}}[B-1]$ are uniquely determined. In fact, there are up to $2^{B-1}$ vectors, modulo global phase (\textbf{T1}), reflection and conjugation (\textbf{T3}) that satisfy the constraints in \eqref{eq:var1} and \eqref{eq:var2} \citep{pinilla2021banraw}.
	
	\textbf{Step 2: } The second row of \eqref{eq:piramid} yields the following system of equations
	\begin{align}
		\lvert \mathbf{S}[p,1] \rvert = \frac{1}{N}\left\lvert \sum_{\ell=0}^{B-2} \tilde{\mathbf{x}}[\ell + 1]\overline{\tilde{\mathbf{x}}[\ell]} e^{2\pi i \ell p/N} \right\rvert, p=0,\dots,N-1.
		\label{eq:var3}
	\end{align}
	Fix one of the possible solutions for $\tilde{\mathbf{x}}[1]$ from \textbf{Step 1}. The cross-product in \eqref{eq:var3} implies the presence of ambiguities \textbf{T2} and \textbf{T4} in the entries $\overline{\tilde{\mathbf{x}}[1]}\tilde{\mathbf{x}}[2],\dots, \tilde{\mathbf{x}}[B-1]\overline{\tilde{\mathbf{x}}[B-2]}$. Since $\tilde{\mathbf{x}}[0]$ is known, it follows from Lemma \ref{lem:uniqueMeas} that for almost all signals $\overline{\tilde{\mathbf{x}}[1]}\tilde{\mathbf{x}}[2],\dots, \tilde{\mathbf{x}}[B-1]\overline{\tilde{\mathbf{x}}[B-2]}$ are uniquely determined.  
	
	\textbf{Step 3: } Now that $\tilde{\mathbf{x}}[0]$, and $\tilde{\mathbf{x}}[1]$ are known, $\tilde{\mathbf{x}}[2]$ can be estimated from \textbf{Step 2}. Thus, with known $\overline{\tilde{\mathbf{x}}[0]}\tilde{\mathbf{x}}[2]$, Lemma \ref{lem:uniqueMeas} implies that for almost all signals $\overline{\tilde{\mathbf{x}}[1]}\tilde{\mathbf{x}}[3],\dots,\overline{\tilde{\mathbf{x}}[B-3]}\tilde{\mathbf{x}}[B-1]$ are uniquely determined. However, at this stage, there are still $2^{B-1}$ possible solutions from \textbf{Step 2}. Despite a large number of possible solutions, we now prove that, at this step, there is only one out of the $2^{B-1}$ vectors in \textbf{Step 2} that is consistent with the constraints in \eqref{eq:var1}, \eqref{eq:var2}, and \eqref{eq:var3}, up to trivial ambiguities. First, from \textbf{Step 1}, $\lvert\tilde{\mathbf{x}}[0]\rvert,\dots,\lvert\tilde{\mathbf{x}}[B-1]\rvert$ are uniquely determined. Therefore, from known $\{\lvert\tilde{\mathbf{x}}[\ell]\rvert\}_{\ell=0}^{B-1}$, it implies that $\lvert \overline{\tilde{\mathbf{x}}[0]}\tilde{\mathbf{x}}[1] \rvert,\lvert \overline{\tilde{\mathbf{x}}[1]}\tilde{\mathbf{x}}[2] \rvert,\dots, \lvert \tilde{\mathbf{x}}[B-1]\overline{\tilde{\mathbf{x}}[B-2]} \rvert $ are also known. Using this knowledge and $\{\lvert\mathbf{S}[p,1]\rvert\}_{p=0}^{N-1}$ in \eqref{eq:var3} of Step 2, it follows from Lemma \ref{lem:uniqueFourier} that $\overline{\tilde{\mathbf{x}}[0]}\tilde{\mathbf{x}}[1],\overline{\tilde{\mathbf{x}}[1]}\tilde{\mathbf{x}}[2],\dots, \tilde{\mathbf{x}}[B-1]\overline{\tilde{\mathbf{x}}[B-2]}$ are uniquely determined for almost all signals. This now leads to a unique selection (up to trivial ambiguities) of $\tilde{\mathbf{x}}[1]$ in \textbf{Step 2} since $\tilde{\mathbf{x}}[0]$ is known. Consequently, $\tilde{\mathbf{x}}[2]$ is uniquely determined in this \textbf{Step 3}. Additionally, for a unique estimation of the signal, the existence of at least $\tilde{\mathbf{x}}[0],\tilde{\mathbf{x}}[1],\tilde{\mathbf{x}}[2],$ and $\tilde{\mathbf{x}}[B-1]$ is required (arising from the cross-product in \eqref{eq:var3}). This leads to $N\geq 4$. 
	
	\textbf{Step $B-1$: } From previous $B-2$ steps, the entries $\tilde{\mathbf{x}}[0],\dots,\tilde{\mathbf{x}}[B-2]$ are uniquely determined (up to trivial ambiguities). Hence, it follows from Lemma \ref{lem:uniqueMeas} that $\tilde{\mathbf{x}}[B-1]$ is also uniquely determined.
	
	The total number of measurements required for unique construction is concluded as follows. In \textbf{Step 0}, only one measurement (i.e., $\tilde{\mathbf{x}}[0]$) is needed. Then, according to Lemma \ref{lem:uniqueMeas}, \textbf{Step 1} requires $2B-2$ measurements. Finally, from Lemma \ref{lem:uniqueFourier}, $B$ more measurements are needed in \textbf{Step 2}. No further measurements are used in \textbf{Step 3}. This brings the total number of measurements required to $1+2B\minus2+B = 3B\minus1$. In addition, if we have access to the spectrum $\lvert \tilde{\mathbf{x}} \rvert$, then we can uniquely determine $\tilde{\mathbf{x}}$ at \textbf{Step 1} with $N\geq 3$ i.e. with one less entry in $\tilde{\mathbf{x}}$. Here, no further measurements are required from \textbf{Step 2} onward. 
	Therefore, summing up the measurements in \textbf{Step 0} and \textbf{Step 1} yields $1+2B-2 = 2B-1$. 
\end{proof}


Note that Proposition~\ref{prop:uniqueness} states that all delay steps are not needed to recover the signal. Therefore, it is desired to develop a method that also works for such partial measurements. 
Step 5 of the proof 
reveals that the first and the $(B-1)$-th rows of AF in \eqref{eq:Ambiguity} must be perfectly preserved in order to ensure uniqueness (up to trivial ambiguities). 

We note that the signals $\mathbf{x}[n]$, $\mathbf{x}[n]e^{\textrm{i}\phi}$, $\mathbf{x}[n-a]$, $\mathbf{x}[-n]$, and $e^{\textrm{i}bn}\mathbf{x}[n]$ yield the same magnitude measurements for any constant $\phi$, $a$, $b \in \mathbb{R}$ and therefore these constants cannot be recovered by any algorithm. This naturally leads to the following measure of the relative error between the true signal $\mathbf{x}$ and any $\mathbf{w}\in \mathbb{C}^{N}$.
\begin{definition}
	\label{def:dis}The distance between the two vectors $\mathbf{x} \in \mathbb{C}^N$ and $\mathbf{w} \in \mathbb{C}^N$ is defined as 
	\begin{equation}
		\text{dist}(\mathbf{x},\mathbf{w}):= \min_{\mathbf{z}\in \mathcal{T}(\mathbf{x})}\lVert \mathbf{w}-\mathbf{z}\rVert_{2},
		\label{eq:distance}
	\end{equation}
	where the set $\mathcal{T}(\mathbf{x})=\{\mathbf{z}\in \mathbb{C}^{N} \hspace{0.2em}:\hspace{0.2em} \mathbf{z}[n]=e^{i\beta}e^{i b n}\mathbf{x}[\epsilon n - a] \text{ for } \beta,b\in \mathbb{R}, \text{ and }\epsilon=\pm 1, a\in \mathbb{Z} \}$ contains vectors with all possible trivial ambiguities. If $\text{dist}(\mathbf{x},\mathbf{w})=0$ and the uniqueness conditions of Proposition~\ref{prop:uniqueness} are met, then, for almost all signals, $\mathbf{x}$ and $\mathbf{w}$ are equal \emph{up to trivial ambiguities}. 
\end{definition}
In Appendix~\ref{app:distance}, we prove that the minimum of $\lVert \mathbf{w}-\mathbf{z}\rVert_{2}$ in \eqref{eq:distance} exists because $\mathcal{T}(\mathbf{x})$ is a closed set. 

\subsection{Uniqueness for time-limited signals}
We analogously specify the time-limited signals via the Definition~\ref{def:timelimited} as follows.
\begin{definition}[$S$-time-limitedness]
	A signal $\mathbf{x}\in \mathbb{C}^{N}$ is defined to be $S$-time-limited signal if it 
	contains $N-S$ consecutive zeros. That is, there exists $k$ such that $\mathbf{x}[k]=\cdots=\mathbf{x}[N+k-S-1]=0$.
	\label{def:timelimited}
\end{definition}
Then, following Proposition~\ref{prop:uniqueness}, the corresponding result for time-limited waveforms is stated by the Corollary~\ref{coro:time} below. The notion of \textit{almost all} signals here is the same as in Proposition~\ref{prop:uniqueness}. 
\begin{corollary}
	Assume $\mathbf{x}\in \mathbb{C}^{N}$ is a $S$-band-limited signal for some $S\leq N/2$. Then, for $N\geq 4$, almost all signals are uniquely determined from at least $3S-1$ measurements of their AF $\mathbf{A}[p,k]$, up to trivial ambiguities. If, in addition, we have access to the signal's power and $N \geq 3$, then $2S-1$ measurements suffice.
	\label{coro:time}
\end{corollary}
\begin{proof}
	Recall that, according to \eqref{eq:Ambiguity}, we have
	\begin{align}
		\mathbf{A}[p,k] = \left\lvert \sum_{n=0}^{N-1} \mathbf{x}[n] \overline{\mathbf{x}[n-p]}e^{-2\mathrm{i}\pi nk/N} \right\rvert^{2}.
		\label{eq:ambiTime}
	\end{align}
	Assume that $S = N/2$, $N$ is even, that $\mathbf{x}[n] \not= 0$ for $n = 0\dots,S-1$, and that $\mathbf{x}[n] = 0$ for $n = N/2,\dots, N-1$. If the signal's nonzero coefficients are not in the interval $0,\dots, N/2-1$, then we can cyclically reindex the signal without affecting the proof. If $N$ is odd, then one should replace $N/2$ by $\lfloor N/2 \rfloor$ everywhere in the sequel. Clearly, the proof carries through for any $S \leq N/2$.
	
	It follows from \eqref{eq:newModel} that the bandlimitedness of the signal forms a ``inverted pyramid" structure. From \eqref{eq:ambiTime}, the cross product $ \mathbf{x}[n] \overline{\mathbf{x}[n-p]}$ forms an ``inverted pyramid" structure as a consequence of the time-limitedness of the signal $\mathbf{x}$. Then, each row represents fixed $p$ (as row) and varying $n$ (as column) of $\overline{\mathbf{x}[n- p]}\mathbf{x}[n]$ for $p=0,\dots,N/2-1$ as follows:
	\begin{align}
		&\lvert \mathbf{x}[0]\rvert^{2},\lvert \mathbf{x}[1]\rvert^{2},\dots,\lvert \mathbf{x}[S-1]\rvert^{2},0,\dots,0 \nonumber\\
		&0,\overline{\mathbf{x}[0]}\mathbf{x}[1],\overline{\mathbf{x}[1]}\mathbf{x}[2],\dots,\overline{\mathbf{x}[S-2]}\mathbf{x}[S-1],0,\dots,0 \nonumber\\
		&0,0,\overline{\mathbf{x}[0]}\mathbf{x}[2],\overline{\mathbf{x}[1]}\mathbf{x}[3],\dots,\overline{\mathbf{x}[S-3]}\mathbf{x}[S-1],0,\dots,0 \nonumber\\
		&\hspace{10em}\vdots\nonumber\\
		&0,0,\dots,0,\overline{\mathbf{x}[0]}\mathbf{x}[S-1],0,\dots,0,0,\dots,0\hspace{5em} \nonumber\\
		&\mathbf{x}[0]\overline{\mathbf{x}[S-1]},0,0,\dots,0\hspace{5em} \nonumber\\
		&\mathbf{x}[0]\overline{\mathbf{x}[S-2]},\mathbf{x}[1]\overline{\mathbf{x}[S-1]},0,0,\dots,0\hspace{5em}\nonumber\\
		&\hspace{10em}\vdots \nonumber\\
		&\mathbf{x}[0]\overline{\mathbf{x}[1]},\mathbf{x}[1]\overline{\mathbf{x}[2]},\dots\mathbf{x}[S-2]\overline{\mathbf{x}[S-1]},0,0,\dots,0.
		\label{eq:piramidTime}
	\end{align}
	Then, $\mathbf{A}[p,k]$ as in \eqref{eq:ambiTime} is a subsample of the Fourier transform of each one of the pyramid's rows.
	
	Therefore, performing an analogous construction procedure as in Proposition~\ref{prop:uniqueness} on \eqref{eq:piramidTime}, we have that the signal $\mathbf{x}$, with $N\geq 4$, can be uniquely determined from $3S-1$ measurements. If the power $\lvert \mathbf{x} \rvert$ of the signal is additionally available, we can similarly uniquely determine $\mathbf{x}$ with $N\geq 3$ from only $2S-1$ measurements.
\end{proof}
The proof of Corollary \ref{coro:time} is also a construction procedure as in the proof of Proposition~\ref{prop:uniqueness}, and that the first and the $(S-1)$-th rows of the AF in \eqref{eq:Ambiguity} must be perfectly preserved in order to ensure uniqueness (up to trivial ambiguities). 

\subsection{AF-based PR optimization}
\label{subsec:opt}
It is instructive to note that the the AF-based PR requires recovering a signal from phaseless quadratic random measurements. Among prior works on similar problems, a popular approach is to minimize the intensity least-squares objective 
\citep{candes2015wirtinger}. A few recent works have further shown that this technique 
leads to better reconstruction under noisy scenarios \citep{zhang2016reshaped}. However, the amplitude least-squares cost function is non-smooth and thus may lead to a biased descent direction \citep{pinilla2018phase}. In \citep{zhang2016reshaped}, the non-smoothness was addressed by introducing truncation parameters into the gradient step in order to eliminate the errors in the estimated descent direction. However, this procedure may modify the search direction and increase the sample complexity of the PR problem \citep{pinilla2018phase}. The smoothing strategy recently proposed in \citep{pinilla2018phase} overcomes these shortcomings. 

\begin{figure}[ht]
	\centering
	\includegraphics[width=1\textwidth]{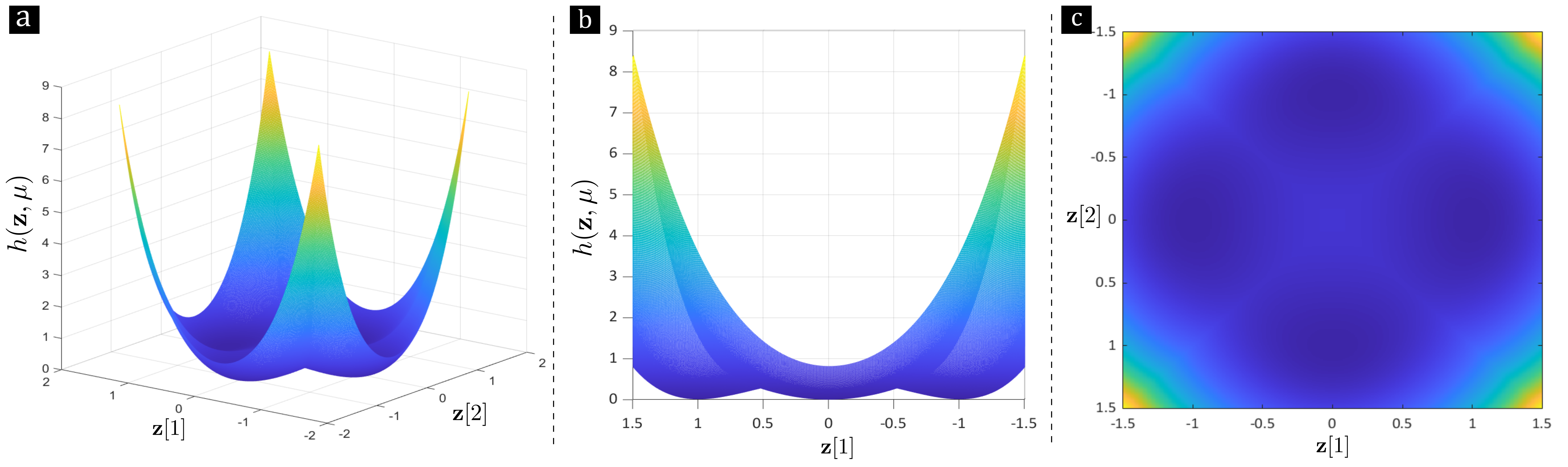}
	\caption{(a) The cost function $h(\textbf{z},\mu)$ in the optimization problem \eqref{eq:auxproblem}, when evaluated at $\mu = 0$ and $\mathbf{A}$ computed for $\textbf{x}=[1;0]$. (b) Side-view of the cost function $h(\textbf{z},0)$ from $\mathbf{z}[1]$ dimension. The only local minima are also global minima, located at $\textbf{z}=\pm \textbf{x}$ and $\textbf{z}=\pm \hat{\textbf{x}}$ with $\hat{\textbf{x}}[n] = \textbf{x}[n-1]$, corresponding to the ambiguities \textbf{T1}-\textbf{T4}. There are four saddle points near to $\pm [0.5;0.5]$, and $\pm [-0.5;0.5]$. (c) The top view of the same function $h(\textbf{z},0)$ plotted with respect to the values along each dimension of the 2-D variable $\textbf{z}$.}
	\label{fig:landscape}
\end{figure}

Following \citep{pinilla2018phase}, define a function $\varphi_{\mu}: \mathbb{R}\rightarrow \mathbb{R}_{++}$ as $$\varphi_{\mu}(w) := \sqrt{w^2+\mu^{2}},$$ where $\mu\in \mathbb{R}_{++}$ is a tunable parameter. The first stage of our algorithm is to recover the underlying signal $\mathbf{z}$ by considering the smooth version of amplitude least-squares objective as follows
\begin{align}
	\underset{\mathbf{z}\in \mathbb{C}^{N}}{\textrm{minimize}}\; h(\mathbf{z},\mu) = \underset{\mathbf{z}\in \mathbb{C}^{N}}{\textrm{minimize}}\; \frac{1}{N^{2}}\sum_{k,p=0}^{N-1} \ell_{k,p}(\mathbf{z},\mu),
	\label{eq:auxproblem}
\end{align}
where
\begin{equation}
	\ell_{k,p}(\mathbf{z},\mu) := \left\lbrack\varphi_{\mu}\left(\left\lvert\sum_{n=0}^{N-1} \mathbf{z}[n]\overline{\mathbf{z}[n-p]}e^{-2\pi \mathrm{i}nk/N} \right \rvert\right) -\sqrt{\mathbf{A}[p,k]}\right\rbrack^{2}.
	\label{eq:ellfunction}
\end{equation}
Note that when $\mu=0$, \eqref{eq:ellfunction} reduces to a non-smooth formulation. 
Fig. \ref{fig:landscape} shows 
that $h(\textbf{z},\mu)$ has more minimizers and saddle points than the cost functions of traditional PR problems \citep{sun2018geometric} (i.e., where only $\pm \textbf{z}$ are minimizers) because of the increased number of ambiguities in AF-based PR.

\section{Reconstruction Algorithm}
\label{sec:algorithm}
To solve \eqref{eq:auxproblem}, we now propose a trust region algorithm based on the Cauchy point, which lies on the gradient and minimizes the quadratic cost function by iteratively refining the trust region of the solution. We suggest minimizing the objective in \eqref{eq:auxproblem} by employing a Cauchy point update rule, namely, by iteratively applying
\begin{align}
	\mathbf{x}^{(r+1)}:=\mathbf{x}^{(r)} + \alpha^{(r)}\mathbf{b}^{(r)},
	\label{eq:updateforx}
\end{align}
where $\alpha^{(r)}$ is the step size at iteration $r$ and gradient descent direction vector is
\begin{align}
	\mathbf{b}^{(r)} := \argmin_{\mathbf{b}\in \mathbb{C}^{n}} &\hspace{1.5em}h(\mathbf{x}^{(r)},\mu^{(r)}) + 2\mathcal{R}\left(\mathbf{b}^{H}\mathbf{d}^{(r)}\right) \nonumber\\
	\textrm{subject to} & \hspace{1.5em}\lVert \mathbf{b} \rVert_{2}\leq \mu^{(r)},
	\label{eq:trust}
\end{align}
where $\mathbf{d}^{(r)}$ is the gradient of $h(\mathbf{z},\mu)$ with respect to $\overline{\mathbf{z}}$ at iteration $r$. The solution to \eqref{eq:trust} is \citep[Section 4.1]{nocedal2006numerical}
\begin{align}
	\mathbf{b}^{(r)} = - \frac{\mu^{(r)}}{\lVert \mathbf{d}^{(r)} \rVert_{2}} \mathbf{d}^{(r)}.
\end{align}

To directly compute the gradient $\mathbf{d}^{(r)}$, we employ Wirtinger derivatives \citep{hunger2007introduction}, which are partial first-order derivatives of complex variables with respect to a real variable. Define 
\begin{equation}
	\mathbf{f}_{k}^{H} := \left[\omega^{-0(k-1)},\omega^{-1(k-1)},\cdots,\omega^{-(N-1)(k-1)}\right].
	\label{eq:vectora}
\end{equation}
The Wirtinger derivative of $h(\mathbf{z},\mu)$ in \eqref{eq:auxproblem} with respect to $\overline{\mathbf{z}[\ell]}$ is
\begin{align}
	\hspace{-0.3em}\frac{\partial h(\mathbf{z},\mu)}{\partial \overline{\mathbf{z}[\ell]}} := &\frac{1}{N^{2}}\sum_{k,p=0}^{N-1}\left( \mathbf{f}_{k}^{H}\mathbf{g}_{p} - \upsilon_{k,p} \right) \mathbf{z}[\ell-p]e^{2\pi i\ell k/N} +\frac{1}{N^{2}}\sum_{k,p=0}^{N-1}\left( \mathbf{f}_{k}^{T}\overline{\mathbf{g}_{p}} - \upsilon_{k,p} \right) \mathbf{z}[\ell+p]e^{-2\pi \mathrm{i}(\ell+p) k/N},
	\label{eq:stochStep}
\end{align}
where $\upsilon_{k,p}:= \sqrt{\mathbf{A}[p,k]}\frac{\mathbf{f}_{k}^{H}\mathbf{g}_{p}}{\varphi_{\mu}\left(\left\lvert\mathbf{f}_{k}^{H}\mathbf{g}_{p} \right \rvert\right)}$ and 
$\mathbf{g}_{p} := \left[\mathbf{z}[0]\overline{\mathbf{z}[p]},\cdots,\mathbf{z}[N-1]\overline{\mathbf{z}[N-1+p]} \right]^{T}$.
The gradient $\mathbf{d}^{(r)}$ becomes
\begin{align}
	\mathbf{d}^{(r)} := \left[\frac{\partial h(\mathbf{x}^{(r)},\mu)}{\partial \overline{\mathbf{z}[0]}},\cdots,\frac{\partial h(\mathbf{x}^{(r)},\mu)}{\partial \overline{\mathbf{z}[N-1]}}\right]^{T}.
	\label{eq:gradient}
\end{align}

The sum in \eqref{eq:stochStep} contains $N^{2}$ inner products of vectors with size $N$ and, therefore, the computational time to compute \eqref{eq:stochStep} rapidly scales proportionally to $N^{3}$. To address the computational complexity and memory requirements for large $N$, we adopt a block stochastic gradient descent strategy. That is, instead of calculating \eqref{eq:stochStep}, we choose only a random subset of the sum for each iteration $r$. Define a set $\Gamma_{(r)}$ chosen uniformly and independently at random at each iteration $r$ from subsets of $\{1,\cdots,N \}^{2}$ with cardinality $Q\in \{1,\cdots,N^{2}\}$. Then, the new gradient vector is
\begin{align}
	\mathbf{d}_{\Gamma_{(r)}}[\ell]=&\sum_{p,k\in \Gamma_{(r)}}\left( \mathbf{f}_{k}^{H}\mathbf{g}^{(r)}_{p} - \upsilon_{k,p,t} \right) \mathbf{x}^{(r)}[\ell-p]e^{2\pi i\ell k/N} +\sum_{p,k\in \Gamma_{(r)}} \left( \mathbf{f}_{k}^{T}\overline{\mathbf{g}_{p}}^{(r)} - \upsilon_{k,p} \right) \mathbf{x}^{(r)}[\ell+p]e^{-2\pi \mathrm{i}(\ell+p) k/N}.
	\label{eq:functionh}
\end{align}
Specifically, we achieve this by uniformly sampling the gradient in \eqref{eq:gradient} using a minibatch data of size $Q$ for each update step, such that the expectation $\mathbb{E}_{\Gamma_{(r)}}[\mathbf{d}_{\Gamma_{(r)}}]$ is \eqref{eq:stochStep} \citep[page 130]{spall2005introduction}. Our algorithm is summarized in Algorithm~\ref{alg:smothing} below. 
\begin{algorithm}[!t]
	\caption{Estimating band/time-limited signal from AF-based PR}
	\label{alg:smothing}
	\begin{algorithmic}[1]
		\Statex \textbf{Input: }Data $\left\lbrace\mathbf{A}[p,k]:k,p=0,\cdots,N-1 \right\rbrace$, constants $\gamma_{1},\gamma,\alpha\in(0,1)$, $\mu^{(0)}\geq 0$, cardinality $Q\in \{1,\cdots,N^{2}\}$, tolerance $\epsilon>0$, and number of iterations $R$
		\Statex \textbf{Output: } Estimate $\mathbf{x}^{(R)}$ of $\mathbf{x}$
		
		\State Initial point $\mathbf{x}^{(0)}$ (from Algorithm 2) 
		\Statex{}
		\While{$\left\lVert\mathbf{d}_{\Gamma_{(r)}}\right\rVert_{2}^{2}\geq \epsilon$ or $r\leq R$}
		\State Choose $\Gamma_{(r)}$ uniformly at random from the subsets of $\{1,\cdots,N \}^{2}$ with cardinality $Q$ 
	
	\State{\begin{align*}
			\mathbf{d}_{\Gamma_{(r)}}[\ell]\leftarrow&\sum_{p,k\in \Gamma_{(r)}}\left( \mathbf{f}_{k}^{H}\mathbf{g}^{(r)}_{p} - \upsilon_{k,p,t} \right) \mathbf{x}^{(r)}[\ell-p]e^{2\pi i\ell k/N} +\sum_{p,k\in \Gamma_{(r)}} \left( \mathbf{f}_{k}^{T}\overline{\mathbf{g}_{p}}^{(r)} - \upsilon_{k,p} \right) \mathbf{x}^{(r)}[\ell+p]e^{-2\pi \mathrm{i}(\ell+p) k/N}
	\end{align*}}
	\Statex{}
	\State{$\mathbf{g}^{(r)}_{p} \leftarrow \left[\mathbf{x}^{(r)}[0]\overline{\mathbf{x}[p]}^{(r)},\cdots,\mathbf{x}^{(r)}[N-1]\overline{\mathbf{x}[N-1+p]}^{(r)} \right]^{T}$}
	\State{$\upsilon_{k,p,t} \leftarrow \sqrt{\mathbf{A}[p,k]}\frac{\mathbf{f}_{k}^{H}\mathbf{g}^{(r)}_{p}}{\varphi_{\mu^{(r)}}\left(\left\lvert\mathbf{f}_{k}^{H}\mathbf{g}^{(r)}_{p} \right \rvert\right)} $}
	\Statex{}
	\State{$\displaystyle\mathbf{x}^{(r+1)}\leftarrow\mathbf{x}^{(r)} + \alpha^{(r)} \mathbf{b}_{\Gamma_{(r)}} = \mathbf{x}^{(r)} - \alpha^{(r)}\frac{\mu^{(r)}}{\lVert \mathbf{d}_{\Gamma_{(r)}} \rVert_{2}} \mathbf{d}_{\Gamma_{(r)}}$}
	\If{$\displaystyle\left\lVert\mathbf{d}_{\Gamma_{(r)}}\right\rVert_{2}\geq \gamma \mu^{(r)}$}
	\State{$\mu^{(r+1)}\leftarrow\mu^{(r)}$}
	\Else
	\State{ $\mu^{(r+1)} \leftarrow \gamma_{1}\mu^{(r)}$}
	\EndIf
	\State{$t\leftarrow t+1$}
	\EndWhile
	\State{\textbf{return: } $\mathbf{x}^{(R)}$}
\end{algorithmic}
\end{algorithm}

As mentioned in Section~\ref{subsec:opt}, choosing $\mu>0$ leads to a smooth cost function. This prevents a bias in the update direction and we are able to construct a descent rule for $\mu$ (line 8-12 of Algorithm \ref{alg:smothing}) in order to guarantee convergence to a first-order optimal point (that is, a point with zero gradient) in the vicinity of the solution. Note that we initialize the algorithm through a minimally iterative spectral procedure described in the next section. The following Theorem~\ref{theo:contraction} states the convergence conditions of Algorithm~\ref{alg:smothing}.
\begin{theorem} 
Let $\mathbf{x}$ be $S$-time-limited or $B$-band-limited for some $S\leq N/2$ or $B\leq N/2$, respectively, and $\mathbf{A}$ in \eqref{eq:Ambiguity} is complete. Suppose that $\Gamma_{(r)}$ is sampled uniformly at random from all subsets of $\{1,\cdots,N \}^{2}$ with cardinality $Q$, independently for each iteration $r$. Then, for almost all signals, Algorithm \ref{alg:smothing} with step size $\alpha\in (0,\frac{2}{U}] $ with $U>0$ the Lipschitz constant of Wirtinger derivative in \eqref{eq:stochStep}, constant $\gamma\in (0,1)$, and starting point $(\mathbf{x}^{(0)},\mu^{(0)})$, converges to a critical point of $h(\mathbf{z},\mu)$ in \eqref{eq:auxproblem} with $\lim_{r\rightarrow \infty} \mu^{(r)} = 0$ and 
\begin{align}
		\lim_{r\rightarrow \infty}\left\lVert\mathbb{E}_{\Gamma_{(r)}}\left[\mathbf{d}_{\Gamma_{(r)}}\right]\right\rVert^{2}_{2}\leq \gamma \lim_{r\rightarrow \infty} \mu^{(r)}\left(\frac{2h(\mathbf{x}^{(0)},\mu^{(0)})}{t} + \left(\hat{D} + \frac{2h(\mathbf{x}^{(0)},\mu^{(0)})}{\hat{D}U}\right)\frac{\sigma}{\sqrt{t}}\right)^{1/2},
		\label{eq:convergence}
\end{align}
for all $\mathbf{x}^{(r)}$ satisfying $\text{dist}(\mathbf{x},\mathbf{x}^{(r)})\leq \rho$ for some sufficiently small constant $\rho>0$, and the positive constants $\sigma,\hat{D}$ depend on $U$. 
\label{theo:contraction}
\end{theorem}
\begin{proof}
See Appendix~\ref{app:prooftheo4}.
\end{proof}
The proof of Theorem~\ref{theo:contraction} in Appendix~\ref{app:prooftheo4} exploits the AF properties \textbf{P1} and \textbf{P2} to ensure that the gradient of $h(\mathbf{x},\mu)$ is continuous. This allows us to prove that Algorithm \ref{alg:smothing} converges to a critical point as stated in \eqref{eq:convergence}. The properties \textbf{P3} and \textbf{P4} are satisfied by including the ambiguities \textbf{T1}-\textbf{T4}. Note that the convergence above is consistent with the previously known results in the Fourier-based PR such as in \citep{pauwels2018fienup}. In generalized PR problems (e.g., with Gaussian sampling), the convergence is stronger because it has only one ambiguity \textbf{T1}. 

\begin{remark}
The convergence guarantee of Algorithm~\ref{alg:smothing} in Theorem~\ref{theo:contraction} contrasts with prior works that lack algorithms to solve the AF-based PR. When compared to other bivariate PR problems such as FROG \citep{pinilla2019frequency} and STFT \citep{jaganathan2016stft}, it is more challenging to provide this guarantee for AF-based PR because of the presence of the conjugate term and the resulting additional ambiguities. The key to the proof of Theorem~\ref{theo:contraction} is the update rule for the smoothing value $\mu^{(r)}$ in Lines 7-8 of Algorithm~\ref{alg:smothing}. This rule controls the magnitude of the gradient descent direction vector $\mathbf{d}_{\Gamma^{(r)}}$ leading to its minimization with the iterative decrease in $\mu^{(r)}$.
\end{remark}

Some recent works have established finite sample guarantees for conventional PR under randomized sampling \citep{candes2013phaselift, cai2022sample}. In \citep{candes2013phaselift}, the number of samples for convex PR scales with the signal dimension when the measurements are drawn independently and uniformly at random from the unit sphere. Under the same random sampling model, \citep{cai2022sample} recently showed that sparse PR  is possible given sufficient samples. However, directly adapting these results for AF-based PR is challenging because the structured sampling via autocorrelation-type measurements is different from independent uniform random sampling. Further, current results \citep{cai2022sample} for sparse signals consider only real-valued signals. 

We now prove that our proposed iterative algorithm converges in a finite number of iterations despite the structured nature of the measurements. The following Corollary~\ref{corr:finite} states this guarantee regarding the necessary number of iterations required for Algorithm~\ref{alg:smothing} to satisfy the stopping criterion in Line 2. 

\begin{corollary}
    Under the setup of Theorem \ref{theo:contraction}, we obtain that Algorithm \ref{alg:smothing} requires at least
\begin{align}
	r > \frac{\gamma^{2} (\mu^{(0)})^{2}}{\epsilon^{2}} \left(2h(\mathbf{x}^{(0)},\mu^{(0)}) + \left(\hat{D} + \frac{2\sigma h(\mathbf{x}^{(0)},\mu^{(0)})}{\hat{D}U}\right)\right),
	\label{eq:finiteIterations1}
\end{align}
iterations to satisfy the stopping criterion $\left\lVert\mathbf{d}_{\Gamma_{(r)}}\right\rVert_{2}^{2}< \epsilon$ for a given $\epsilon>0$.
\label{corr:finite}
\end{corollary}
\begin{proof}
    The proof follows directly from \eqref{eq:convergence}, 
    wherein the minimum number of required iterations for the magnitude of the gradient to reach a value of $\epsilon>0$ (per Line 2 in Algorithm \ref{alg:smothing}) is readily obtained. 
\end{proof}

\begin{remark}
Finite convergence guarantees are rare in PR literature. 
In this context, \citep{balan2009painless} studied fast PR algorithms using tight frames. 
Reconstruction complexity grows at most cubically with the signal dimension thereby offering a speedup over previous methods. An algorithm in \citep{balan2009painless} was shown to reconstruct an $n$-dimensional signal in $\mathcal{O}(n)$ operations. 
In \citep{balan2010signal}, reconstructing discrete-time signals from absolute values of short-time Fourier coefficients requires factorizing the $z$-transform of the spectral autocorrelation function of the underlying unknown signal. This entails only finitely many possible choices and is compatible with the measurements. 
In this context, Corollary~\ref{corr:finite} above guarantees convergence within a tolerance level for AF-based PR. Note that this guarantee is sufficient for radar waveform design problem, which is usually carried out offline in contrast with the real-time conventional PR scenarios such as those mentioned in \citep{balan2009painless}. 
\end{remark}

\color{black}
\section{Initialization Algorithm}
\label{sec:initialization}
The unconstrained minimization of the function in (\ref{eq:auxproblem}) is a non-convex problem for which there is no guarantee that an arbitrary initialization will converge to a global minimum. Fig. \ref{fig:landscape} suggests the existence of spurious stationary points (i.e. saddle points) in $h(\textbf{z},\mu)$. Therefore, we devise a method to initialize the gradient iterations. This strategy approximates the signal $\mathbf{x}$ from the AF as the leading eigenvector (with appropriate normalization) of a carefully designed matrix that approximates the correlation matrix of the signal $\mathbf{x}$. 

Instead of directly dealing with the AF in \eqref{eq:Ambiguity}, we consider the acquired data in a transformed domain by taking its 1-D discrete Fourier transform (DFT) with respect to the frequency variable (normalized by $1/N$). Our measurement model is then
\begin{align}
\mathbf{Y}[p,\ell] &= \frac{1}{N}\sum_{k=0}^{N-1}\mathbf{A}[p,k]e^{-2\pi \mathrm{i} k\ell/N} = \frac{1}{N}\sum_{k,n,m=0}^{N-1}\mathbf{x}[n]\overline{\mathbf{x}[n-p]}\mathbf{x}[m-p]\overline{\mathbf{x}[m]}e^{-2\pi \mathrm{i}k\frac{(m-n-\ell)}{N}}\nonumber\\
&=\sum_{n=0}^{N-1}\mathbf{x}[n]\overline{\mathbf{x}[n-p]}\mathbf{x}[n+\ell-p]\overline{\mathbf{x}[n+\ell]},
\label{eq:system1}
\end{align}
where $p,\ell=0,\cdots,N-1$. Observe that, for fixed $p$, $\mathbf{Y}[p,\ell]$ is the autocorrelation of $\mathbf{x}\odot\overline{\mathbf{x}_{p}}$, where $\mathbf{x}_{p}[n]=\mathbf{x}[n-p]$.

Define $\mathbf{D}_{p}\in \mathbb{C}^{N\times N}$ be a diagonal matrix composed of the entries of $\mathbf{x}_{p}$, and let $\mathbf{C}_{\ell}$ be a circulant matrix that shifts the entries of a vector by $\ell$ locations, namely, $(\mathbf{C}_{\ell}\mathbf{x})[n]=\mathbf{x}[n+\ell]$. Then, the matrix $\mathbf{X}:=\mathbf{x}\mathbf{x}^{H}$ is linearly mapped to $\mathbf{Y}[p,\ell]$ as follows:
\begin{align}
\mathbf{Y}[p,\ell] &= \left(\overline{\mathbf{D}}_{-\ell+p}\mathbf{D}_{p}\mathbf{C}_{\ell}\mathbf{x}\right)^{H}\mathbf{x} = \mathbf{x}^{H}\mathbf{A}_{p,\ell}\mathbf{x}\nonumber\\
&=\textrm{Tr}(\mathbf{X}\mathbf{A}_{p,\ell}),
\label{eq:systemIniti}
\end{align}
where $\mathbf{A}_{p,\ell} = \mathbf{C}_{-\ell}\overline{\mathbf{D}}_{p}\mathbf{D}_{-\ell+p}$. Observe that $\mathbf{C}_{\ell}^{T}=\mathbf{C}_{-\ell}$. Thus, 
\begin{align}
\mathbf{y}_{\ell} = \mathbf{G}_{\ell}\mathbf{x}_{\ell},
\label{eq:initiFinal}
\end{align}
for a fixed $\ell\in \{0,\cdots,N-1\}$, where $\mathbf{y}_{\ell}[n]=\mathbf{Y}[n,\ell]$ and $\mathbf{x}_{\ell} = \text{diag}(\mathbf{X},\ell)$. The $(p,n)$-th entry of the matrix $\mathbf{G}_{\ell}\in \mathbb{C}^{\lceil\frac{N}{L} \rceil\times N}$ is
\begin{align}
\mathbf{G}_{\ell}[p,n] := \overline{\mathbf{x}_{p}[n]}\mathbf{x}_{p}[n+\ell].
\label{eq:matrixG}
\end{align}
From \eqref{eq:matrixG} it follows that $\mathbf{G}_{\ell}$ is a circulant matrix. Therefore, $\mathbf{G}_{\ell}$ is invertible if and only if the DFT of its first column, i.e. $\overline{\mathbf{x}}_{p}\odot (\mathbf{C}_{\ell}\mathbf{x}_{p})$, is non-vanishing.

Using \eqref{eq:initiFinal}, we propose a method to estimate the signal $\mathbf{x}$ from measurements \eqref{eq:Ambiguity} using an alternating method: fix $\mathbf{G}_{\ell}$, solve for $\mathbf{x}_{\ell}$, update $\mathbf{G}_{\ell}$ and so forth. After a few iterations of this two-step procedure, the output is used to initialize the gradient Algorithm~\ref{alg:smothing}. This alternating scheme is summarized in Algorithm \ref{alg:initialization}. We begin with the initial point
\begin{align}
\mathbf{x}_{\textrm{init}}[p]:=\mathbf{v}[p] \exp(i\mathbf{\theta}[p]),
\label{eq:initialvector}
\end{align}where $\mathbf{\theta}[r]\in [0,2\pi)$ is chosen uniformly at random for all $r\in \{0,\cdots,N-1\}$. The $r$-th entry of $\mathbf{v}$ corresponds to the summation of the measured AF over the frequency axis:
\begin{align}
\mathbf{v}[p] &:= \frac{1}{N}\sum_{k=0}^{N-1}\mathbf{A}[p,k] = \frac{1}{N}\sum_{k=0}^{N-1}\left\lvert\sum_{n=0}^{N-1} \mathbf{x}[n]\overline{\mathbf{x}[n-p]}e^{-2\pi \mathrm{i}nk/N} \right \rvert^{2} \nonumber\\
&:=\sum_{n=0}^{N-1} \lvert\mathbf{x}[n]\rvert^{2}\lvert \mathbf{x}[n-p]\rvert^{2}.
\label{eq:vectorv}
\end{align}
Once $\mathbf{x}_{\textrm{init}}$ is constructed, the vectors $\mathbf{x}_{\ell}^{(r)}$ at $t=0$ are constructed as 
\begin{align}
\mathbf{x}_{\ell}^{(0)}=\text{diag}(\mathbf{X}_{0}^{(0)},\ell),
\label{eq:initialguess}
\end{align}
where 
\begin{equation}
\mathbf{X}_{0}^{(0)}=\mathbf{x}_{\textrm{init}}\mathbf{x}_{\textrm{init}}^{H}.
\label{eq:X0}
\end{equation}Then, from \eqref{eq:initialguess}, we proceed with an alternating procedure between estimating the matrix $\mathbf{G}_{\ell}$, and updating the vector $\mathbf{x}_{\ell}$ as follows.

\begin{itemize}

\item \textit{Update rule for $\mathbf{G}_{\ell}$: }In order to update $\mathbf{G}_{\ell}$, we update the matrix $\mathbf{X}_{0}^{(r)}$ as
\begin{align}
	\text{diag}(\mathbf{X}_{0}^{(r)},\ell) = \mathbf{x}_{\ell}^{(r)}.
	\label{eq:solZ0}
\end{align}

Observe that if $\mathbf{x}_\ell^{(r)}$ is close to $\mathbf{x}_\ell$ for all $\ell$, then $\mathbf{X}_0^{(r)}$ is close to $\mathbf{x}\mathbf{x}^H$. Assume $\mathbf{w}^{(r)}$ to be the leading (unit-norm) eigenvector of the matrix $\mathbf{X}_{0}^{(r)}$ constructed in \eqref{eq:solZ0}. From \eqref{eq:matrixG}, each matrix $\mathbf{G}^{(r)}_{\ell}$ at iteration $r$ is 
\begin{align}
	\mathbf{G}_{\ell}^{(r)}[p,n] = \overline{\mathbf{x}_{p}}^{(r)}[n]\mathbf{x}^{(r)}_{p}[n+\ell],
	\label{eq:matrixG1}
\end{align}
where $\mathbf{x}^{(r)}_{p}[n]= \mathbf{w}^{(r)}[n-p]$. \vspace{0.5em}

\item \textit{Optimization with respect to $\mathbf{x}_{\ell}$: } For a fixed $\mathbf{G}_{\ell}^{(r-1)}$, we estimate $\mathbf{x}^{(r)}_{\ell}$ at iteration $r$ by solving the linear least-squares (LS) problem 
\begin{align}
	\underset{\mathbf{p}_{\ell}\in \mathbb{C}^{N}}{\textrm{minimize}}\; & \hspace{1em} \lVert \mathbf{y}_{\ell}-\mathbf{G}^{(r-1)}_{\ell}\mathbf{p}_{\ell} \rVert_{2}^{2}.
	\label{eq:problemInitiz}
\end{align}
The relationship between the vectors $\mathbf{x}^{(r)}_{\ell}$ is ignored at this stage. If $\mathbf{G}_{\ell}^{(r-1)}$ is invertible, then the solution to this problem is given by $(\mathbf{G}_{\ell}^{(r-1)})^{-1}\mathbf{y}_{\ell}$. Since $\mathbf{G}_{\ell}^{(r-1)}$ is a circulant matrix, it is invertible if and only if the DFT of $\overline{\mathbf{x}}^{(r-1)}\odot (\mathbf{C}_{\ell}\mathbf{x}^{(r-1)})$ is non-vanishing. In general, this condition cannot be ensured. Thus, we propose a surrogate proximal optimization problem to estimate $\mathbf{x}_{\ell}^{(r)}$ by
\begin{align}
	\underset{\mathbf{p}_{\ell}\in \mathbb{C}^{N}}{\textrm{minimize}}\; & \hspace{1em} \lVert \mathbf{y}_{\ell}-\mathbf{G}^{(r-1)}_{\ell}\mathbf{p}_{\ell} \rVert_{2}^{2} + \frac{1}{2\lambda_{(r)}}\lVert \mathbf{p}_{\ell} - \mathbf{x}_{\ell}^{(r-1)} \rVert_{2}^{2},
	\label{eq:problemInitiz1}
\end{align}
where $\lambda_{(r)}>0$ is a regularization parameter. In practice, $\lambda_{(r)}$ is a tunable parameter \citep{parikh2014proximal}. Specifically, in this work, the value of $\lambda_{(r)}$ was determined using a cross-validation strategy such that each simulation uses the value that results in the smallest relative error according to \eqref{eq:distance}. The surrogate optimization problem in \eqref{eq:problemInitiz1} is strongly convex~\citep{parikh2014proximal}, and admits the following closed form solution
\begin{align}
	\mathbf{x}_{\ell}^{(r)} = \mathbf{B}_{\ell,t}^{-1} \mathbf{e}_{\ell,t},
	\label{eq:solZl}
\end{align}
where
\begin{align}
	\mathbf{B}_{\ell,t} &= \left(\mathbf{G}^{(r-1)}_{\ell}\right)^{H}\left(\mathbf{G}^{(r-1)}_{\ell}\right) + \frac{1}{2\lambda}\mathbf{I}, \nonumber\\
	\mathbf{e}_{\ell,t}&=\left(\mathbf{G}^{(r)}_{\ell}\right)^{H}\mathbf{y}_{\ell} + \frac{1}{2\lambda}\mathbf{x}^{(r-1)}_{\ell},
	\label{eq:auxMatrices}
\end{align}
with $\mathbf{I}\in \mathbb{R}^{N\times N}$ the identity matrix. Clearly $\mathbf{B}_{\ell,t}$ in \eqref{eq:auxMatrices} is always invertible. The update step for each $\mathbf{x}_{\ell}^{(r)}$ is computed in line 9 of Algorithm \ref{alg:initialization}.
\end{itemize}

\begin{algorithm}[t]
\caption{Initialization of AF-based PR}
\label{alg:initialization}
\begin{algorithmic}[1]
	\Statex{\textbf{Input: }Measurements $\mathbf{A}[p,k]$, number of iterations $R$, and $\lambda>0$}
	\Statex{\textbf{Output:} Estimate $\mathbf{x}^{(0)}$ of $\mathbf{x}$}
	\State{\textbf{Initialize: }$\mathbf{x}_{\textrm{init}}[p]\leftarrow\mathbf{v}[p] \exp(i\mathbf{\theta}[p])$, and $\displaystyle\mathbf{v}[p]\leftarrow\frac{1}{N}\sum_{k=0}^{N-1}\mathbf{A}[p,k]$, $\mathbf{\theta}[p]\in [0,2\pi)$ is chosen uniformly and independently at random}
	\State{Compute $\mathbf{Y}[p,\ell]$ the 1-D inverse DFT with respect to $k$}
	\Statex{of $\mathbf{A}[p,k]$}
	\For{$t=1$ to $R$}
	\State{Construct $\mathbf{G}^{(r)}_{\ell}$ according to \eqref{eq:matrixG1}.}
	\State{ $\mathbf{B}_{\ell,t}\leftarrow (\mathbf{G}^{(r)}_{\ell})^{H}(\mathbf{G}^{(r)}_{\ell}) + \frac{1}{2\lambda}\mathbf{I}$}
	\State{ $\mathbf{e}_{\ell,t}\leftarrow(\mathbf{G}^{(r)}_{\ell})^{H}\mathbf{y}_{\ell} + \frac{1}{2\lambda}\mathbf{x}^{(r-1)}_{\ell}$.}
	\State{Construct the matrix $\mathbf{X}^{(r)}_{0}$ such that
		$\text{diag}(\mathbf{X}^{(r)}_{0},\ell)=
		\mathbf{B}_{\ell,t}^{-1} \mathbf{e}_{\ell,t}, \hspace{1em} \ell=0,\cdots,N-1$
	}
	\State{Set $\mathbf{w}^{(r)}$ be the leading (unit-norm) eigenvector of $\mathbf{X}^{(r)}_{0}$}
	\State{
		$\mathbf{x}^{(r)}_{p}[n]\leftarrow \mathbf{w}^{(r)}[n-p]$}
	\EndFor
	\State{
		$\mathbf{x}^{(0)} \leftarrow\sqrt[4]{\sum_{n\in \mathcal{S}}\left(\mathbf{B}_{0,T}^{-1} \mathbf{e}_{0,T}\right)[n]}\mathbf{w}^{(R)}$
		where $\mathcal{S}:=\left\lbrace n: \left(\mathbf{B}_{0,T}^{-1} \mathbf{e}_{0,T}\right)[n]>0\right\rbrace$}
	\State{\textbf{return: }$\mathbf{x}^{(0)}$}
\end{algorithmic}
\end{algorithm}

Finally, in order to estimate $\mathbf{x}$, the (unit-norm) principal eigenvector of $\mathbf{X}_{0}^{(R)}$ is normalized by
\begin{align}
\beta=\sqrt[4]{\sum_{n\in \mathcal{S}}\left(\mathbf{B}_{0,T}^{-1} \mathbf{e}_{0,T}\right)[n]},
\label{eq:normestimated}
\end{align}
where $\mathcal{S}:=\left\lbrace n: \left(\mathbf{B}_{0,T}^{-1} \mathbf{e}_{0,T}\right)[n]>0\right\rbrace$ is the set with indices that correspond to the positive entries of the vector $\mathbf{B}_{0,T}^{-1} \mathbf{e}_{0,T}$. Observe that \eqref{eq:normestimated} results from the fact that $\displaystyle \sum_{n=0}^{N-1}\text{diag}(\mathbf{X},0)[n]=\lVert\mathbf{x}\rVert^{4}_{2}$. 

The following Theorem~\ref{theo:initialization} establishes that the resulting $\mathbf{x}^{(0)}$ from the update rules in \eqref{eq:matrixG1} and \eqref{eq:solZl} leads to a close estimation of the real unknown signal $\mathbf{x}$ for complete radar AF. 
\begin{theorem}
Assume that $\mathbf{A}$ in \eqref{eq:Ambiguity} is complete and the regularization parameter $\lambda_{(r)}$ satisfies $\lambda_{(r)}\sigma^{2}_{\textrm{min}}(\mathbf{G}^{(r-1)})>1/2$ for all $t>0$ where $\sigma_{\textrm{min}}(\mathbf{G}^{(r-1)})$ is the smallest singular value of $\mathbf{G}_{\ell}^{(r-1)}$ greater than zero. Then, initializing with $\mathbf{x}_{\textrm{init}}$ in \eqref{eq:initialvector}, the estimated vector $\mathbf{x}^{(0)}$ in Algorithm \ref{alg:initialization} satisfies
\begin{equation}
	\text{dist}(\mathbf{x}_{\ell},\mathbf{x}_{\ell}^{(r)}) < \tau \hspace{0.5em} \text{dist}(\mathbf{x}_{\ell},\mathbf{x}_{\ell}^{(0)}),
\end{equation}
for some $\tau\in (0,1)$.
\label{theo:initialization}
\end{theorem}
\begin{proof}
See Appendix~\ref{app:initia}.
\end{proof}

The initialization procedures in previous studies \citep{bendory2018non} usually relied on constructing a \textit{fixed} matrix to approximate the correlation matrix $\mathbf{x}\mathbf{x}^{H}$ of the signal itself, and then extracting its principal eigenvector (with appropriate normalization). However, in our initialization procedure, such a matrix $\mathbf{G}_{\ell}$ is obtained iteratively. In \citep{pinilla2019frequency}, initialization similar to ours was employed for FROG PR but it lacked any performance guarantees.

\section{Numerical Experiments}
\label{sec:numerical}
We validated our models and methods through extensive numerical experiments. We examine the ability of Algorithm~\ref{alg:smothing} to recover the band- and time-limited signals from complete and incomplete data in both noiseless and noisy settings. The radar function is incomplete when only few shifts or Fourier frequencies are considered. Here, we define the SNR$ = 10\log_{10}(\lVert \mathbf{A} \rVert^{2}_{\mathcal{F}}/\lVert \mathbf{\sigma} \rVert^{2}_{\text{2}})$, where $\mathbf{\sigma}$ is the variance of the noise. The signals used in the simulations were constructed as follows. For all tests, we built a set of $\left\lceil \frac{N-1}{2} \right\rceil$-band-limited and time-limited signals that conform to a Gaussian power spectrum. 
Specifically, each signal ($N=128$ grid points) is produced via the Fourier transform of a complex vector with a Gaussian-shaped amplitude. 
Next, we multiply the obtained power spectrum by a uniformly distributed random phase. We used the inverse Fourier transform of this signal as the underlying (original) signal. Note that we have normalize the Doppler and delays to $[0,1]$.

\begin{figure*}[ht]
	\centering
	\includegraphics[width=0.9\linewidth]{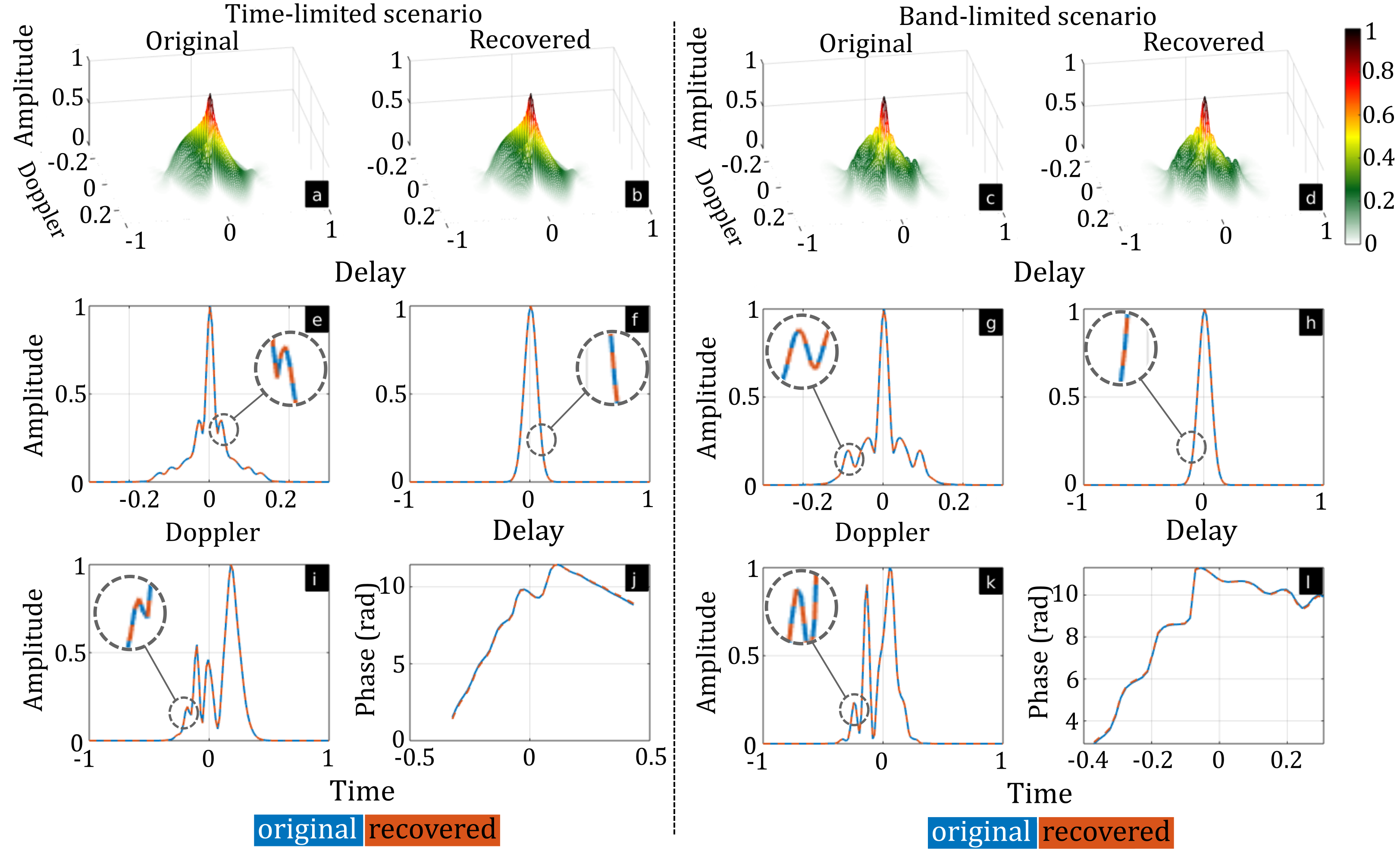}
	\caption{Reconstructed time- and band-limited signals with their AFs in the absence of noise. In both cases, the attained relative error as defined by \eqref{eq:distance} is $1\times 10^{-6}$. For time-limited [band-limited] signal, (a) [(c)] and (b) [(d)] show the original and recovered AFs, respectively; (e) [(g)] and (f) [(h)] are 1-D slices of the AFs at zero delay and Doppler, respectively; (i) [(k)] and (j) [(l)] are the, respectively, magnitude and phase of recovered (red) signal juxtaposed with the original (blue).}
	\label{fig:complete_sinnoiseresults}
\end{figure*}

Throughout the experiments, we used the following parameters for Algorithm \ref{alg:smothing}: $\gamma_{1}=0.1$, $\gamma=0.1$, $\alpha = 0.6$, $\mu_{0}=65$, and $\epsilon=1\times 10^{-10}$\footnote{All simulations were implemented in Matlab R2019a on an Intel Core i7 $3.41$ Ghz CPU with $32$ GB RAM.}. The number of indices chosen uniformly at random were set to $Q=N$.

\begin{figure*}[t!]
\centering
\includegraphics[width=0.9\linewidth]{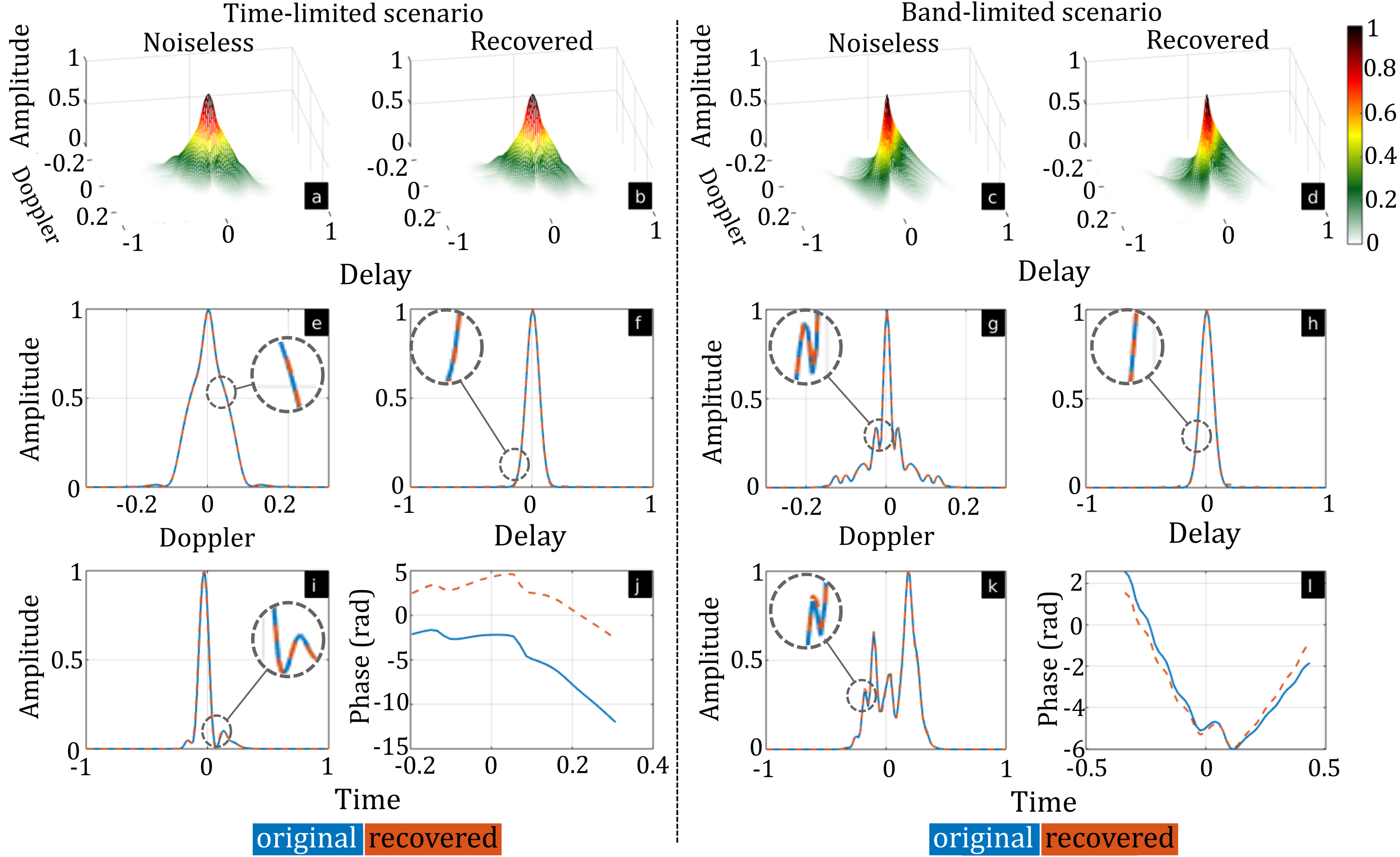}
\caption{As in Fig.~\ref{fig:complete_sinnoiseresults} but in the presence of noise with SNR = $20$ dB. The attained relative error 
	is $5\times 10^{-2}$ for both time- and band-limited signals. 
}
\label{fig:complete_noiseresults}
\end{figure*}

\begin{figure*}[t!]
\centering
\includegraphics[width=0.9\linewidth]{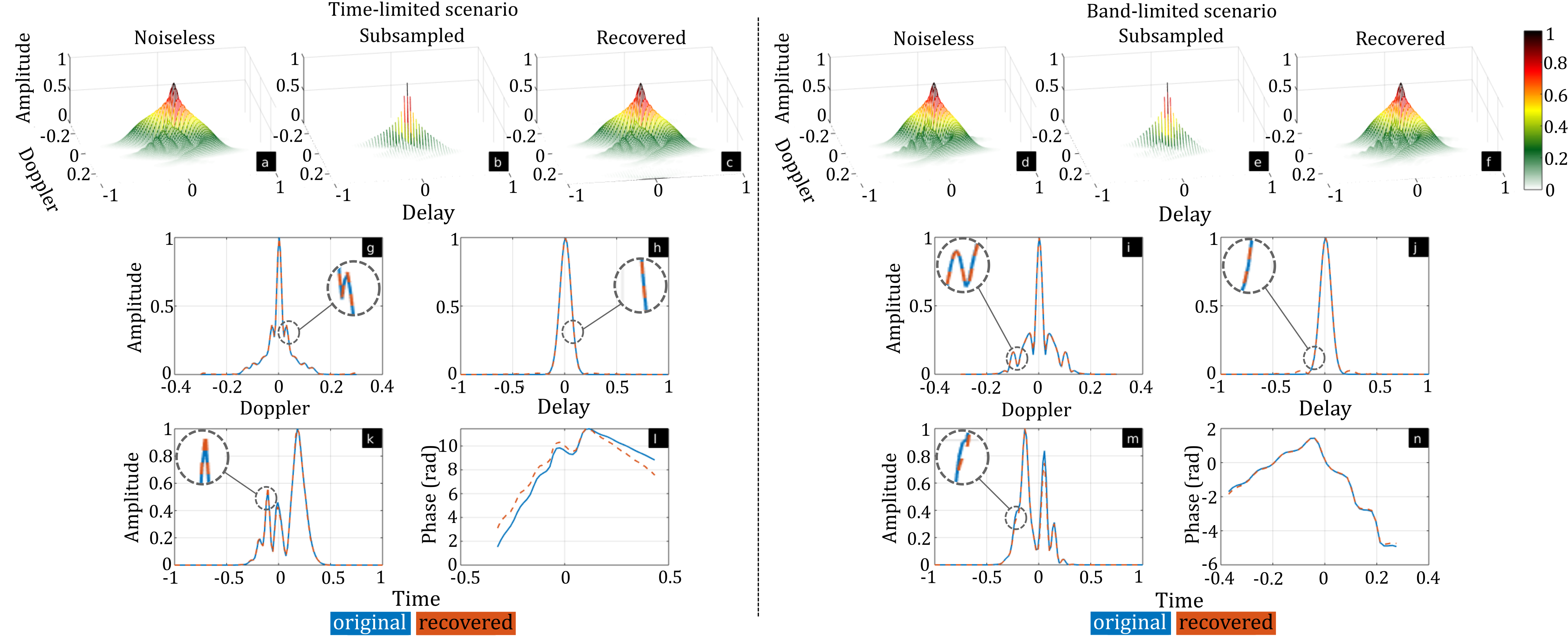}
\caption{As in Fig.~\ref{fig:complete_sinnoiseresults} but in the presence of noise with SNR = $20$ dB and with 50\% of the delays uniformly removed from the AF. The attained relative error is $5\times 10^{-2}$ for both time- and band-limited signals. Additionally, For time-limited [band-limited] signal, (a) [(d)], (b) [(e)], and (c) [(f)] show the original, undersampled, and recovered AFs, respectively; (g) [(i)] and (h) [(j)] are 1-D slices of the AFs at zero delay and Doppler, respectively; (k) [(m)] and (l) [(n)] are the, respectively, magnitude and phase of recovered (red) signal juxtaposed with the original (blue). 
}
\label{fig:incomplete_noisyresults}
\end{figure*}

\begin{figure*}[ht]
\centering
\includegraphics[width=0.9\linewidth]{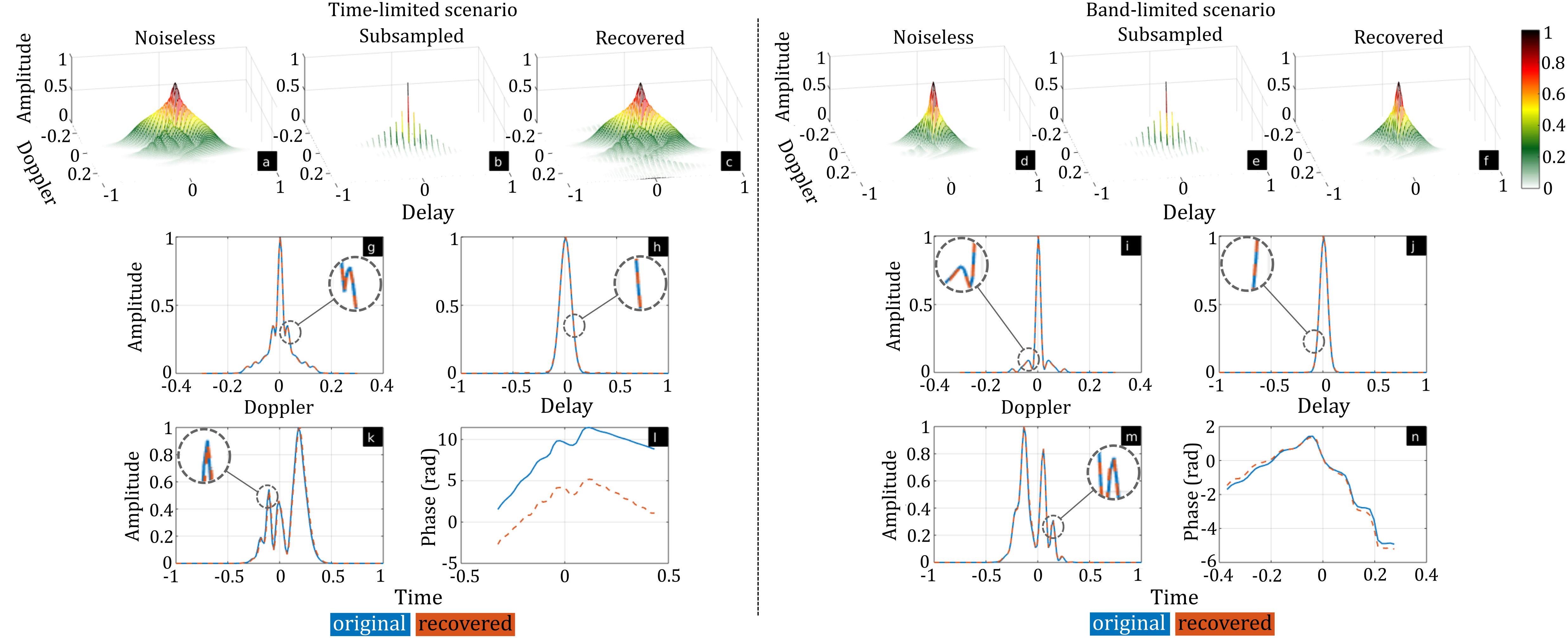}
\caption{As in Fig.~\ref{fig:incomplete_noisyresults} but with 75\% of the delays uniformly removed from the AF. 
	The attained relative error 
	is $5\times 10^{-2}$ for both time- and band-limited signals.
}
\label{fig:incomplete_noisyresults1}
\end{figure*}
\noindent\textbf{Signal reconstruction from complete AF}: Figs.~\ref{fig:complete_sinnoiseresults} and \ref{fig:complete_noiseresults} show the recovery of time and band-limited signals from a given complete AF in noiseless and noisy settings, respectively. For the latter, the radar AF trace is corrupted by Gaussian noise with SNR = $20$ dB. Here, the ``noisy'' AF implies that AF is not perfectly designed thus allowing us to evaluate the robustness of Algorithm~\ref{alg:smothing}. A comparison of estimated signals with the original using the error metric as defined by \eqref{eq:distance} in the results of Figs.~\ref{fig:complete_sinnoiseresults} and \ref{fig:complete_noiseresults} suggests near-perfect recovery using Algorithm~\ref{alg:smothing}.

\noindent\textbf{Signal reconstruction from uniformly undersampled AF}: Figs.~\ref{fig:incomplete_noisyresults} and \ref{fig:incomplete_noisyresults1} show the estimated time- and band-limited signals from a noisy AF in which 50\% and 75\% of the delays were removed, respectively. Here, the delays were performed uniformly. For instance, in case of 50\% removal, every two delays starting from the first one are preserved. Similarly, in case of 75\% removal, one out of every four delays starting from the first one is preserved; this implies a total of $32$ delays are used from the available $128$ points. These results numerically validate Proposition~\ref{prop:uniqueness} and Corollary~\ref{coro:time}) that all delays may not be required to estimate the underlying signal. We observe that Algorithm \ref{alg:smothing} is able to return a close estimation of the signal even when the incomplete AF is assumed imperfectly designed in Figs.~\ref{fig:incomplete_noisyresults} and \ref{fig:incomplete_noisyresults1}. This further suggests robust performance of Algorithm \ref{alg:smothing}.

\begin{figure*}[t!]
\centering
\includegraphics[width=0.9\linewidth]{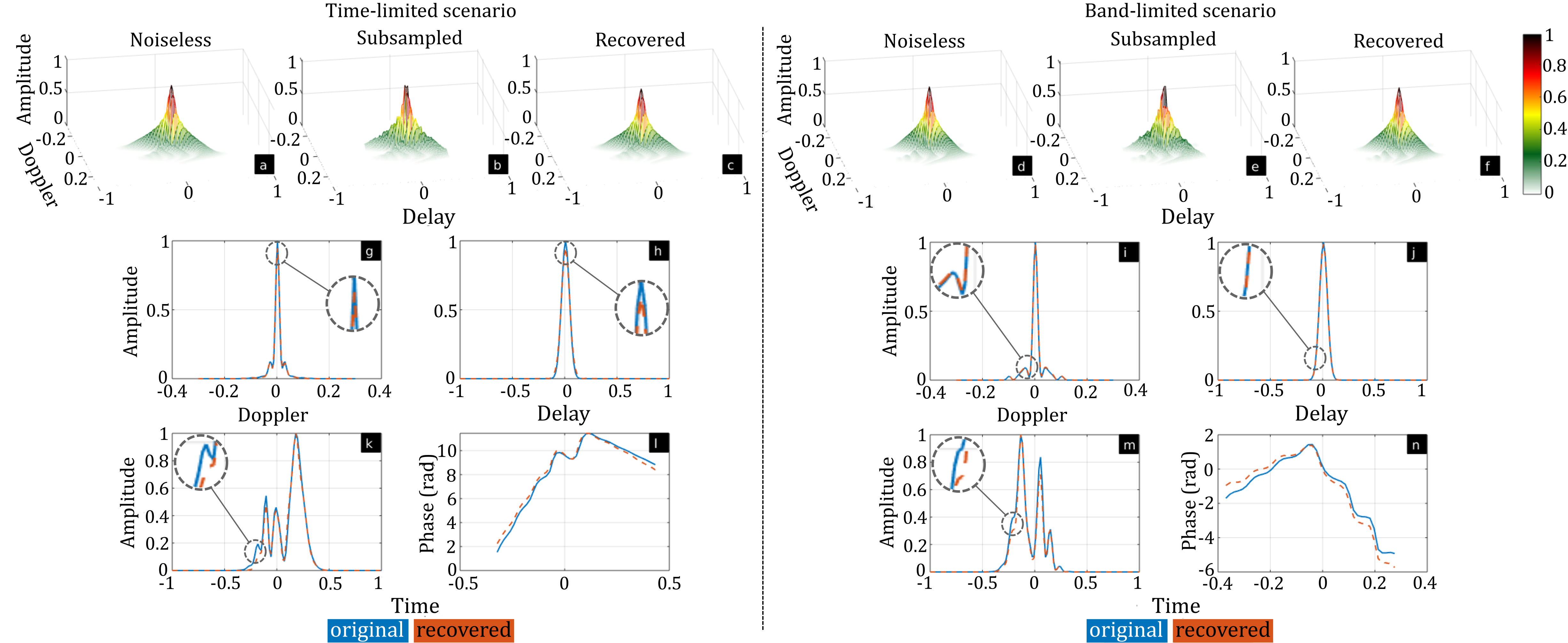}
\caption{As in Fig.~\ref{fig:incomplete_noisyresults} but with 50\% of the delays non-uniformly removed from the AF. 
	The attained relative error 
	is $9\times 10^{-2}$ for both time- and band-limited signals. 
}
\label{fig:incomplete_noisyresults2}
\end{figure*}

\begin{figure*}[t!]
\centering
\includegraphics[width=0.9\linewidth]{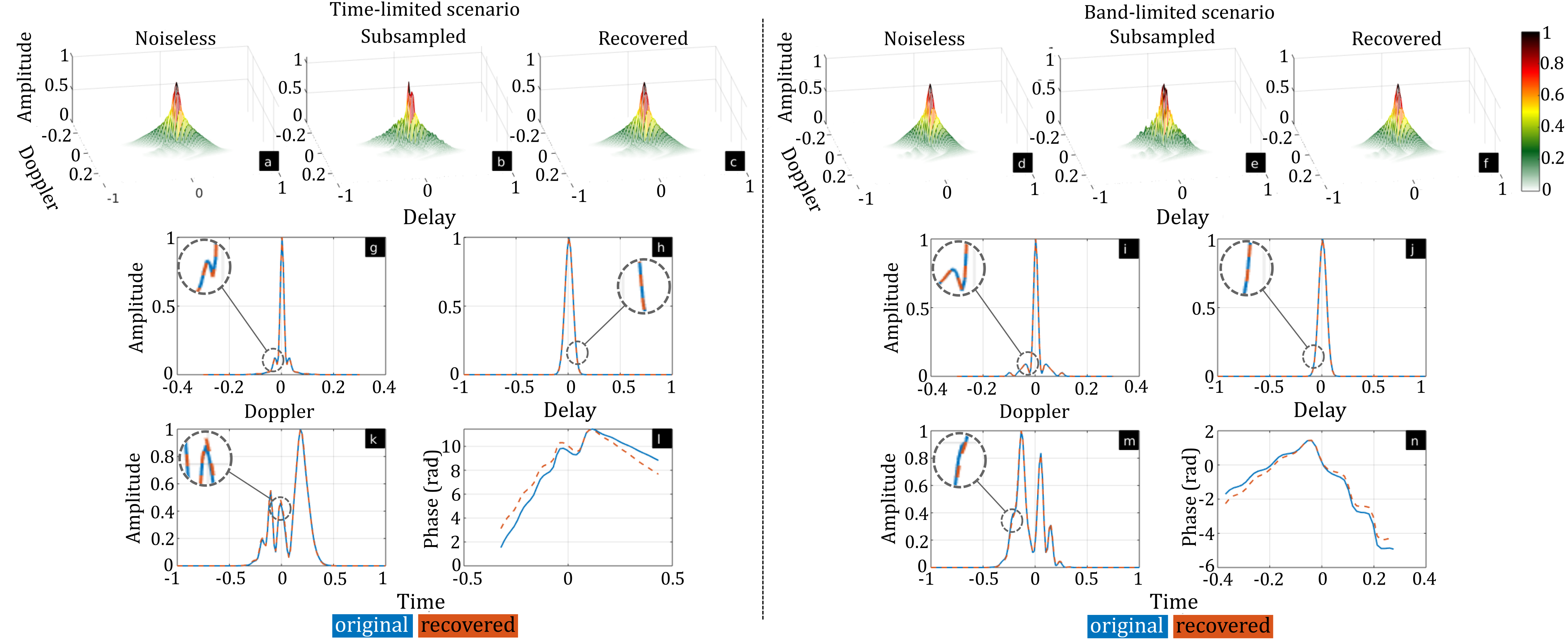}
\caption{As in Fig.~\ref{fig:incomplete_noisyresults} but with 50\% of the Doppler frequencies non-uniformly removed from the AF. 
	The attained relative error 
	is $6\times 10^{-2}$ for both time- and band-limited signals. 
}
\label{fig:incomplete_noisyresults3}
\end{figure*}

\noindent\textbf{Signal reconstruction from randomly undersampled AF}: 
Next, we evaluated the performance of Algorithm~\ref{alg:smothing} when some delays as well as the Doppler frequencies in the AF were non-uniformly removed. Figs.~\ref{fig:incomplete_noisyresults2} and \ref{fig:incomplete_noisyresults3} show the recovery performance when 25\% of the first and last (i.e., a total of 50\%) delays and frequencies, respectively, of the AF were set to zero. 
A higher relative error in the results of Fig.~\ref{fig:incomplete_noisyresults2} suggests that a non-uniform selection of the removed delays reduces the ability of Algorithm~\ref{alg:smothing}. 
On the other hand, the error does not increase significantly when frequencies are non-uniformly removed (Fig.~\ref{fig:incomplete_noisyresults3}).

\begin{figure*}[t!]
\centering
\includegraphics[width=0.9\linewidth]{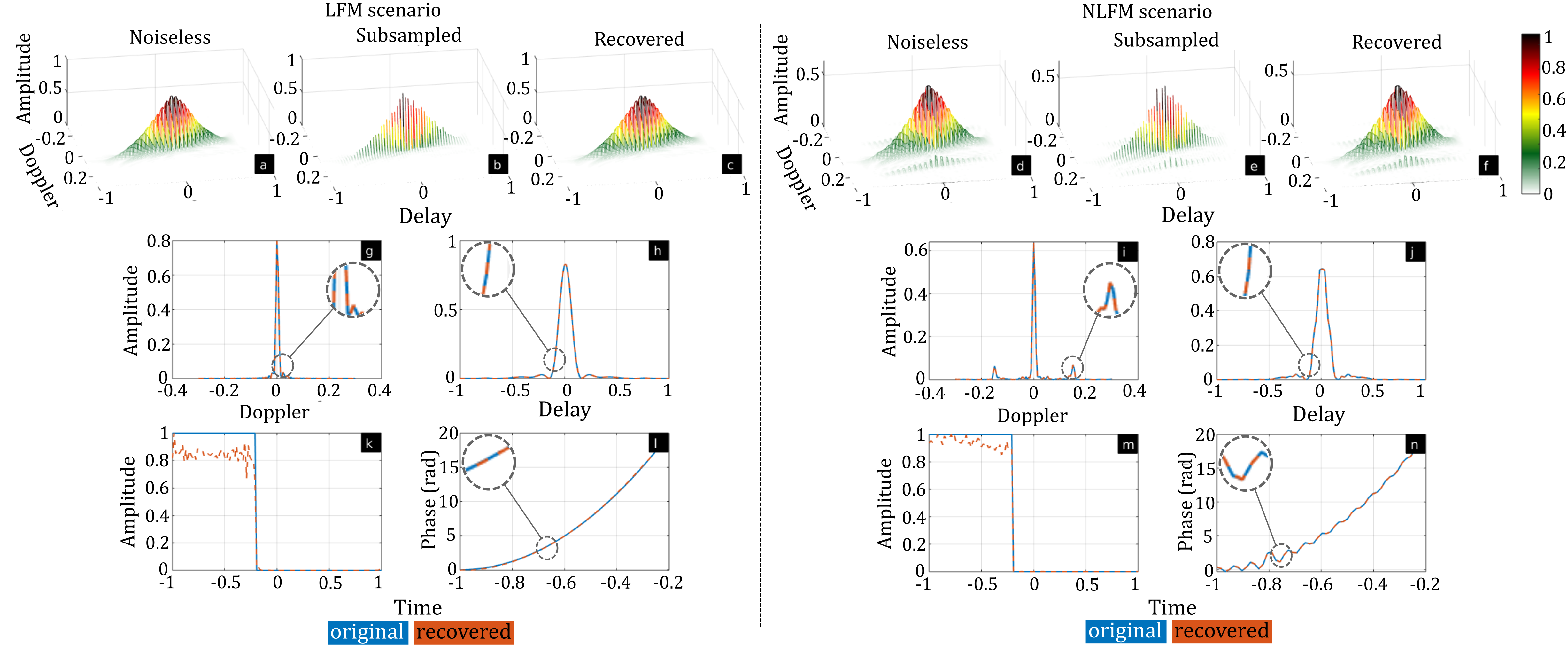}
\caption{As in Fig.~\ref{fig:incomplete_noisyresults} but time- and band-limited cases replaced by LFM and NLFM signals.
	The attained relative error 
	is $6\times 10^{-2}$ for both time- and band-limited signals. 
}
\label{fig:incomplete_noisyresults4}
\end{figure*}

\begin{figure*}[t!]
\centering
\includegraphics[width=0.9\linewidth]{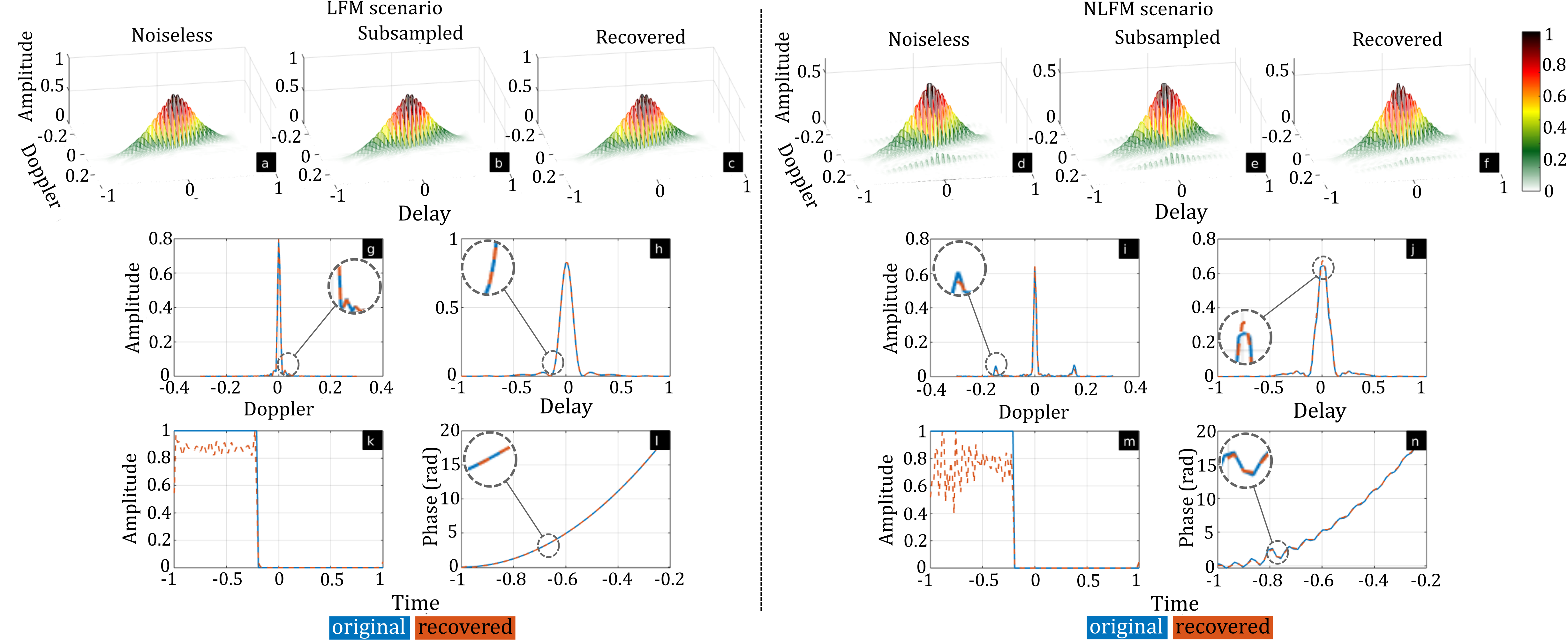}
\caption{As in Fig.~\ref{fig:incomplete_noisyresults4} but with 19\% of the first and last Doppler frequencies of the AFs removed.
	The attained relative error 
	is $9\times 10^{-2}$ for both time- and band-limited signals. 
}
\label{fig:incomplete_noisyresults5}
\end{figure*}

\noindent\textbf{Modulated signals}: So far, we considered that the underlying signals are only time- or band-limited. We now impose an additional constraint of modulation, i.e. LFM and NLFM, to investigate the performance of Algorithm \ref{alg:smothing}. These signals are very commonly used by radar systems \citep{peebles1998radar}. 
We model these signals as
\begin{align}
\mathbf{x}[n] = \mathbf{a}[n]e^{\mathrm{j}\pi \varphi[n]},
\end{align}
where 
\begin{equation}
\mathbf{a}[n] = \left \lbrace \begin{array}{ll}
	1 & 0\leq \Delta tn \leq T \\
	0 & \text{otherwise}
\end{array} \right..
\end{equation}
is a rectangular envelope, $T$ as the duration of the pulse, $\Delta t$ as the sampling size in time, and 
\begin{align}
\varphi[n] &= \pi k (\Delta tn)^{2}, \hspace{0.5em}&(\text{LFM}) \nonumber\\
\varphi[n] &= \pi k t^{2} + \sum_{l=1}^{L}\alpha_{l}\cos(2\pi l \Delta tn/T) \hspace{0.5em}&(\text{NLFM}),
\end{align}
with $k=\frac{\Delta f}{T}$ such that $\Delta f$ is the swept bandwidth and $L>0$ is an integer. In this experiment, we use 
$\alpha_{l} = \frac{0.4T}{l}$, $\Delta f = 128\times 10^{3}$, and $\Delta t = 0.4\times 10^{-6}$. 

We considered two noisy scenarios each with $SNR = 20$dB. Fig.~\ref{fig:incomplete_noisyresults4} shows the recovery performance for both LFM and NLFM signals when the 50\% of the delays are uniformly removed from the the AF. In Fig.~\ref{fig:incomplete_noisyresults5}, 19\% of the first and last Fourier frequencies of the AF were removed. These results suggest that Algorithm \ref{alg:smothing} accurately estimates the phase of the pulses while the reconstructed magnitudes present some artifacts. This limitation comes from the fact that the LFM/NLFM AF is significantly wide such that the removed information is enough to degrade the reconstruction quality.

\begin{figure}[ht]
\centering
\includegraphics[width=0.6\linewidth]{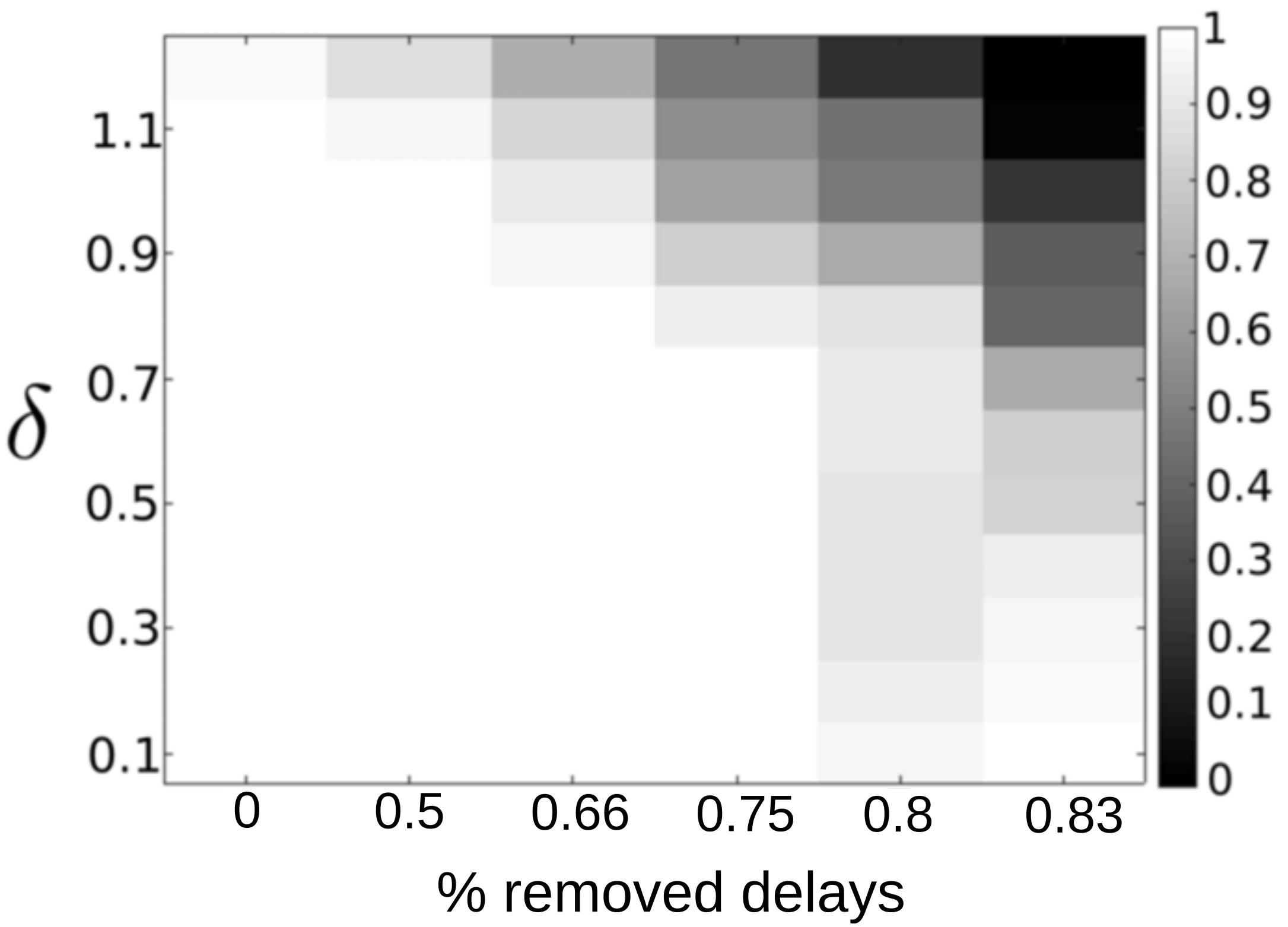}
\caption{Empirical success rate of Algorithm \ref{alg:smothing} as a function of \% removed delays (uniformly) and $\delta$ in the absence of noise.}
\label{fig:orthoini}
\end{figure}

\noindent\textbf{Statistical performance}: We evaluated the success rate of Algorithm~\ref{alg:smothing} when the AF was incomplete. To this end, we initialized the algorithm at $\mathbf{x}^{(0)} = \mathbf{x}+\delta\zeta$, where $\delta$ is a fixed constant and $\zeta$ takes values on $\{-1,1\}$ with equal probability, while a percentage of the removed delays were set to zero. A trial was declared successful when the returned estimate attains a relative error (defined by \eqref{eq:distance}) that is smaller than $10^{-6}$. We numerically determined the empirical success rate over 100 trials. Fig.~\ref{fig:orthoini} summarizes these results for the case of time-limited signals and shows a high success rate of Algorithm~\ref{alg:smothing} even when a significant number of delays are removed from the AF.

\begin{figure}[ht]
\centering
\includegraphics[width=0.6\linewidth]{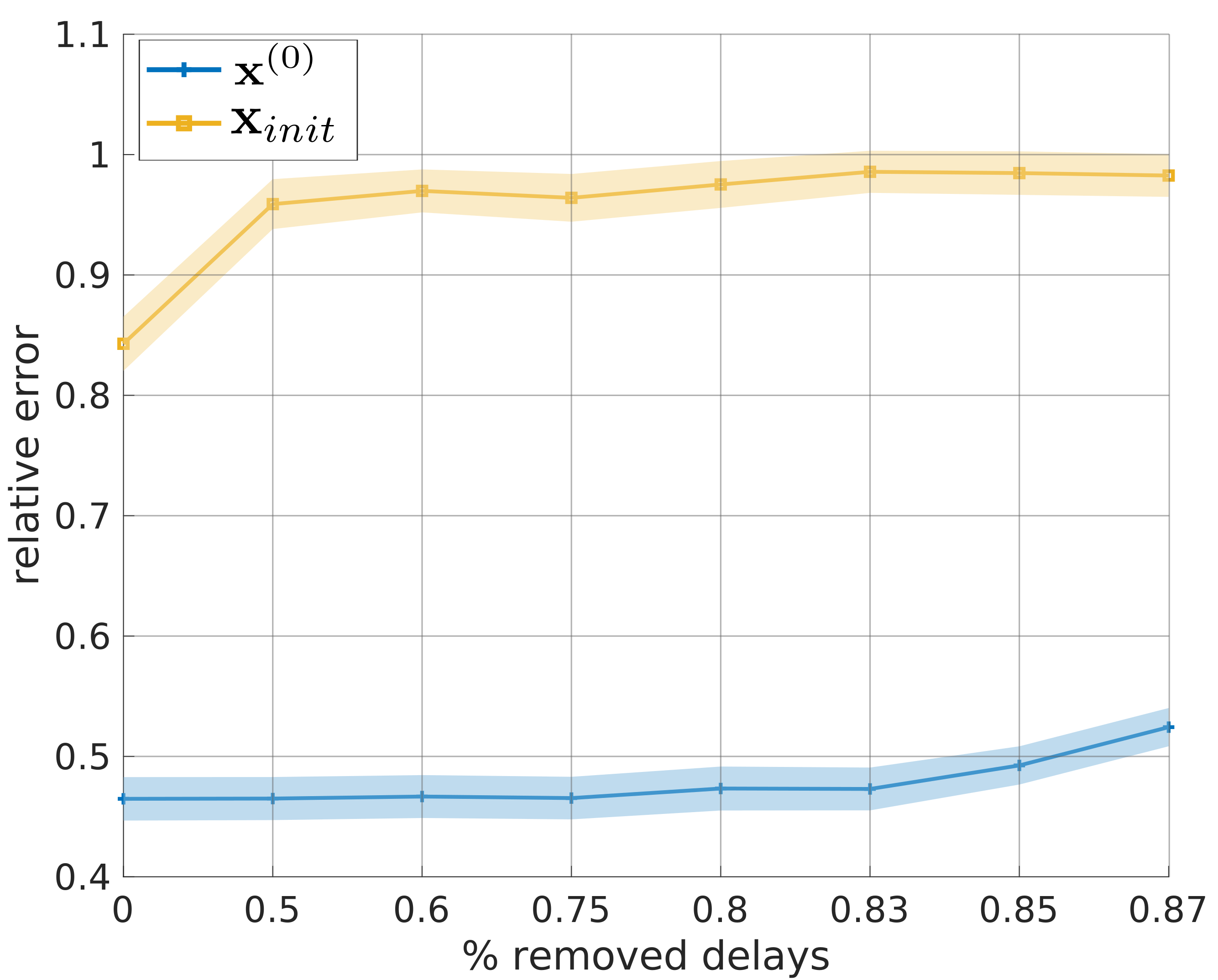}
\caption{Relative error comparison between the initial vector $\mathbf{x}_{\textrm{init}}$ as defined in \eqref{eq:initialvector}, and the returned initial guess $\mathbf{x}^{(0)}$ for different percentages of (uniformly) removed delays in a noiseless setting. The relative error was averaged over 100 trials. The shaded background region represents the variance of the relative error over 100 trials, while the solid lines are the mean.}
\label{fig:initerrors}
\end{figure}

\begin{figure}[ht]
\centering
\includegraphics[width=0.65\linewidth]{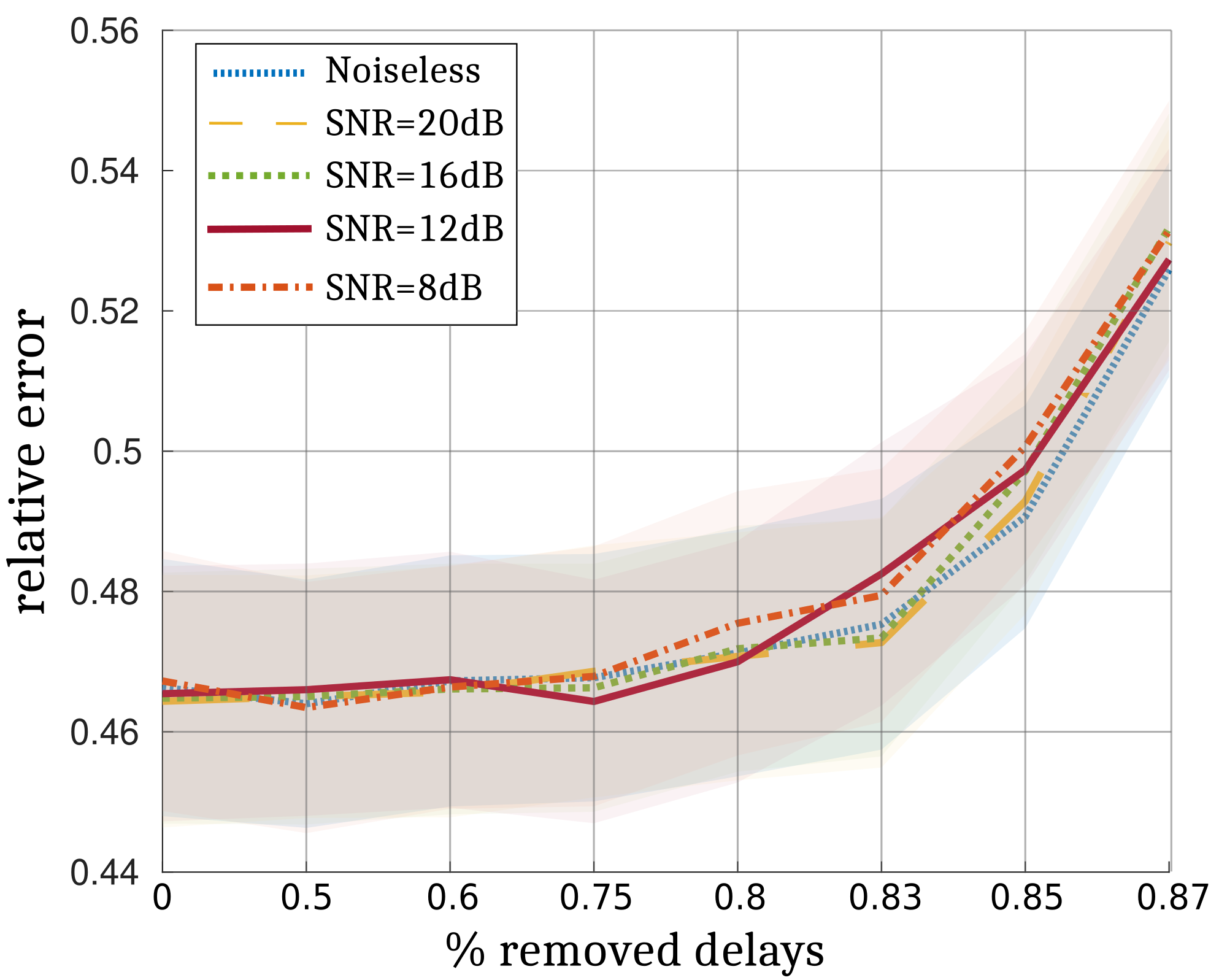}
\caption{Performance of the proposed initialization (Algorithm \ref{alg:initialization}) for different SNR values as the (uniformly) removed delays are increased. The relative error was averaged over 100 trials. The shaded background represents the variance of the relative error over 100 trials, while the solid lines are the mean. }
\label{fig:init}
\end{figure}

\noindent\textbf{Initialization procedure error}: Finally, we examined the impact of our proposed initialization procedure in Algorithm~\ref{alg:initialization} under noiseless (Fig.~\ref{fig:initerrors}) and noisy (Fig.~\ref{fig:init})scenarios. We compare the relative error between the initial vector in \eqref{eq:initialvector} and the returned solution $\mathbf{x}^{(0)}$ of the proposed method. The number of iterations to attain the vector $\mathbf{x}^{(0)}$ using the designed initialization was fixed to $R=2$. We averaged the relative error over 100 trials. The results for both settings show that the proposed initialization algorithm outperforms $\mathbf{x}_{\textrm{init}}$.

\section{Summary}
\label{sec:conclusion}
There is a large body of literature on radar waveform design, including adapting existing waveforms for specialized tasks pertaining to detection, estimation, and tracking. Several non-AF-based waveform design techniques also exist (see \citep{he2012waveform} for a survey), in which the waveform is optimized to obey the specific SNR or sidelobe levels. However, modern agile radar systems require arbitrarily designing the transmit waveform to meet very specific AF performance requirements. Whereas prior works apply classical GSA to Sussman's least-squares synthesis \citep{sussman1962least} of waveforms from their AFs, their convergence and recovery guarantees are inferior. Hence, it is beneficial to investigate recent advances in PR for AF-based signal recovery.

In this context, we analytically demonstrated that time/band-limited signals can be perfectly estimated (up to trivial ambiguities) from their (phaseless) AF. Our proposed trust-region gradient method estimated these signals under complete/incomplete and noisy/noiseless scenarios. We verified the recovery with sufficient accuracy for complete AF. In particular, our new spectral initialization method suggested substantial performance enhancement.

In the case of incomplete data, whereas Proposition~\ref{prop:uniqueness} and Corollary~\ref{coro:time} suggested that full AF is not required to guarantee uniqueness, more work is required in future to better estimate the pulses from incomplete data. Numerical results indicate that signals with wider AFs may not be retrieved accurately in undersampled scenario. 
The initialization strategy employed may also be improved. The AF-based PR optimization problem is highly non-convex for which we use an alternating approach. Hence, the presence of saddle points and local minimum should be avoided. We fixed the algorithmic parameters through hit-and-trial but they may also be learned from the kind of signals. Finally, as we mentioned in the beginning, we remark that AF is not unique in definition. Therefore, generalization of this work to other more complex definitions will provide critical insights in apply PR theory to AF-based signal design.

\appendix
\section{Proof that $\mathcal{T}(\mathbf{x})$ is closed}
\label{app:distance}
If $\mathcal{T}(\mathbf{x})$ is closed, then it guarantees the existence of $\mathbf{z}\in \mathcal{T}(\mathbf{x})$ such that 
\begin{align}
	\text{dist}(\mathbf{x},\mathbf{w}) = \lVert \mathbf{w} - \mathbf{z} \rVert_{2},
	\label{eq:wellDefined}
\end{align}
for any $\mathbf{w}\in \mathbb{C}^{N}$. 
\begin{lemma}
    The set 
    \begin{align}
	\mathcal{T}(\mathbf{x})=\{\mathbf{z}\in \mathbb{C}^{N} \hspace{0.2em}:\hspace{0.2em} \mathbf{z}[n]=e^{i\beta}e^{i b n}\mathbf{x}[\epsilon n - a] \text{ for } \beta,b\in \mathbb{R}, \text{ and }\epsilon=\pm 1, a\in \mathbb{Z} \},
	\label{eq:levelSet}
\end{align}
    is closed for $\mathbf{x}\in \mathbb{C}^{N}$.
\end{lemma}
 
\begin{proof}
	Observe that a given element $\mathbf{z}$ in the set $\mathcal{T}(\mathbf{x})$ is the product between an orthogonal matrix and the signal $\mathbf{x}$. This is concluded as follows.
	\begin{itemize}
		\item The time inversion on $\mathbf{x}$ is achieved by the product of an orthogonal matrix $\mathbf{S}_{1}\in \mathbb{R}^{N\times N}$ (which depends on $\epsilon\pm 1$), and $\mathbf{x}$. When $\epsilon=1$, $\mathbf{S}_{1}$ is an identity matrix of size $N$; otherwise, it is a permutation matrix that models time reversal, i.e. $\mathbf{x}[-n]$.
		
		\item A constant time shift of $a$ on the signal $\mathbf{x}$ according to $\mathcal{T}(\mathbf{x})$ is modelled by the product of a circulant and orthogonal) matrix $\mathbf{S}_{2}\in \mathbb{R}^{N\times N}$, which depends on the translation parameter $a$, and $\mathbf{x}$. 
		
		\item The time scale ambiguity in $\mathcal{T}(\mathbf{x})$ is modelled by the product of a diagonal (and, therefore, orthogonal) matrix $\mathbf{S}_{3}\in \mathbb{C}^{N\times N}$, which depends on the parameter $b$, and $\mathbf{x}$. The matrix $\mathbf{S}_{3}$ is orthogonal because it takes values from the complex unit circle.
		
		\item The global phase shift is a constant taken from the complex unit circle governed by $\beta$.
	\end{itemize}
	It follows from above that a given element 
	\begin{align}
		\mathbf{z} = \mathbf{Q}\mathbf{x}, 
		\label{eq:auxExpression}
	\end{align}
	where 
	the orthogonal matrix 
	\begin{align}
		\mathbf{Q} = e^{i\beta}\mathbf{S}_{s}\mathbf{S}_{r}\mathbf{S}_{u} \in \mathbb{C}^{N\times N},
		\label{eq:matrixQ}
	\end{align}
	for $s,r,u\in \{1,2,3\}$. Note that the values of $s,r,u$, which determine the product order, directly depend on $\epsilon$, $b$, and $a$.
	
	To prove that the set $\mathcal{T}(\mathbf{x})$ is closed,  
	take a sequence $(\mathbf{z}_{n})\subset \mathcal{T}(\mathbf{x})$ that converges to $\mathbf{z}_{*}$. We have to show that $\mathbf{z}_{*}\in \mathcal{T}(\mathbf{x})$. Since $\mathbf{z}_{n}\rightarrow \mathbf{z}_{*}$, then for any $\delta>0$, there exists a natural number $n(\delta)$ such that
	\begin{align}
		\lVert \mathbf{z}_{n} - \mathbf{z}_{*} \rVert_{2} < \delta, \hspace{0.5em} \forall n>n(\delta).
		\label{eq:closeSet}
	\end{align}
	Each $\mathbf{z}_{n}$ belongs to $\mathcal{T}(\mathbf{x})$. Hence, from \eqref{eq:auxExpression}, there exists $\mathbf{Q}_{n}\in \mathbb{C}^{N\times N}$ orthogonal matrix with the form of \eqref{eq:matrixQ} such that $\mathbf{z}_{n}=\mathbf{Q}_{n}\mathbf{x}$. Consequently, from \eqref{eq:closeSet}, for any $\delta>0$
	\begin{align}
		\lVert \mathbf{z}_{n} - \mathbf{z}_{*} \rVert_{2} = \lVert \mathbf{x} - \mathbf{Q}_{n}^{H}\mathbf{z}_{*} \rVert_{2} < \delta , \hspace{0.5em} \forall n>n(\delta),
	\end{align}
	for some natural number $n(\delta)$ because $\mathbf{Q}_{n}$ is orthogonal. The above equation \textit{ipso facto} means that the sequence $\mathbf{r}_{n}=\mathbf{Q}_{n}^{H}\mathbf{z}_{*}$ converges to $\mathbf{x}$. 
	This leads to the existence of $\mathbf{Q}_{*}\in \mathbb{C}^{N\times N}$ following the form of \eqref{eq:matrixQ} such that $\mathbf{x}=\mathbf{Q}_{*}^{H}\mathbf{z}_{*}$. Equivalently, we obtain that $\mathbf{z}_{*} = \mathbf{Q}_{*}\mathbf{x}$ implying that $\mathbf{z}_{*}\in \mathcal{T}(\mathbf{x})$ because $\mathbf{Q}_{*}$ follows the form of \eqref{eq:matrixQ}. This completes the proof.
\end{proof}

\section{Proof of Theorem \ref{theo:contraction}}
\label{app:prooftheo4}
\subsection{Preliminaries to the Proof}
\label{app:lemma}
Define the search set as
\begin{align}
\mathcal{J}:=\{ \mathbf{z}\in \mathbb{C}^{N},\; \text{$B$-bandlimited}: \text{dist}(\mathbf{x},\mathbf{z})\leq \rho, B\leq N/2 \},
\label{eq:covergenceset}
\end{align}
for some small constant $\rho>0$. Recall that $\mathbf{z}$ is a $B$-bandlimited signal if there exists $k$ such that $\tilde{\mathbf{z}}[k]=\cdots=\tilde{\mathbf{z}}[N+k+B-1]=0$, where $\tilde{\mathbf{z}}$ is the Fourier transform of $\mathbf{z}$. The bandlimitedness guarantees that we have unique solution, according to Proposition~\ref{prop:uniqueness}. We first prove the positivity of the product $\left\lvert\mathbf{f}_{k}^{H}\mathbf{g}_{p}(\mathbf{z})\right\rvert$ over the set $\mathcal{J}$ in the following Lemma~\ref{lem:generic}.
\begin{lemma}
\label{lem:generic}
Let $\mathbf{z}\in \mathcal{J}$ where $\mathcal{J}$ as defined in \eqref{eq:covergenceset}. Then, for almost all $\mathbf{z}\in \mathcal{J}$ the following holds
\begin{align}
	\left\lvert\mathbf{f}_{k}^{H}\mathbf{g}_{p}(\mathbf{z})\right\rvert>0,
	\label{eq:boundConvergence}
\end{align}
for all $k,p\in \{0,\cdots,N-1\}$, with
\begin{align}
	\mathbf{g}_{p}(\mathbf{z}) :=& \left[\mathbf{z}[0]\overline{\mathbf{z}[p]},\cdots,\mathbf{z}[N-1]\overline{\mathbf{z}[N-1+p]} \right]^{T}.
	\label{eq:auxvector}
\end{align}
\end{lemma}
\begin{proof}
We prove this lemma by contradiction. Assume $\left\lvert\mathbf{f}_{k}^{H}\mathbf{g}_{p}(\mathbf{z})\right\rvert=0$. Then, from \eqref{eq:Ambiguity} we have that
\begin{align}
	&\left\lvert\mathbf{f}_{k}^{H}\mathbf{g}_{p}(\mathbf{z})\right\rvert^{2} = \left\lvert\sum_{n=0}^{N-1} \mathbf{z}[n]\overline{\mathbf{z}[n-p]}e^{-2\pi \mathrm{i}nk/N} \right \rvert^{2}=\sum_{n,m=0}^{N-1}\left(\mathbf{z}[n]\overline{\mathbf{z}[n-p]}\mathbf{z}[m-p]\overline{\mathbf{z}[m]}\right)e^{\frac{2\pi i(m-n)k}{N}}=0.
	\label{eq:quadraticpolynomical}
\end{align}
Note that \eqref{eq:quadraticpolynomical} is a quartic polynomial equation with respect to the entries of $\mathbf{z}$. However, for almost all signals $\mathbf{z}\in \mathcal{J}$ the left hand side of \eqref{eq:quadraticpolynomical} will not be equal to zero which leads to a contradiction. In addition, considering the cross-product in \eqref{eq:auxvector} and the Fourier transform via $\mathbf{f}_{k}$ in \eqref{eq:boundConvergence}, it follows that any of $\mathbf{z}[n]e^{\textrm{i}\phi}$, $\mathbf{z}[n-a]$, $\mathbf{z}[-n]$, and $e^{\textrm{i}bn}\mathbf{z}[n]$ will satisfy \eqref{eq:boundConvergence}.
\end{proof}

In order to prove Theorem \ref{theo:contraction}, the function $h(\mathbf{z},\mu)$ in \eqref{eq:auxproblem} must satisfy the four requirements stated in the following Lemma~\ref{lem:assumption}. These conditions are used in the analysis of convergence for stochastic gradient methods \citep{ghadimi2013stochastic}. 
\begin{lemma}
\label{lem:assumption}
The function $h(\mathbf{z},\mu)$ in \eqref{eq:auxproblem} and its Wirtinger derivative in \eqref{eq:gradient} satisfy the following properties.
\begin{description}
	\item[\textbf{C1}]\hspace{0.5em} The cost function $h(\mathbf{z},\mu)$ in \eqref{eq:auxproblem} is bounded below.
	
	\item[\textbf{C2}]\hspace{0.5em} The set $\mathcal{J}$ as defined in \eqref{eq:covergenceset} is closed and bounded.
	
	\item[\textbf{C3}]\hspace{0.5em} There exists a constant $U>0$, such that
	\begin{equation}
		\left\lVert \frac{\partial h(\mathbf{z}_{1},\mu)}{\partial \overline{\mathbf{z}}} - \frac{\partial h(\mathbf{z}_{2},\mu)}{\partial \overline{\mathbf{z}}} \right\rVert_{2}\leq U\left\lVert\mathbf{z}_{1}-\mathbf{z}_{2} \right\rVert_{2},
		\label{eq:assumption}
	\end{equation}
	holds for all $\mathbf{z}_{1},\mathbf{z}_{2}\in \mathcal{J}$.
	
	\item[\textbf{C4}]\hspace{0.5em} For all $\mathbf{z}\in \mathcal{J}$
	\begin{align}
		\mathbb{E}_{\Gamma_{(r)}}\left[\left\lVert \mathbf{d}_{\Gamma_{(r)}} - \frac{\partial h(\mathbf{z},\mu)}{\partial \overline{\mathbf{z}}} \right\rVert^{2}_{2}\right]\leq \zeta^{2},
		\label{eq:variance}
	\end{align}
	for some $\zeta>0$, where $\mathbf{d}_{\Gamma_{(r)}}$ is as in line 7 of Algorithm \ref{alg:smothing}.
\end{description}
\end{lemma}
\begin{proof}

\begin{itemize}
	\item To prove \textbf{C1}, from the definition of $h(\mathbf{z},\mu)$ in \eqref{eq:auxproblem}, it follows that $h(\mathbf{z},\mu)\geq 0$ and thus bounded below.
	
	\item To prove \textbf{C2}, we show that $\mathcal{J}$ in \eqref{eq:covergenceset} is bounded and its complement $\mathcal{J}^{c}$ is open. To prove $\mathcal{J}$ is bounded, pick $\mathbf{z}\in \mathcal{J}$. This implies that $\text{dist}(\mathbf{x},\mathbf{z})\leq \rho$. As a consequence from \eqref{eq:wellDefined} there exits $\mathbf{w}\in \mathcal{T}(\mathbf{x})$ such that 
	\begin{align}
		\lVert \mathbf{w} - \mathbf{z} \rVert_{2} \leq \rho.
		\label{eq:}
	\end{align}
	From the above expression we obtain by applying the triangle inequality an upper bound of $\lVert \mathbf{z} \rVert_{2}$ as
	\begin{align}
		\lVert \mathbf{z} \rVert_{2} - \lVert \mathbf{w} \rVert_{2} &\leq \rho \nonumber\\
		\lVert \mathbf{z} \rVert_{2} & \leq \rho + \lVert \mathbf{w} \rVert_{2}.
		\label{eq:boundedNorm}
	\end{align}
	Because we know that the target signal $\mathbf{x}$ is $B$-bandlimited with $B\leq N/2$, this means that $\mathbf{x}$ and any $\mathbf{w}$ of its trivial transformation in $\mathcal{T}(\mathbf{x})$ has a bounded $\ell_{2}$-norm (Fourier transform exits). Therefore, from \eqref{eq:boundedNorm} we obtain that $\lVert \mathbf{z} \rVert_{2}$ is bounded, leading to the boundedness of $\mathcal{J}$.
	
	Now, we prove $\mathcal{J}^{c}$ is an open set. To that end, from \eqref{eq:levelSet} we have that $\mathcal{J}^{c}$ is given by
	\begin{align}
		\mathcal{J}^{c} = \{ \mathbf{z}\in \mathbb{C}^{N},\; \text{$B$-bandlimited}: \text{dist}(\mathbf{x},\mathbf{z})> \rho, B> N/2 \}.
	\end{align}
	Take $\mathbf{z}_{0}\in \mathcal{J}^{c}$. Define $\mathbf{w}_{0}\in \mathbb{C}^{N}$ such that
	\begin{equation}
		\mathbf{w}_{0}[n] = \left\lbrace \begin{array}{ll}
			\mathbf{z}_{0}[n] & \text{ for }n\leq N/2\\
			0 & \text{ otherwise.}
		\end{array} \right.		
	\end{equation} 
	Observe that $\mathbf{w}_{0}$ is the closest $B$-bandlimited signal to $\mathbf{z}_{0}$ with $B\leq N/2$ i.e. the closest signal to $\mathbf{z}_{0}$ in $\mathcal{J}$. Observe that since $\mathbf{z}_{0}\in \mathcal{J}^{c}$ then we have that $\text{dist}(\mathbf{x},\mathbf{z}_{0}) > \rho$ which implies that $\text{dist}(\mathbf{x},\mathbf{z}_{0}) = \rho + \zeta$ for some $\zeta>0$. Take $\kappa = \text{dist}(\mathbf{z}_{0},\mathbf{w}_{0})$, and $\epsilon = \min\{\kappa/2,\zeta\}$. Then the neighbourhood $\mathcal{U}$ of $\mathbf{z}_{0}$ given as 
	\begin{align}
		\mathcal{U} = \{ \mathbf{r}\in \mathbb{C}^{N}: \text{dist}(\mathbf{z}_{0},\mathbf{r}) < \epsilon \},
	\end{align}
	does not contain any element from $\mathcal{J}$, because the closest element in $\mathcal{J}$ is found at least a distance of $2\epsilon$. Now, we claim that $\mathcal{U}\subset \mathcal{J}^{c}$. To prove this claim, pick $\mathbf{r}_{0}\in \mathcal{U}$. Because of carefully chosen $\epsilon$, we have that $\mathbf{r}_{0}$ is a $B$-bandlimited signal with $B>N/2$. We now show that $\mathbf{r}_{0} \in \mathcal{J}^{c}$ by stating that $\text{dist}(\mathbf{x},\mathbf{r}_{0})>\rho$. To that end, observe that the triangle inequality gives us
	\begin{align}
		\text{dist}(\mathbf{x},\mathbf{z}_{0}) &\leq \text{dist}(\mathbf{x},\mathbf{r}_{0}) + \text{dist}(\mathbf{z}_{0},\mathbf{r}_{0}) \nonumber\\
		\rho + \zeta &\leq \text{dist}(\mathbf{x},\mathbf{r}_{0}) + \text{dist}(\mathbf{z}_{0},\mathbf{r}_{0}) \nonumber\\
		\rho + \zeta &< \text{dist}(\mathbf{x},\mathbf{r}_{0}) + \epsilon \nonumber\\
		\rho + \zeta - \epsilon &< \text{dist}(\mathbf{x},\mathbf{r}_{0}).
		\label{eq:triangle}
	\end{align}
	By definition of $\epsilon$ we have that $\zeta - \epsilon \geq 0$, and because $\rho> 0$ from \eqref{eq:triangle} we obtain
	\begin{align}
		\rho < \text{dist}(\mathbf{x},\mathbf{r}_{0}),
	\end{align}
	as desired. Hence, $\mathbf{r}_{0}\in \mathcal{J}^{c}$, and we have $\mathcal{U}\subset  \mathcal{J}^{c}$. Because $\mathbf{r}_{0}$ was arbitrary we have that $\mathcal{J}^{c}$ is an open set, and thus $\mathcal{J}$ is closed.
	
	\item Regarding \textbf{C3}, it follows from \eqref{eq:stochStep} that $\ell$-th entry of $\frac{\partial h(\mathbf{z},\mu)}{\partial \overline{\mathbf{z}}}$ is
	\begin{align}
		\frac{\partial h(\mathbf{z},\mu)}{\partial \overline{\mathbf{z}}}[\ell] &= \frac{1}{N^{2}}\sum_{k,p=0}^{N-1}\left( \mathbf{f}_{k}^{H}\mathbf{g}_{p}(\mathbf{z}) - \upsilon_{k,p} \right) \mathbf{z}[\ell-p]e^{2\pi i\ell k/N} +\frac{1}{N^{2}}\sum_{k,p=0}^{N-1}\left( \mathbf{f}_{k}^{T}\overline{\mathbf{g}_{p}} - \upsilon_{k,p} \right) \mathbf{z}[\ell+p]e^{-2\pi \mathrm{i}(\ell+p) k/N}.
	\end{align}
	Define $\mathbf{D}_{p}(\mathbf{z})\in \mathbb{C}^{N\times N}$ as a diagonal matrix whose main diagonal is formed by the vector $\mathbf{z}_{p}=[\mathbf{z}_{p}[0], \dots, \mathbf{z}_{p}[N-1]]^{T}=[\mathbf{z}[-p], \dots, \mathbf{z}[N-1-p]]^{T}$. Thus,
	\begin{align}
		\frac{\partial h(\mathbf{z},\mu)}{\partial \overline{\mathbf{z}}} = \frac{1}{N^{2}}\sum_{p,k=0}^{N-1}f_{k,p}(\mathbf{z}) + g_{k,p}(\mathbf{z}),
		\label{eq:convergence3}
	\end{align}
	where 
	$f_{k,p}(\mathbf{z})=\rho_{k,p}(\mathbf{z})\mathbf{D}_{p}(\mathbf{z})\mathbf{f}_{k}$, 
	$g_{k,p}(\mathbf{z})=\omega^{-kp}\overline{\rho_{k,p}}(\mathbf{z})\mathbf{D}_{-p}(\mathbf{z})\mathbf{f}_{k}$, 
	and
	\begin{align}
		\rho_{k,p}(\mathbf{z})= \mathbf{f}_{k}^{H}\mathbf{g}_{p}(\mathbf{z}) -\sqrt{\mathbf{A}[p,k]}\frac{\mathbf{f}_{k}^{H}\mathbf{g}_{p}(\mathbf{z})}{\varphi_{\mu}\left(\left\lvert\mathbf{f}_{k}^{H}\mathbf{g}_{p}(\mathbf{z}) \right \rvert\right)}.
		\label{eq:rhokp}
	\end{align}
	
	We now establish that, for all $\mathbf{z}_{1},\mathbf{z}_{2}\in \mathcal{J}$ and some constants $r_{k,p},s_{k,p}>0$, any $f_{k,p}(\mathbf{z})$ and $g_{k,p}(\mathbf{z})$ satisfy
	\begin{align}
		\left\lVert f_{k,p}(\mathbf{z}_{1}) - f_{k,p}(\mathbf{z}_{2}) \right\rVert_{2} \leq r_{k,p} \lVert \mathbf{z}_{1} - \mathbf{z}_{2} \rVert_{2},
		\label{eq:functionf}
	\end{align}
	and
	\begin{align}
		\left\lVert g_{k,p}(\mathbf{z}_{1}) - g_{k,p}(\mathbf{z}_{2}) \right\rVert_{2} \leq s_{k,p} \lVert \mathbf{z}_{1} - \mathbf{z}_{2} \rVert_{2}.
		\label{eq:functiong}
	\end{align}
	We show only \eqref{eq:functionf}; the proof for $g_{k,p}(\mathbf{z})$ follows \textit{mutatis mutandis}. 
	
	Considering the fact that $\mathbf{D}_{p}(\mathbf{z}_{1})$ and $\mathbf{D}_{p}(\mathbf{z}_{2})$ are diagonal matrices and $\lVert \mathbf{f}_{k} \rVert_{2}=\sqrt{N}$, it follows from the definition of $f_{k,p}(\mathbf{z})$ that, for any $\mathbf{z}_{1},\mathbf{z}_{2}\in \mathcal{J}$,
	\begin{equation}
		\frac{1}{\sqrt{N}}\left\lVert f_{k,p}(\mathbf{z}_{1}) - f_{k,p}(\mathbf{z}_{2}) \right\rVert_{2}\leq \left\lVert \rho_{k,p}(\mathbf{z}_{1})\mathbf{z}_{1} - \rho_{k,p}(\mathbf{z}_{2})\mathbf{z}_{2} \right\rVert_{2},\label{eq:ine1}
	\end{equation}
	Using \eqref{eq:rhokp} in the right-hand-side of \eqref{eq:ine1} yields
	\begin{align}
		\left\lVert \rho_{k,p}(\mathbf{z}_{1})\mathbf{z}_{1} - \rho_{k,p}(\mathbf{z}_{2})\mathbf{z}_{2} \right\rVert_{2} &= \left\lVert \rho_{k,p}(\mathbf{z}_{1})(\mathbf{z}_{1} + \mathbf{z}_{2}-\mathbf{z}_{2}) - \rho_{k,p}(\mathbf{z}_{2})\mathbf{z}_{2} \right\rVert_{2} \nonumber\\
		&= \left\lVert \rho_{k,p}(\mathbf{z}_{1})(\mathbf{z}_{1}-\mathbf{z}_{2}) + \rho_{k,p}(\mathbf{z}_{1})\mathbf{z}_{2}- \rho_{k,p}(\mathbf{z}_{2})\mathbf{z}_{2} \right\rVert_{2} \nonumber\\
		&\leq \lvert \rho_{k,p}(\mathbf{z}_{1}) \rvert \left\lVert \mathbf{z}_{1}-\mathbf{z}_{2}\right\rVert_{2} + \left\lVert \mathbf{z}_{2} \right\rVert_{2} \lvert \rho_{k,p}(\mathbf{z}_{1}) - \rho_{k,p}(\mathbf{z}_{2})\rvert \nonumber\\
		&= \frac{\left\lvert\mathbf{f}_{k}^{H}\mathbf{g}_{p}(\mathbf{z}_{1})\right\rvert}{\varphi_{\mu}\left(\left\lvert\mathbf{f}_{k}^{H}\mathbf{g}_{p}(\mathbf{z}) \right \rvert\right)} \left \lvert \varphi_{\mu}\left(\left\lvert\mathbf{f}_{k}^{H}\mathbf{g}_{p}(\mathbf{z}_{1}) \right \rvert\right)-\sqrt{\mathbf{A}[p,k]} \right\rvert \left\lVert \mathbf{z}_{1}-\mathbf{z}_{2}\right\rVert_{2} \nonumber\\
		&+ \left\lVert \mathbf{z}_{2} \right\rVert_{2} \lvert \rho_{k,p}(\mathbf{z}_{1}) - \rho_{k,p}(\mathbf{z}_{2})\rvert.
		\label{eq:extraEquation}
	\end{align}
	Then, combining \eqref{eq:extraEquation} with the inequality $\varphi_{\mu}\left(\left\lvert\mathbf{f}_{k}^{H}\mathbf{g}_{p}(\mathbf{z}_{1}) \right \rvert\right)\geq \mu$, \eqref{eq:ine1} becomes
	\begin{align}
		\frac{1}{\sqrt{N}}\left\lVert f_{k,p}(\mathbf{z}_{1}) - f_{k,p}(\mathbf{z}_{2}) \right\rVert_{2} &\leq \frac{\left\lvert\mathbf{f}_{k}^{H}\mathbf{g}_{p}(\mathbf{z}_{1})\right\rvert}{\mu}\left(\varphi_{\mu}\left(\left\lvert\mathbf{f}_{k}^{H}\mathbf{g}_{p}(\mathbf{z}_{1}) \right \rvert\right)+\sqrt{\mathbf{A}[p,k]} \right) \left\lVert \mathbf{z}_{1} - \mathbf{z}_{2} \right\rVert_{2} \nonumber\\
		&+\left\lVert \mathbf{z}_{2} \right\rVert_{2}\underbrace{\left\lvert \rho_{k,p}(\mathbf{z}_{1})- \rho_{k,p}(\mathbf{z}_{2}) \right\rvert}_{=p_{1}}.
		\label{eq:ine2}
	\end{align}
	The term $p_{1}$ in \eqref{eq:ine2} is upper bounded as
	
	\begin{align}
		p_{1} &= \left \lvert \mathbf{f}_{k}^{H}\mathbf{g}_{p}(\mathbf{z}_{1}) \left( 1- \frac{\sqrt{\mathbf{A}[p,k]}}{\varphi_{\mu}\left(\left\lvert\mathbf{f}_{k}^{H}\mathbf{g}_{p}(\mathbf{z}_{1}) \right \rvert\right)} \right) - \mathbf{f}_{k}^{H}\mathbf{g}_{p}(\mathbf{z}_{2}) \left( 1- \frac{\sqrt{\mathbf{A}[p,k]}}{\varphi_{\mu}\left(\left\lvert\mathbf{f}_{k}^{H}\mathbf{g}_{p}(\mathbf{z}_{2}) \right \rvert\right)} \right)\right \rvert \nonumber\\
		&\leq \left\lvert \mathbf{f}_{k}^{H}\mathbf{g}_{p}(\mathbf{z}_{1}) - \mathbf{f}_{k}^{H}\mathbf{g}_{p}(\mathbf{z}_{2}) \right\rvert +\sqrt{\mathbf{A}[p,k]}\left\lvert\frac{\mathbf{f}_{k}^{H}\mathbf{g}_{p}(\mathbf{z}_{1})}{\varphi_{\mu}\left(\left\lvert\mathbf{f}_{k}^{H}\mathbf{g}_{p}(\mathbf{z}_{1}) \right \rvert\right)} - \frac{\mathbf{f}_{k}^{H}\mathbf{g}_{p}(\mathbf{z}_{2})}{\varphi_{\mu}\left(\left\lvert\mathbf{f}_{k}^{H}\mathbf{g}_{p}(\mathbf{z}_{2}) \right \rvert\right)}\right\rvert\nonumber\\
		&= \left\lvert \mathbf{f}_{k}^{H}\mathbf{g}_{p}(\mathbf{z}_{1}) - \mathbf{f}_{k}^{H}\mathbf{g}_{p}(\mathbf{z}_{2}) \right\rvert + \sqrt{\mathbf{A}[p,k]}\left\lvert\frac{\mathbf{f}_{k}^{H}\mathbf{g}_{p}(\mathbf{z}_{1}) + \mathbf{f}_{k}^{H}\mathbf{g}_{p}(\mathbf{z}_{2}) -\mathbf{f}_{k}^{H}\mathbf{g}_{p}(\mathbf{z}_{2})}{\varphi_{\mu}\left(\left\lvert\mathbf{f}_{k}^{H}\mathbf{g}_{p}(\mathbf{z}_{1}) \right \rvert\right)} - \frac{\mathbf{f}_{k}^{H}\mathbf{g}_{p}(\mathbf{z}_{2})}{\varphi_{\mu}\left(\left\lvert\mathbf{f}_{k}^{H}\mathbf{g}_{p}(\mathbf{z}_{2}) \right \rvert\right)}\right\rvert \nonumber\\
		&\leq \left\lvert \mathbf{f}_{k}^{H}\mathbf{g}_{p}(\mathbf{z}_{1}) - \mathbf{f}_{k}^{H}\mathbf{g}_{p}(\mathbf{z}_{2}) \right\rvert + \sqrt{\mathbf{A}[p,k]}\left\lvert\frac{\mathbf{f}_{k}^{H}\mathbf{g}_{p}(\mathbf{z}_{1}) -\mathbf{f}_{k}^{H}\mathbf{g}_{p}(\mathbf{z}_{2})}{\varphi_{\mu}\left(\left\lvert\mathbf{f}_{k}^{H}\mathbf{g}_{p}(\mathbf{z}_{1}) \right \rvert\right)} \right \rvert \nonumber\\
		&+\sqrt{\mathbf{A}[p,k]}\left \lvert \frac{\mathbf{f}_{k}^{H}\mathbf{g}_{p}(\mathbf{z}_{2})}{\varphi_{\mu}\left(\left\lvert\mathbf{f}_{k}^{H}\mathbf{g}_{p}(\mathbf{z}_{1}) \right \rvert\right)} - \frac{\mathbf{f}_{k}^{H}\mathbf{g}_{p}(\mathbf{z}_{2})}{\varphi_{\mu}\left(\left\lvert\mathbf{f}_{k}^{H}\mathbf{g}_{p}(\mathbf{z}_{2}) \right \rvert\right)}\right\rvert \nonumber\\
		&\leq \left\lvert \mathbf{f}_{k}^{H}\mathbf{g}_{p}(\mathbf{z}_{1}) - \mathbf{f}_{k}^{H}\mathbf{g}_{p}(\mathbf{z}_{2}) \right\rvert +\frac{\sqrt{\mathbf{A}[p,k]}}{\mu^{2}}\varphi_{\mu}\left(\left\lvert\mathbf{f}_{k}^{H}\mathbf{g}_{p}(\mathbf{z}_{2}) \right \rvert\right)\left\lvert \mathbf{f}_{k}^{H}\mathbf{g}_{p}(\mathbf{z}_{1}) - \mathbf{f}_{k}^{H}\mathbf{g}_{p}(\mathbf{z}_{2}) \right\rvert \nonumber\\
		&+\frac{\sqrt{\mathbf{A}[p,k]}}{\mu^{2}}\left\lvert \mathbf{f}_{k}^{H}\mathbf{g}_{p}(\mathbf{z}_{2})\right\rvert \left\lvert \varphi_{\mu}\left(\left\lvert\mathbf{f}_{k}^{H}\mathbf{g}_{p}(\mathbf{z}_{1}) \right \rvert\right) - \varphi_{\mu}\left(\left\lvert\mathbf{f}_{k}^{H}\mathbf{g}_{p}(\mathbf{z}_{2}) \right \rvert\right) \right\rvert.
		\label{eq:ine3}
	\end{align}
	
	Recall that, by Heine-Borel Theorem \citep{rudin1991functional}, $\mathcal{J}$ is a closed bounded set, and thus compact. Since $\varphi_{\mu}(\cdot)$ is a continuous function, there exists a constant $M_{\varphi_{\mu}}$ such that $\varphi_{\mu}\left(\left\lvert\mathbf{f}_{k}^{H}\mathbf{g}_{p}(\mathbf{z}) \right \rvert\right)\leq M_{\varphi_{\mu}}$ for all $\mathbf{z}\in \mathcal{J}$. Also, from \citep[Lemma 2]{pinilla2018phase} we have that $\varphi_{\mu}(\cdot)$ is a 1-Lipschitz function. Combining this with \eqref{eq:ine3} yields
	\begin{align}
		p_{1}&\leq \left\lvert \mathbf{f}_{k}^{H}\mathbf{g}_{p}(\mathbf{z}_{1}) - \mathbf{f}_{k}^{H}\mathbf{g}_{p}(\mathbf{z}_{2}) \right\rvert +\frac{\sqrt{\mathbf{A}[p,k]}M_{\varphi_{\mu}}}{\mu^{2}}\left\lvert \mathbf{f}_{k}^{H}\mathbf{g}_{p}(\mathbf{z}_{1}) - \mathbf{f}_{k}^{H}\mathbf{g}_{p}(\mathbf{z}_{2}) \right\rvert \nonumber\\
		&+\frac{\sqrt{\mathbf{A}[p,k]}}{\mu^{2}}\left\lvert \mathbf{f}_{k}^{H}\mathbf{g}_{p}(\mathbf{z}_{2})\right\rvert \left\lvert \left\lvert\mathbf{f}_{k}^{H}\mathbf{g}_{p}(\mathbf{z}_{1}) \right \rvert - \left\lvert\mathbf{f}_{k}^{H}\mathbf{g}_{p}(\mathbf{z}_{2}) \right \rvert \right\rvert\nonumber\\
		&
		\;\;\;\;\;\leq \left(\frac{\sqrt{\mathbf{A}[p,k]}M_{\varphi_{\mu}}}{\mu^{2}}+ 1\right)\left\lvert \mathbf{f}_{k}^{H}\mathbf{g}_{p}(\mathbf{z}_{1}) - \mathbf{f}_{k}^{H}\mathbf{g}_{p}(\mathbf{z}_{2}) \right\rvert +\frac{\sqrt{\mathbf{A}[p,k]}}{\mu^{2}}\left\lvert \mathbf{f}_{k}^{H}\mathbf{g}_{p}(\mathbf{z}_{2})\right\rvert \left\lvert\mathbf{f}_{k}^{H}\mathbf{g}_{p}(\mathbf{z}_{1}) - \mathbf{f}_{k}^{H}\mathbf{g}_{p}(\mathbf{z}_{2}) \right\rvert,
		\label{eq:ine4}
	\end{align}
	where 
	the second inequality follows from the reverse triangle inequality. Using \eqref{eq:ine4} in \eqref{eq:ine2} gives
	\begin{align}
		&\frac{1}{\sqrt{N}}\left\lVert f_{k,p}(\mathbf{z}_{1}) - f_{k,p}(\mathbf{z}_{2}) \right\rVert_{2} \nonumber\\
		\leq& \frac{\left\lvert\mathbf{f}_{k}^{H}\mathbf{g}_{p}(\mathbf{z}_{1})\right\rvert}{\mu}\left(M_{\varphi_{\mu}} +\sqrt{\mathbf{A}[p,k]}\right) \left\lVert \mathbf{z}_{1} - \mathbf{z}_{2} \right\rVert_{2} + \left\lVert \mathbf{z}_{2} \right\rVert_{2}\left(\frac{\sqrt{\mathbf{A}[p,k]}M_{\varphi_{\mu}}}{\mu^{2}}+ 1\right)\left\lvert \mathbf{f}_{k}^{H}\mathbf{g}_{p}(\mathbf{z}_{1}) - \mathbf{f}_{k}^{H}\mathbf{g}_{p}(\mathbf{z}_{2}) \right\rvert \nonumber\\
		+&\frac{\left\lVert \mathbf{z}_{2} \right\rVert_{2}\sqrt{\mathbf{A}[p,k]}}{\mu^{2}}\left\lvert \mathbf{f}_{k}^{H}\mathbf{g}_{p}(\mathbf{z}_{2})\right\rvert \left\lvert\mathbf{f}_{k}^{H}\mathbf{g}_{p}(\mathbf{z}_{1}) - \mathbf{f}_{k}^{H}\mathbf{g}_{p}(\mathbf{z}_{2}) \right\rvert.
		\label{eq:ine5}
	\end{align}
	
	Observe that the upper bound in \eqref{eq:ine5} directly depends on a term of the form $\mathbf{f}_{k}^{H}\mathbf{g}_{p}(\mathbf{z})$ for some $\mathbf{z}\in \mathcal{J}$, which might be zero. However, from Lemma \ref{lem:generic}, $\left\lvert\mathbf{f}_{k}^{H}\mathbf{g}_{p}(\mathbf{z})\right\rvert>0$ or, equivalently, $\mathbf{f}_{k}^{H}\mathbf{g}_{p}(\mathbf{z})\not = 0$, for almost all $\mathbf{z}\in \mathcal{J}$. 
	
	We now proceed to bound the term $\left\lvert\mathbf{f}_{k}^{H}\mathbf{g}_{p}(\mathbf{z})\right\rvert$. From \eqref{eq:Ambiguity}, 
	\begin{align}
		\left\lvert\mathbf{f}_{k}^{H}\mathbf{g}_{p}(\mathbf{z})\right\rvert &= \left\lvert\sum_{n=0}^{N-1} \mathbf{z}[n]\overline{\mathbf{z}[n-p]}e^{-2\pi \mathrm{i}nk/N} \right \rvert \leq \sum_{n=0}^{N-1} \left\lvert\mathbf{z}[n]\overline{\mathbf{z}[n-p]}\right \rvert \leq N\lVert \mathbf{z} \rVert_{2},
		\label{eq:ine6}
	\end{align}
	where the second inequality follows from the norm inequalities $\lVert \mathbf{z} \rVert_{1}\leq \sqrt{N}\lVert \mathbf{z} \rVert_{2}$ and $\lVert \mathbf{z} \rVert_{2} \leq \sqrt{N}\lVert \mathbf{z} \rVert_{\infty}$. Given the cross-product $\left\lvert\mathbf{f}_{k}^{H}\mathbf{g}_{p}(\mathbf{z})\right\rvert$ and the Fourier transform within it, the above bound is also valid for any trivial ambiguity \textbf{T1}-\textbf{T4} in $\mathbf{z}$ (as we also remarked in Lemma \ref{lem:generic}). Combining \eqref{eq:ine5} and \eqref{eq:ine6} we get
	\begin{align}
		\frac{1}{\sqrt{N}}\left\lVert f_{k,p}(\mathbf{z}_{1}) - f_{k,p}(\mathbf{z}_{2}) \right\rVert_{2} &\leq \frac{N\lVert \mathbf{z}_{1} \rVert_{2}}{\mu}\left(M_{\varphi_{\mu}} +\sqrt{\mathbf{A}[p,k]}\right) \left\lVert \mathbf{z}_{1} - \mathbf{z}_{2} \right\rVert_{2}\nonumber\\
		&+\left\lVert \mathbf{z}_{2} \right\rVert_{2}\left(\frac{\sqrt{\mathbf{A}[p,k]}M_{\varphi_{\mu}}}{\mu^{2}}+ 1\right)\left\lvert \mathbf{f}_{k}^{H}\mathbf{g}_{p}(\mathbf{z}_{1}) - \mathbf{f}_{k}^{H}\mathbf{g}_{p}(\mathbf{z}_{2}) \right\rvert \nonumber\\
		&+\frac{N\lVert \mathbf{z}_{2}\rVert^{2}_{2}\sqrt{\mathbf{A}[p,k]}}{\mu^{2}} \left\lvert\mathbf{f}_{k}^{H}\mathbf{g}_{p}(\mathbf{z}_{1}) - \mathbf{f}_{k}^{H}\mathbf{g}_{p}(\mathbf{z}_{2}) \right\rvert.
		\label{eq:ine7}
	\end{align}
	
	Next, we bound the term $\left\lvert\mathbf{f}_{k}^{H}\mathbf{g}_{p}(\mathbf{z}_{1}) - \mathbf{f}_{k}^{H}\mathbf{g}_{p}(\mathbf{z}_{2}) \right\rvert$ in \eqref{eq:ine7}. Again, from \eqref{eq:Ambiguity} 
	\begin{align}
		\left\lvert\mathbf{f}_{k}^{H}\mathbf{g}_{p}(\mathbf{z}_{1}) - \mathbf{f}_{k}^{H}\mathbf{g}_{p}(\mathbf{z}_{2}) \right\rvert &\leq \sum_{n=0}^{N-1} \left\lvert\mathbf{z}_{1}[n]\overline{\mathbf{z}_{1}[n-p]} - \mathbf{z}_{2}[n]\overline{\mathbf{z}_{2}[n-p]}\right \rvert \leq N\left(\lVert \mathbf{z}_{1}\rVert_{2} + \lVert \mathbf{z}_{2}\rVert_{2}\right) \lVert \mathbf{z}_{1} - \mathbf{z}_{2}\rVert_{2},
		\label{eq:ine8}
	\end{align}
	where the second inequality again relies on the norm inequalities. 
	Combining \eqref{eq:ine7} and \eqref{eq:ine8} produces
	\begin{align}
		\left\lVert f_{k,p}(\mathbf{z}_{1}) - f_{k,p}(\mathbf{z}_{2}) \right\rVert_{2} \leq r_{k,p} \lVert \mathbf{z}_{1} - \mathbf{z}_{2} \rVert_{2},
		\label{eq:secondresult}
	\end{align}
	where the constant $r_{k,p} $ is 
	\begin{align}
		r_{k,p} &= \frac{N\sqrt{N}\lVert \mathbf{z}_{1} \rVert_{2}}{\mu}\left(M_{\varphi_{\mu}} +\sqrt{\mathbf{A}[p,k]}\right) \nonumber\\
		&+N\sqrt{N}\left(\lVert \mathbf{z}_{1}\rVert_{2} + \lVert \mathbf{z}_{2}\rVert_{2}\right)\lVert \mathbf{z}_{2}\rVert_{2}\left(\frac{\sqrt{\mathbf{A}[p,k]}M_{\varphi_{\mu}}}{\mu^{2}}+ 1\right) +N^{2}\sqrt{N}\left(\lVert \mathbf{z}_{1}\rVert_{2} + \lVert \mathbf{z}_{2}\rVert_{2}\right)\frac{\lVert \mathbf{z}_{2}\rVert^{2}_{2}\sqrt{\mathbf{A}[p,k]}}{\mu^{2}}.
		\label{eq:ine9}
	\end{align}
	Since the set $\mathcal{J}$ is bounded, then $\lVert \mathbf{z} \rVert_{2}<\infty$ for all $\mathbf{z}\in \mathcal{J}$. 
	It follows from the constant boundedness property of \textbf{P1} that $0<r_{k,p}<\infty$, and from \eqref{eq:secondresult} the result holds. 
	This yields the Lipschitz constant $U$ of the gradient in \eqref{eq:assumption} as $U=\sum_{k,p=0}^{N-1}r_{k,p} \propto \sum_{k,p=0}^{N-1} \sqrt{\mathbf{A}[p,k]}$. From the constant volume property of \textbf{P2}, $U$ is indeed a constant.  
    Since the value of $r_{k,p}$ in \eqref{eq:ine9} relies on the bound in \eqref{eq:ine6}, computed $U$ is valid for any trivial ambiguity \textbf{T1}-\textbf{T4} (which are also related to \textbf{P3} and \textbf{P4}).
	
	\item To prove \textbf{C4}, 
	observe that
	\begin{align}
		&\mathbb{E}_{\Gamma_{(r)}}\left[\left\lVert \mathbf{d}_{\Gamma_{(r)}} - \frac{\partial h(\mathbf{z},\mu)}{\partial \overline{\mathbf{z}}} \right\rVert^{2}_{2}\right] \leq \mathbb{E}_{\Gamma_{(r)}}\left[ 2\left\lVert \mathbf{d}_{\Gamma_{(r)}}\right\rVert^{2}_{2}\right] + 2\left\lVert\frac{\partial h(\mathbf{z},\mu)}{\partial \overline{\mathbf{z}}}\right\rVert^{2}_{2},
		\label{eq:final1}
	\end{align}
	where the inequality follows from 
	$\left\lVert \mathbf{w}_{1} + \mathbf{w}_{2} \right\rVert^{2}_{2}\leq 2 \left(\left\lVert \mathbf{w}_{1} \right\rVert^{2}_{2}+\left\lVert \mathbf{w}_{2} \right\rVert^{2}_{2}\right)$ for any $\mathbf{w}_{1},\mathbf{w}_{2}\in \mathbb{C}^{N}$. Combining \eqref{eq:assumption} and \eqref{eq:final1} yields
	\begin{align}
		\mathbb{E}_{\Gamma_{(r)}}\left[\left\lVert \mathbf{d}_{\Gamma_{(r)}} - \frac{\partial h(\mathbf{z},\mu)}{\partial \overline{\mathbf{z}}} \right\rVert^{2}_{2}\right] &\leq \mathbb{E}_{\Gamma_{(r)}}\left[ 2\left\lVert \mathbf{d}_{\Gamma_{(r)}}\right\rVert^{2}_{2}\right] + 2U\left\lVert\mathbf{z}\right\rVert^{2}_{2},
		\label{eq:final3}
	\end{align}
	for some $U>0$. Recall that $\Gamma_{(r)}$ is sampled uniformly at random from all subsets of $\{1,\cdots,N \}\times \{1\cdots,N\}$ with cardinality $Q$. From the definition of $\mathbf{d}_{\Gamma_{(r)}}$ in line 7 of Algorithm \ref{alg:smothing}, we conclude that 
	\begin{align}
		\mathbb{E}_{\Gamma_{(r)}}\left[ 2\left\lVert \mathbf{d}_{\Gamma_{(r)}}\right\rVert^{2}_{2}\right] &\leq \frac{4Q}{N^{2}}\sum_{p,k=0}^{N-1} \left\lVert f_{k,p}(\mathbf{z}) + g_{k,p}(\mathbf{z})\right\rVert^{2}_{2} \leq \frac{8Q}{N^{2}}\sum_{p,k=0}^{N-1} \left\lVert f_{k,p}(\mathbf{z})\right\rVert^{2}_{2} + \left\lVert g_{k,p}(\mathbf{z})\right\rVert^{2}_{2},
	\end{align}
	where we use the inequality $\left\lVert \mathbf{w}_{1} + \mathbf{w}_{2} \right\rVert^{2}_{2}\leq 2 \left(\left\lVert \mathbf{w}_{1} \right\rVert^{2}_{2}+\left\lVert \mathbf{w}_{2} \right\rVert^{2}_{2}\right)$ for any $\mathbf{w}_{1},\mathbf{w}_{2}\in \mathbb{C}^{N}$. 
	Since $f_{k,p}(\mathbf{z})$ and $g_{k,p}(\mathbf{z})$ satisfy \eqref{eq:functionf} and \eqref{eq:functiong}, respectively. Hence, for some constants $r_{k,p},s_{k,p}>0$,
	\begin{align}
		\mathbb{E}_{\Gamma_{(r)}}\left[ 2\left\lVert \mathbf{d}_{\Gamma_{(r)}}\right\rVert^{2}_{2}\right] \leq \frac{8Q\left\lVert\mathbf{z}\right\rVert^{2}_{2}}{N^{2}}\sum_{p,k=0}^{N-1} r^{2}_{k,p} + s^{2}_{k,p}.
		\label{eq:final2}
	\end{align}
	
	Thus, combining \eqref{eq:final3} and \eqref{eq:final2} yields
	\begin{align}
		\mathbb{E}_{\Gamma_{(r)}}\left[\left\lVert \mathbf{d}_{\Gamma_{(r)}} - \frac{\partial h(\mathbf{z},\mu)}{\partial \overline{\mathbf{z}}} \right\rVert^{2}_{2}\right] \leq \zeta^{2},
		\label{eq:final4}
	\end{align}
	where 
	\begin{align}
		\zeta = \left\lVert\mathbf{z}\right\rVert_{2}\sqrt{\frac{8Q}{N^{2}}\sum_{p,k=0}^{N-1} r^{2}_{k,p} + s^{2}_{k,p} + 2U}.
	\end{align}
	Note that $\zeta<\infty$ because the set $\mathcal{J}$ is bounded. This completes the proof. 
\end{itemize}
\end{proof}

\subsection{Proof of the Theorem}
Denote the set $\mathcal{K}_{1}:=\{t|\mu^{(r+1)}=\gamma_{1}\mu^{(r)} \}$, where $\gamma_{1}\in (0,1)$ is a tunable parameter \citep{zhang2009smoothing}. If the set $\mathcal{K}_{1}$ is finite, then it follows from the lines 7-8 of Algorithm \ref{alg:smothing} that there exists an integer $\grave{t}$ such that for all $t>\grave{t}$
\begin{align}
\left\lVert \mathbf{d}_{\Gamma_{(r)}} \right\rVert_{2}\geq \gamma \mu^{(\grave{t})},
\end{align}
with $\gamma\in (0,1)$. Taking $\grave{\mu}=\mu^{(\grave{t})}$, the optimization problem \eqref{eq:auxproblem} reduces to
\begin{align}
\underset{\mathbf{z}\in \mathbb{C}^{N}}{\textrm{minimize}}\; h(\mathbf{z},\grave{\mu}).
\label{eq:resultantOptiProblem}
\end{align}
Using the properties stated in Lemma \ref{lem:assumption}, it follows from \citep[Theorem 2.1]{ghadimi2013stochastic} that
\begin{align}
		\lim_{r\rightarrow \infty} \left\lVert\frac{\partial h(\mathbf{x}^{(r)},\mu^{(r)})}{\partial \overline{\mathbf{z}}}\right\rVert_{2}^{2}=\lim_{r\rightarrow \infty} \left\lVert\mathbb{E}_{\Gamma_{(r)}}\left[\mathbf{d}_{\Gamma_{(r)}}\right] \right\rVert_{2}^{2}\leq \lim_{r\rightarrow \infty} \left(\frac{2h(\mathbf{x}^{(0)},\mu^{(0)})}{t} + \left(\hat{D} + \frac{2h(\mathbf{x}^{(0)},\mu^{(0)})}{\hat{D}U}\right)\frac{\sigma}{\sqrt{t}}\right),
		\label{eq:conjugate1}
\end{align}
for some positive constants $\sigma,\hat{D}$. Clearly, having $\left\lVert\frac{\partial h(\mathbf{x}^{(r)},\mu^{(r)})}{\partial \overline{\mathbf{z}}}\right\rVert_{2}^{2}\rightarrow 0$ from \eqref{eq:conjugate1} contradicts the assumption $\left\lVert \mathbf{d}_{\Gamma_{(r)}} \right\rVert_{2}\geq \gamma \mu^{(\grave{t})}$, for all $t>\grave{t}$. This shows that $\mathcal{K}_{1}$ must be infinite and $\displaystyle\lim_{r\rightarrow \infty} \mu^{(r)} = 0$. 

Given that $\mathcal{K}_{1}$ is infinite, we deduce that
\begin{align}
		\lim_{r\rightarrow \infty} \left\lVert\mathbb{E}_{\Gamma_{(r)}}\left[\mathbf{d}_{\Gamma_{(r)}}\right] \right\rVert_{2}^{2} \leq \lim_{r\rightarrow \infty} \mathbb{E}_{\Gamma_{(r)}}\left[\left\lVert\mathbf{d}_{\Gamma_{(r)}}\right\rVert_{2}^{2}\right] \leq \gamma \lim_{r\rightarrow \infty} \mu^{(r)}\left(\frac{2h(\mathbf{x}^{(0)},\mu^{(0)})}{t} + \left(\hat{D} + \frac{2h(\mathbf{x}^{(0)},\mu^{(0)})}{\hat{D}U}\right)\frac{\sigma}{\sqrt{t}}\right)^{1/2},
		\label{eq:gradientMu4}
\end{align}
where the first inequality follows from Jensen's inequality. This completes the proof. 

\section{Proof of Theorem \ref{theo:initialization}}
\label{app:initia}
Note that the function as defined in \eqref{eq:problemInitiz1}, i.e. $$\mathbf{p}_{\ell}\rightarrow \lVert \mathbf{y}_{\ell}-\mathbf{G}^{(r-1)}_{\ell}\mathbf{p}_{\ell} \rVert_{2}^{2} + \frac{1}{2\lambda_{(r)}}\lVert \mathbf{p}_{\ell} - \mathbf{x}_{\ell}^{(r-1)} \rVert_{2}^{2}$$ is $\frac{1}{\lambda_{(r)}}$-strongly convex in $\mathbf{p}_{\ell}$. Here, we allow inaccuracy in the solutions. More precisely, if we define $\mathbf{x}_{\ell,*}^{(r)}$ to be the exact minimizer of \eqref{eq:problemInitiz1}, the standard optimality conditions for strongly convex minimization imply that there exists a sequence of additive accuracy parameters $\epsilon_{(r)}\geq 0$ such that the numerically attained solution $\mathbf{x}^{(r)}_{\ell}$ satisfies
\begin{align}
\lVert \mathbf{y}_{\ell}-\mathbf{G}^{(r-1)}_{\ell}\mathbf{x}_{\ell}^{(r)} \rVert_{2}^{2} + \frac{1}{2\lambda_{(r)}}\lVert \mathbf{x}^{(r)}_{\ell} - \mathbf{x}_{\ell}^{(r-1)} \rVert_{2}^{2} &\leq \lVert \mathbf{y}_{\ell}-\mathbf{G}^{(r-1)}_{\ell}\mathbf{x}_{\ell,*}^{(r)} \rVert_{2}^{2} + \frac{1}{2\lambda_{(r)}}\lVert \mathbf{x}_{\ell,*}^{(r)} - \mathbf{x}_{\ell}^{(r-1)} \rVert_{2}^{2} + \epsilon_{(r)} \nonumber\\
&\leq \lVert \mathbf{y}_{\ell}-\mathbf{G}^{(r-1)}_{\ell}\mathbf{x}_{\ell} \rVert_{2}^{2} + \frac{1}{2\lambda_{(r)}}\lVert \mathbf{x}_{\ell} - \mathbf{x}_{\ell}^{(r-1)} \rVert_{2}^{2} + \epsilon_{(r)},
\label{eq:solving1}
\end{align}
where $\mathbf{x}_{\ell}$ is the unknown desired signal and $\frac{1}{2\lambda_{(r)}}\lVert \mathbf{x}^{(r)}_{\ell}-\mathbf{x}_{\ell} \rVert\leq \epsilon_{(r)}$ \citep{duchi2019solving}. Then, from \eqref{eq:solving1},
\begin{align}
\lVert \mathbf{y}_{\ell}-\mathbf{G}^{(r-1)}_{\ell}\mathbf{x}_{\ell}^{(r)} \rVert_{2}^{2} &\leq \lVert \mathbf{y}_{\ell}-\mathbf{G}^{(r-1)}_{\ell}\mathbf{x}_{\ell} \rVert_{2}^{2} + \frac{1}{2\lambda_{(r)}}\lVert \mathbf{x}_{\ell} - \mathbf{x}_{\ell}^{(r-1)} \rVert_{2}^{2} + \epsilon_{(r)}.
\label{eq:solving2}
\end{align}

Recall that the radar PR problem has several ambiguities. In fact, the feasible set $\mathcal{A}_{\ell}$ for each $\ell$ is 
\begin{align}
\mathcal{A}_{\ell}= \{\mathbf{z}\in \mathbb{C}^{N} \hspace{0.2em}:\hspace{0.2em} \mathbf{z}[n]=e^{i\beta}e^{i b n}\mathbf{x}_{\ell}[\epsilon n - a] \text{ for } \beta,b\in \mathbb{R}, \text{ and }\epsilon=\pm 1, a\in \mathbb{Z} \}.
\end{align}
Thus, performing a parallel derivation with $\mathbf{q}^{(r-1)}_{\ell}\in \mathcal{A}_{\ell}$ replacing $\mathbf{x}_{\ell}^{(r-1)}$ and simplifying, we have from \eqref{eq:solving1} and \eqref{eq:solving2} that
\begin{align}
\lVert \mathbf{y}_{\ell}-\mathbf{G}^{(r-1)}_{\ell}\mathbf{x}_{\ell}^{(r)} \rVert_{2}^{2} &\leq \lVert \mathbf{y}_{\ell}-\mathbf{G}^{(r-1)}_{\ell}\mathbf{x}_{\ell} \rVert_{2}^{2} + \frac{1}{2\lambda_{(r)}}\lVert \mathbf{x}_{\ell} + \mathbf{q}_{\ell}^{(r-1)} \rVert_{2}^{2} + \epsilon_{(r)}.
\label{eq:solving3}
\end{align}
We rewrite the term at the left-hand-side of the inequality in \eqref{eq:solving3} as
\begin{align}
\lVert \mathbf{y}_{\ell}-\mathbf{G}^{(r-1)}_{\ell}\mathbf{x}_{\ell}^{(r)} \rVert_{2}^{2} &= \lVert \mathbf{y}_{\ell}-\mathbf{G}^{(r-1)}_{\ell}\mathbf{x}_{\ell}^{(r)} -\mathbf{G}_{\ell}^{(r-1)}\mathbf{x}_{\ell}+\mathbf{G}_{\ell}^{(r-1)}\mathbf{x}_{\ell}\rVert_{2}^{2} \nonumber\\
&=\lVert \mathbf{y}_{\ell}-\mathbf{G}^{(r-1)}_{\ell}\mathbf{x}_{\ell}\rVert^{2}_{2}+ \lVert \mathbf{G}^{(r-1)}_{\ell}\mathbf{z}_{\ell}^{(r)} -\mathbf{G}_{\ell}^{(r-1)}\mathbf{x}_{\ell}\rVert_{2}^{2} \nonumber\\
&\;\;\;\; - 2\mathcal{R}\left( \left(\mathbf{y}_{\ell}-\mathbf{G}^{(r-1)}_{\ell}\mathbf{x}_{\ell}\right)^{H}\left(\mathbf{G}^{(r-1)}_{\ell}\mathbf{z}_{\ell}^{(r)} -\mathbf{G}_{\ell}^{(r-1)}\mathbf{x}_{\ell}\right) \right).
\end{align}
From the above equality we conclude that
\begin{align}
	\lVert \mathbf{G}^{(r-1)}_{\ell}\mathbf{z}_{\ell}^{(r)} -\mathbf{G}_{\ell}^{(r-1)}\mathbf{x}_{\ell}\rVert_{2}^{2} &\leq \lVert \mathbf{y}_{\ell}-\mathbf{G}^{(r-1)}_{\ell}\mathbf{x}_{\ell}^{(r)} \rVert_{2}^{2} + 2\left\lvert \left(\mathbf{y}_{\ell}-\mathbf{G}^{(r-1)}_{\ell}\mathbf{x}_{\ell}\right)^{H}\left(\mathbf{G}^{(r-1)}_{\ell}\mathbf{x}_{\ell}^{(r)} -\mathbf{G}_{\ell}^{(r-1)}\mathbf{x}_{\ell}\right) \right\rvert.
\end{align}	
Note that $\rho_{1}^{(r-1)}\lVert \mathbf{x}_{\ell}^{(r)} - \mathbf{x}_{\ell} \rVert_{2}^{2}\leq \lVert \mathbf{G}^{(r-1)}_{\ell}\mathbf{z}_{\ell}^{(r)} -\mathbf{G}_{\ell}^{(r-1)}\mathbf{x}_{\ell}\rVert_{2}^{2}$, where $\rho_{1}^{(r-1)}$ is the squared smallest singular value of $\mathbf{G}_{\ell}^{(r-1)}$ greater than zero and $\rho_{2}^{(r)} = \sigma_{\textrm{max}}\left(\mathbf{G}_{\ell}-\mathbf{G}_{\ell}^{(r-1)} \right)\sigma_{\textrm{max}}\left(\mathbf{G}_{\ell}^{(r-1)} \right)$. Therefore,
\begin{align}
	 \rho_{1}^{(r-1)}\lVert \mathbf{x}_{\ell}^{(r)} - \mathbf{x}_{\ell} \rVert_{2}^{2} &\leq \lVert \mathbf{y}_{\ell}-\mathbf{G}^{(r-1)}_{\ell}\mathbf{x}_{\ell}^{(r)} \rVert_{2}^{2} + 2\rho_{2}^{(r)}\lVert \mathbf{x}_{\ell} \rVert_{2}\lVert \mathbf{x}_{\ell}^{(r)} - \mathbf{x}_{\ell} \rVert_{2}.
	 \label{eq:solving4}
\end{align}
Then, combining \eqref{eq:solving3} and \eqref{eq:solving4} yields
\begin{align}
\rho_{1}^{(r-1)}\lVert \mathbf{x}_{\ell}^{(r)} - \mathbf{x}_{\ell} \rVert_{2}^{2} &\leq \frac{1}{2\lambda_{(r)}}\lVert \mathbf{x}_{\ell} + \mathbf{q}_{\ell}^{(r-1)} \rVert_{2}^{2} +2\rho_{2}^{(r)}\lVert \mathbf{x}_{\ell} \rVert_{2}\lVert \mathbf{x}_{\ell}^{(r)} - \mathbf{x}_{\ell} \rVert_{2} + \lVert \mathbf{y}_{\ell}-\mathbf{G}^{(r-1)}_{\ell}\mathbf{x}_{\ell} \rVert_{2}^{2} + \epsilon_{(r)}.
\label{eq:solving5}
\end{align}

Considering Line 4 of Algorithm \ref{alg:initialization} we have that the construction of matrix $\mathbf{G}_{\ell}^{(r-1)}$ follows \eqref{eq:matrixG1}, therefore, bounding from above the term $\lVert\mathbf{G}_{\ell}-\mathbf{G}_{\ell}^{(r-1)} \rVert_{\mathcal{F}}^{2}$ gives us
\begin{align}
\lVert\mathbf{G}_{\ell}-\mathbf{G}_{\ell}^{(r-1)} \rVert_{\mathcal{F}}^{2} \leq 2N\lambda_{(r)}\epsilon_{(r-1)},
\label{eq:solving6}
\end{align}
because $\frac{1}{2\lambda_{(r-1)}}\lVert \mathbf{x}^{(r-1)}_{\ell}-\mathbf{x}_{\ell} \rVert\leq \epsilon_{(r-1)}$. Combining \eqref{eq:solving5} and \eqref{eq:solving6} produces
\begin{align}
\rho_{1}^{(r-1)}\lVert \mathbf{x}_{\ell}^{(r)} - \mathbf{x}_{\ell} \rVert_{2}^{2} &\leq \frac{1}{2\lambda_{(r)}}\lVert \mathbf{x}_{\ell} + \mathbf{q}_{\ell}^{(r-1)} \rVert_{2}^{2} +2\rho_{2}^{(r)}\lVert \mathbf{x}_{\ell} \rVert_{2}\lVert \mathbf{x}_{\ell}^{(r)} - \mathbf{x}_{\ell} \rVert_{2} + \lVert \mathbf{G}_{\ell}\mathbf{x}_{\ell}-\mathbf{G}^{(r-1)}_{\ell}\mathbf{x}_{\ell} \rVert_{2}^{2} + \epsilon_{(r)} \nonumber\\
&\leq \frac{1}{2\lambda_{(r)}}\lVert \mathbf{x}_{\ell} + \mathbf{q}_{\ell}^{(r-1)} \rVert_{2}^{2} +2\rho_{2}^{(r)}\lVert \mathbf{x}_{\ell} \rVert_{2}\lVert \mathbf{x}_{\ell}^{(r)} - \mathbf{x}_{\ell} \rVert_{2} + 2N\lambda_{(r)}\epsilon_{(r-1)}\lVert\mathbf{x}_{\ell} \rVert_{2}^{2} + \epsilon_{(r)} \nonumber\\
&\leq \frac{1}{2\lambda_{(r)}}\lVert \mathbf{x}_{\ell} + \mathbf{q}_{\ell}^{(r-1)} \rVert_{2}^{2} +4\rho_{2}^{(r)}\lVert \mathbf{x}_{\ell} \rVert_{2}\lambda_{(r)}\epsilon_{(r-1)} + 2N\lambda_{(r)}\epsilon_{(r-1)}\lVert\mathbf{x}_{\ell} \rVert_{2}^{2} + \epsilon_{(r)}.
\label{eq:solving7}
\end{align}
Note that the inequality \eqref{eq:solving7} is valid for all trivial ambiguities of $\mathbf{x}_{\ell}$ for the radar problem. In particular, taking the minimum over the ambiguities of $\mathbf{x}_{\ell}$ on both sides of \eqref{eq:solving7} implies
\begin{align}
\rho_{1}^{(r-1)}\lVert \mathbf{x}_{\ell}^{(r)} - \mathbf{x}_{\ell} \rVert_{2}^{2} &\leq \frac{1}{2\lambda_{(r)}} \text{dist}^{2}(\mathbf{x}_{\ell},\mathbf{x}_{\ell}^{(r-1)}) +4\rho_{2}^{(r)}\lVert \mathbf{x}_{\ell} \rVert_{2}\lambda_{(r)}\epsilon_{(r-1)} + 2N\lambda_{(r)}\epsilon_{(r-1)}\lVert\mathbf{x}_{\ell} \rVert_{2}^{2} + \epsilon_{(r)},
\label{eq:solving8}
\end{align}
following Definition \ref{def:dis}. Inductively applying the inequality \eqref{eq:solving8} when $\epsilon_{(r)}=0$ for all $t$ yields 
\begin{align}
\lVert \mathbf{x}_{\ell}^{(r)} - \mathbf{x}_{\ell} \rVert_{2}&\leq \frac{1}{\sqrt{2\lambda_{(r)}\rho_{1}^{(r-1)}}}\text{dist}(\mathbf{x}_{\ell},\mathbf{x}_{\ell}^{(r-1)}).
\end{align}
By definition, $\text{dist}(\mathbf{x}_{\ell},\mathbf{x}_{\ell}^{(r)})\leq \lVert \mathbf{x}_{\ell}^{(r)} - \mathbf{x}_{\ell} \rVert_{2}$. Hence,
\begin{align}
\text{dist}(\mathbf{x}_{\ell},\mathbf{x}_{\ell}^{(r)})&\leq \frac{1}{\sqrt{2\lambda_{(r)}\rho_{1}^{(r-1)}}}\text{dist}(\mathbf{x}_{\ell},\mathbf{x}_{\ell}^{(r-1)}),
\label{eq:solving9}
\end{align}
 wherein choosing $\lambda_{(r)}$ such that $\lambda_{(r)}\rho_{1}^{(r-1)}> 1/2$ guarantees a reduction in the estimation error of $\mathbf{x}_{\ell}$. Then, from \eqref{eq:solving9}, 
\begin{align}
\text{dist}(\mathbf{x}_{\ell},\mathbf{x}_{\ell}^{(r)}) < \tau \hspace{0.5em} \text{dist}(\mathbf{x}_{\ell},\mathbf{x}_{\ell}^{(0)}),
\label{eq:reductionInit}
\end{align}
where $\tau = \prod_{t} \frac{1}{\sqrt{2\lambda_{(r)}\rho_{1}^{(r-1)}}} < 1$. This completes the proof.

\section*{Data Availability Statement}
Data is available on request by contacting the authors. 

\section*{Acknowledgements}
K. V. M. acknowledges helpful discussions with Prof. Radu Balan of The University of Maryland, College Park regarding finite iteration algorithms mentioned in Section~\ref{sec:algorithm}.

\section*{Funding}
This research was sponsored by the Army Research Office/Laboratory under Grant Number W911NF-21-1-0099, and the VIE project entitled ``Dual blind deconvolution for joint radar-communications processing''. K. V. M. acknowledges support from the National Academies of Sciences, Engineering, and Medicine via the Army Research Laboratory Harry Diamond Distinguished Postdoctoral Fellowship. S. P. acknowledges support from the EMET Research Institute in Colombia. The research was sponsored by the Army Research Laboratory and was accomplished under Cooperative Agreement Number W911NF-21-2-0288. The views and conclusions contained in this document are those of the authors and should not be interpreted as representing the official policies, either expressed or implied, of the Army Research Laboratory or the U.S. Government. The U.S. Government is authorized to reproduce and distribute reprints for Government purposes notwithstanding any copyright notation herein.

\bibliographystyle{abbrvnat}
\bibliography{main}

\end{document}